%% file: bmc_article.tex
\documentclass[twocolumn,table]{bmcart}
\usepackage{amsthm,amsmath, amsfonts}
\usepackage{siunitx}
\usepackage{enumitem}
\usepackage{highlight}
\usepackage[utf8]{inputenc} 
\usepackage{graphicx}

\usepackage[caption=false,font=normalsize,labelfont=sf,textfont=sf]{subfig}
\usepackage{csquotes}
\usepackage{url}
\usepackage[normalem]{ulem}

\usepackage{amsmath,amsfonts}
\usepackage{algorithmic}
\usepackage{array}
\usepackage[caption=false,font=normalsize,labelfont=sf,textfont=sf]{subfig}
\usepackage{siunitx}
\usepackage{enumitem}
\usepackage{tikz}
   \usepackage{pgfplotstable}
    \usepackage{pgfplots}%
    \usepgfplotslibrary{units} 
    \pgfplotsset{compat=1.9}
    \makeatletter
    \pgfplotsset{
     unit code/.code 2 args=
       \begingroup
       \protected@edef\x{\endgroup\si{#2}}\x
    }
\usetikzlibrary{pgfplots.statistics, pgfplots.colorbrewer}

\DeclareMathOperator*{\argmin}{arg\,min}
\startlocaldefs
\endlocaldefs

\begin{document}

\begin{frontmatter}

\begin{fmbox}
\dochead{Research}


\title{Synthesis of Soundfields through Irregular Loudspeaker Arrays Based on Convolutional Neural Networks}


\author[
  corref={aff1}, 
  addressref={aff1},
  email={luca.comanducci@polimi.it}
]{\inits{L.C.}\fnm{Luca} \snm{Comanducci}}
\author[  email={massimiliano.zanoni@polimi.it}    
]{\inits{F.A.}\fnm{Fabio} \snm{Antonacci}}
\author[  email={augusto.sarti@polimi.it}    
]{\inits{A.S.}\fnm{Augusto} \snm{Sarti}}


\address[id=aff1]{%
  \orgdiv{Dipartimento di Elettronica, Infomazione e Bioignegneria (DEIB)}, 
  \orgname{Politecnico di Milano},
  \street{Via Ponzio 34/5},
  \postcode{20133},
  \city{Milano},                              
  \cny{Italy} 
}





\begin{abstractbox}

\begin{abstract} 
Most soundfield synthesis approaches deal with extensive and regular loudspeaker arrays, which are often not suitable for home audio systems, due to  physical space constraints.
In this article we propose a technique for soundfield synthesis through more easily deployable irregular loudspeaker arrays, i.e. where the spacing between loudspeakers is not constant, based on deep learning. The input are the driving signals obtained through a plane wave decomposition-based technique. While the considered driving signals are able to correctly reproduce the soundfield with a regular array, they show degraded performances when using irregular setups. Through a complex-valued Convolutional Neural Network (CNN) we modify the driving signals in order to compensate the errors in the reproduction of the desired soundfield. Since no ground-truth driving signals are available for the compensated ones, we train the model by calculating the loss between the desired soundfield at a number of control points and the one obtained through the driving signals estimated by the network. Numerical results show better reproduction accuracy with respect to the plane wave decomposition-based technique, pressure-matching approach and to linear optimizers for driving signal compensation.
\end{abstract}



\end{abstractbox}

\end{fmbox}

\end{frontmatter}



\section{Introduction}
Soundfield synthesis methods deal with the objective of reproducing a desired pressure field in a target region of space through arrays made of loudspeakers. In recent years the attention towards this field of research has consistently increased due to its potential application in virtual reality, telepresence and gaming.

The first approaches towards soundfield synthesis dealt with extensive loudspeaker setups, driven in order to effectively reproduce an accurate approximation of the desired soundfield. Wave Field Synthesis (WFS)~\cite{berkhout1993acoustic,spors2008theory} is based on the Huygens-Fresnel principle and synthesizes a desired pressure field through a large number of regularly distributed loudspeakers. Ambisonics~\cite{gerzon1973periphony} is based on the analysis of the soundfield in terms of spherical harmonics and reproduces the desired pressure field in a small listening area. In order to enlarge the area where reproduction is accurate, Higher Order Ambisonics (HOA) was introduced~\cite{ward2001reproduction,poletti2005three}. These physically-based approaches reproduce the soundfield with a satisfying quality when regular array geometries are used, such as spherical~\cite{poletti2010sound,kentgens2020translation}, linear~\cite{ahrens2010sound} or circular~\cite{chen20183d}. However their performances severely degrade when using irregular setups. While several techniques were proposed in order to adapt HOA techniques to irregular array setups~\cite{trevino2010high,zotter2013comparison} such as projection decoding methods~\cite{qu2018matching,ge2021partially} and~\cite{zotter2012all} All-round ambisonic panning and decoding(AllRAD) they often require the solution of ill-posed problems.

Optimization-based techniques are more easily applicable to irregular loudspeaker setups. The pressure-matching method~\cite{nelson1994active,gauthier2005sound} is based on the minimization of the reproduction error at a fixed number of positions in the listening area, denoted as control points. The desired driving signals are then obtained through a regularized least squares optimization problem. While this approach is applicable to setups having extremely irregular geometries, the achievable reproduction quality is strongly dependent on the selection of the control points, i.e. by sampling the listening area with a fine grid. Its computational cost, however, increases with the number of selected control points. Mode-matching ~\cite{Samarasinghe2013,betlehem2005theory,ueno2019three} is another optimization-based family of techniques that can be applied to loudspeaker setups having arbitrary geometries. In this case the optimization procedure is based on matching a modal decomposition of the desired soundfield at a single control point. Modal decomposition can be operated using circular or spherical wavefunctions. In doing this, it is needed to limit the decomposition to a maximum mode order, since a too high or small number leads to worse synthesis quality~\cite{ueno2019three}. Several approaches have been proposed to appropriately weight the modes~\cite{betlehem2005theory,ueno2018sound}. Irregular loudspeaker setups have also been considered by intensity-matching methods~\cite{zuo2020intensity,zuo20213d}, where the objective is the minimization of the sound intensity, i.e. particle velocity, in the spherical harmonic domain over a spatial region.

More recently, after its widespread adoption in acoustic signal processing research~\cite{bianco2019machine}, deep learning has also been applied to soundfield synthesis problems~\cite{cobos2022overview} such as the reconstruction of the pressure field at unknown locations~\cite{lluis2020sound,kristoffersen2021deep}.
In~\cite{morgado2018self} the authors proposed a network that is able to convert mono audio recorded using a $360^{\circ}$ video camera into First-Order Ambisonics (FOA). In~\cite{routray2019deep} a network is proposed in order to upscale Ambisonic signals, while in~\cite{gao2022sparse} a learning-based model for frequency  expanding of the Higher Order Ambisonics (HOA) encoding process is presented. Also, in~\cite{zhang2021estimation} the authors propose a technique for the estimation of spherical harmonic coefficients in soundfield recording, using feed-forward neural networks. Finally, in~\cite{Abhayapala2019} the authors present a neural network that is able to calculate the optimal number of driving signals, extracted through a LASSO-based technique. Learning techniques have also been applied to the problem of optimizing the number and placement of sensors in soundfield control scenarios~\cite{koyama2020optimizing}.

Complex-valued neural networks~\cite{lee2022complex,bassey2021survey,trabelsi2018deep,hirose2012complex,yang2020complex} enable to directly treat complex data and have recently been applied to a variety of audio signal processing tasks such as source localization~\cite{HirofumiTSUZUKI2013} and separation~\cite{lee2017fully}. The adoption of such networks enables us to directly treat complex data instead of handling separately the real and imaginary parts such as in~\cite{kristoffersen2021deep}.

In this manuscript, we propose a technique for 2D soundfield synthesis through irregular loudspeaker setups in a free field environment, where the desired driving signals are obtained through a complex-valued Convolutional Neural Network (CNN). Although the proposed method is easily extensible to 3D scenarios, this would involve dealing with 3D CNNs, which would add an increased complexity the computational point of view without enhancing the conceptual reasoning behind the proposed method. For this reason, in this manuscript we decided to focus on 2D deployments and to leave the 3D extension to future works.

Instead of deriving the driving signals from soundfield measurements, the target field is obtained from the Model-based Rendering (MR) method presented in~\cite{bianchi2016model}, based on the plane wave decomposition. While this technique is able to correctly reproduce the soundfield when regular loudspeaker setups are used, irregularities in the reproduced wavefronts appear when the spacing between the loudspeakers becomes uneven. 

Operatively, we generate irregular loudspeaker arrays, by considering regular array setups and randomly removing a number of loudpeakers, simulating configurations where more than half of the loudspeakers are missing, thus paving the way to the use of minimal setups. Through~\cite{bianchi2016model} we compute the driving signals obtained using the irregular setup and feed them into a CNN, giving as output a compensated version of the driving signals.
Differently from what proposed in~\cite{Abhayapala2019} the loss is not based on the  driving signals. Instead, we compute the loss between the ground truth soundfield and the one obtained through the compensated driving signals, which are the output of the network. 

The main contribution of this paper thus, is to provide a first, to the best of our knowledge, application of deep learning to soundfield synthesis when dealing with irregular loudspeaker setups. Such configurations are highly desirable in real world application scenarios, since are more easily deployable in contexts such as home audio. The choice of removing loudspeakers from regular circular and linear setups also goes in this direction, for example a fully regular circular loudspeaker could hardly be deployed in a living room due to the presence of furniture, while the proposed irregularities in the setup could instead accomodate these situations, by removing loudspeakers wherever needed.

In the literature, linear optimizers for loudspeaker driving functions have already been proposed such as Adaptive WaveField  Synthesis (AWFS)~\cite{gauthier2006adaptive,gauthier2007adaptive,gauthier2008adaptivesigproc,gauthier2008adaptiveexp}, where the reproduction error is minimized in a least-mean squares sense. In order to demonstrate the effectiveness of the technique we compare it with AWFS, PM and to a linearly compensated MR both when using simulated and real data.

The rest of this manuscript is organized as follows. In Section~\ref{sec:ii-background} we introduce the notation and present the necessary background related to the $\mathrm{MR}$ and $\mathrm{PM}$ techniques. In Section~\ref{sec:iii-method} we describe the proposed technique for soundfield synthesis using irregular louspeaker arrays. In Section~\ref{sec:iv-results} we present simulation results both when considering a circular and linear loudspeaker array. Finally, in Section~\ref{sec:v-conclusions} we draw some conclusions.

\input{ii-background.tex}

\input{iii-method.tex}

\input{iv-results.tex}

\section{Conclusion}
\label{sec:v-conclusions}
In this manuscript we have proposed a technique for 
soundfield synthesis using irregular loudspeaker arrays. The methodology is based on a deep learning-based approach. More specifically, we consider the driving signals obtained through an already existing soundfield method, based on the plane wave decomposition, and propose a network that is able to modify the driving signals by compensating the errors in the reproduced soundfield due to the irregularity in the loudspeaker setup.
We compare the proposed method with the one used to compute the input driving signals and with pressure-matching, showing that the proposed model is able to obtain better performances in most of the setups.

The obtained results open the possibility of adopting the combination of deep learning and model-based soundfield synthesis for addressing issues arising when irregular loudspeaker arrays are available. For example, a CNN-based pressure matching technique can be devised, by optmizing the driving signals from the knowledge of the soundfield at prescribed control points. Moreover we plan to move to real environments, where noise and reverberation are present, aiming at compensating the environment and mask the noise. Further developments could also entail the application of deep learning and irregular arrays to related problems such as multizone soundfield reproduction in order to create personal audio systems and also conditioning the system in order to be independent of the chosen array setup.


\begin{backmatter}

\section*{Abbreviations}

\end{backmatter}

\section*{Declarations}
\begin{backmatter}
\section*{Availability of data and materials}
The datasets used and/or analysed during the current study are available from the corresponding author on reasonable request. The code used to perform the experiments is fully available at \url{https://github.com/polimi-ispl/deep_learning_soundfield_synthesis_irregular_array}.

\section*{Competing interests}
The authors declare that they have no competing interests.

\section*{Funding}
Not Applicable.

\section*{Authors' contributions}
LC: conceptualization, code implementation, results computation, main writing. FA: conceptualization, writing, reverarch oversee . AS: research oversee and manuscript review.
All authors read and agreed to the submitted version of the manuscript.

\section*{Acknowledgements}
Not applicable.

\section*{Ethics approval and consent to participate}
The authors approve and consent to participate.

\section*{Consent for publication}
The authors consent for publication.



\bibliographystyle{bmc-mathphys} 
\bibliography{bmc_article.bib}      







\section*{Additional Files}
  \subsection*{Additional file 1 --- Sample additional file title}
    Additional file descriptions text (including details of how to
    view the file, if it is in a non-standard format or the file extension).  This might
    refer to a multi-page table or a figure.

  \subsection*{Additional file 2 --- Sample additional file title}
    Additional file descriptions text.

\end{backmatter}
\end{document}

%% file: ii-background.tex
\section{Notation and review of pressure-matching, model-based soundfield synthesis and adaptive wavefield synthesis}
\label{sec:ii-background}
In this section we briefly review three soundfield synthesis techniques related to the proposed approach and we introduce the notation that will be used throughout the rest of the paper. We first introduce the pressure-matching technique and then the model-based soundfield synthesis method, which is used in order to derive the loudspeaker driving signals, that will then be compensated through the proposed method. Finally, we present the adaptive wavefield synthesis technique, which optimizes the WFS driving signals through a linear procedure and will be used in order to compare the performances of the proposed approach.
\subsection{Notation and preliminaries}

Let us consider an arrangement of $L$ loudspeakers, or secondary sources, as often denoted in the soundfield synthesis literature, deployed at positions $\mathbf{r}_l, l=1,\ldots, L$. Let us also consider a set of $A$ points $\mathbf{r}_a, a=1, \ldots, A$ through which we sample the region of the space $\mathcal{A}$, denoted as listening area, where we want to reproduce the soundfield. Let $\mathbf{d}(\omega) =[d_1(\omega), \ldots, d_L(\omega)]^T $ denote the vector containing the driving signals applied to the secondary sources, where $\omega$ is the angular frequency and the superscript $T$ is the transposition. If $g(\mathbf{r}_{a}|\mathbf{r}_l, \omega)$ is the Acoustic Transfer Function (ATF) between secondary source $l$ and point $a$, the vector $\mathbf{g}_a = [g(\mathbf{r}_{a}|\mathbf{r}_1, \omega),\ldots,g(\mathbf{r}_{a}|\mathbf{r}_L, \omega)]^T $ is the juxtaposition of all the ATFs from the secondary sources to the listening point $a$. The synthesized sound pressure can be computed as 
\begin{equation}
   \hat{\mathbf{p}}(\mathbf{r}_a,\omega)= \mathbf{d}^T(\omega) \mathbf{g}_a(\omega)=\sum_{l=1}^{L} d_l(\omega)g(\mathbf{r}_{a}|\mathbf{r}_l, \omega), 
\end{equation}
where in the case of 2D propagation in free space conditions and using the  $e^{j\omega t}$ convention for the Fourier's Transform, $g(\cdot)$ corresponds to the Green's function~\cite{williams1999fourier}
\begin{equation}
\label{eq:green_function}
    g(\mathbf{r}_{a}|\mathbf{r}_l, \omega)= - \frac{j}{4} H_{0}^{(2)} \left(\frac{\omega}{c}\left||\mathbf{r}_a-\mathbf{r}_l\right||\right),
\end{equation}
where $H_0^{(2)}$ is the Hankel function of second kind and  zero order, while $c$ is the speed of sound in air.

The objective of soundfield synthesis techniques can then be defined as retrieving the set of driving signals $\mathbf{d}$ such that 
\begin{equation}
    \argmin_{\mathbf{d}} |\mathbf{p}(\mathbf{r}_a,\omega)-\hat{\mathbf{p}}(\mathbf{r}_a,\omega)|^2, 
\end{equation}
that is, minimizing the error between the reproduced and desired pressure field at the points contained in the listening area. The method through which the driving signals are estimated is what differentiates the various soundfield synthesis techniques.

\subsection{Pressure-matching method}
The pressure-matching technique, formulated as in~\cite{nelson1994active}, is a method for the synthesis of soundfields based on the minimization of the reproduction error at discrete points in the environment, denoted as control points. 

Let us consider a series of control points $\mathbf{r}_i, i=1,\ldots, I$ such that $\mathbf{r}_i \in \mathcal{A}$. In the following, the subscript $\mathrm{cp}$ will indicate that the related term refers only to values measured at the control points. The driving signals to be applied to the secondary sources are obtained by solving the minimization problem
\begin{equation}
    \begin{split}
    \label{eq:pm_formulation}
    \mathbf{d}_{\mathrm{pm}}(\omega)= \arg \min_{\mathbf{d}_{\mathrm{pm}}}&\biggl|\sum_{i=0}^{I-1}\hat{p}_{\mathrm{pm}}(\mathbf{r}_{i},\omega)-p(\mathbf{r}_{i},\omega)\biggr|^2 +\\& \lambda \mathbf{d}_{\mathrm{pm}}^H(\omega)\mathbf{d}_{\mathrm{pm}}(\omega),
\end{split}
\end{equation}
where $\lambda$ is a regularization parameter and $H$ denotes the Hermitian transpose. The solution of~\eqref{eq:pm_formulation} is given by
\begin{equation}
\label{eq:pm_driving_signals}
\mathbf{d}_{\mathrm{pm}}(\omega) = \left(\mathbf{G_{\mathrm{cp}}}^H(\omega)\mathbf{G}_{\mathrm{cp}}(\omega) + \lambda\mathbf{I}_L\right)^{-1} \mathbf{G}_{\mathrm{cp}}^H(\omega) \mathbf{p}_{\mathrm{cp}}(\omega),
\end{equation}
where the entries of  $\mathbf{G}_{\mathrm{cp}}(\omega) \in \mathbb{C}^{I \times L}$, corresponding to the transfer function between secondary sources $\mathbf{r}_a$ and control points $\mathbf{r}_i$ are defined as
\begin{equation}
    (\mathbf{G}_{\mathrm{cp}}(\omega))_{i,l} = g(\mathbf{r}_i|\mathbf{r}_l,\omega),
\end{equation}
and $\mathbf{p}_{\mathrm{cp}} \in \mathbf{C}^{I}$ is a vector corresponding to the ground truth pressure soundfield evaluated at the control points, i.e. $\mathbf{p}_{\mathrm{cp}}(\omega)=[p(\mathbf{r}_{i},\omega), \ldots, p(\mathbf{r}_{I},\omega)]^T$.

While the inversion of a matrix may be computationally expensive, if we consider a single set of secondary sources (i.e. a single loudspeaker array), the pressure-matching technique can be implemented with a more convenient linear computational cost  $\mathcal{O}(IL)$ by rewriting~\eqref{eq:pm_driving_signals} as 
\begin{equation}
    \mathbf{d}_{\mathrm{pm}}(\omega) = \mathbf{C}_{\mathrm{cp}}(\omega)\mathbf{p}_{\mathrm{cp}}(\omega),
\end{equation}
where 
\begin{equation}
    \mathbf{C}_{\mathrm{cp}}(\omega) = \left(\mathbf{G}_{\mathrm{cp}}^H(\omega)\mathbf{G}_{\mathrm{cp}}(\omega) + \lambda\mathbf{I}_L\right)^{-1} \mathbf{G}_{\mathrm{cp}}^H(\omega),
\end{equation}
and $\mathbf{C}_{\mathrm{cp}}(\omega) \in \mathbb{C}^{L\times I}$ is independent on the soundfield.
\subsection{Model-based acoustic rendering based on plane wave decomposition}
\label{sec:model_based_sfs}
The Model-based acoustic Rendering ($\mathrm{MR}$)~\cite{bianchi2016model} technique is based on the decomposition of the soundfield into directional contributions encoded by the Herglotz density function~\cite{colton1998inverse}, which can be converted into driving signals for arbitrary loudspeaker arrangements.

We first summarize how the Herglotz Density function is defined in the case of a point source and then how it has been used in \cite{bianchi2016model} to render the soundfield through circular and linear loudspeaker arrays.
\subsubsection{Herglotz Density Function}
Let us denote as $\mathbf{k}(\theta)$ the wave vector of a plane-wave with direction $\theta$, its norm is defined as $k=||\mathbf{k}(\theta)||\omega/c$ and the corresponding wavenumber as $\hat{\mathbf{k}}(\theta)=[\cos{\theta} \sin{\theta}]^T$.
The pressure soundfield at a point $\mathbf{r} = [x,y]^T$ can be modeled as a superposition of plane waves~\cite{zotkin2009plane, whittaker1903partial}
\begin{equation}
    p(\mathbf{r},\omega) = \frac{1}{2\pi}\int_{0}^{2\pi} e^{j \frac{\omega}{c}(x\cos\theta + y\sin\theta)}\varphi(\theta,\omega)d\theta,
\end{equation}

where $\varphi(\theta,\omega) \in \mathbb{C}$ is the Herglotz density function and it is a function modulating each plane wave component in amplitude and phase~\cite{colton1998inverse}. In the case of an isotropic point source $\mathbf{r}=\rho_z[\cos(\theta_z),\sin(\theta_z)]$, expressed in terms of polar coordinates $\rho_z$ and $\theta_z$, corresponding to radius and azimuth, respectively, $\varphi(\theta,\omega)$ can be defined as~\cite{bianchi2016model}
\begin{equation}
    \label{eq:herglotz_point} 
    \varphi(\theta,\omega) = A(\omega)\sum_{m=-\infty}^{+\infty}j^{-m}\frac{j}{4}H_{m}^{(2)}(\frac{\omega}{c}r)e^{jm(\theta-\theta_z)},
\end{equation}
where $A(\omega)$ is the spectrum of the sound emitted by the source.

\subsubsection{Implementation with circular arrays}
Let us consider a circular array of secondary sources deployed at positions $\mathbf{r}_l$, corresponding to polar coordinates $ \rho_l[\cos{\theta_l} \sin{\theta_l}]^T$, where $\rho_l$ is the radius. Let us also consider a discrete distribution of $N(\omega)$ plane waves with directions $\theta_n, n=1,\ldots,N$, uniformly sampling the $[0, 2\pi)$ interval, where each plane wave is reproduced by the same $L$ loudspeakers, in order to approximate the desired soundfield. We take advantage of the discrete plane wave distribution in order to reproduce the  soundfield by approximating it as~\cite{bianchi2016model}
\begin{equation}
   \hat{p}(\mathbf{r},\omega) = \frac{1}{N} \sum_{n=1}^{N} \varphi(\theta_n,\omega) e^{j \frac{\omega}{c}<\mathbf{r},\hat{\mathbf{k}}(\theta_n)>}.
\end{equation}

The sum in \eqref{eq:herglotz_point} is approximated through a truncation of the modal expansion to order $M$, i.e. ($m=-M,\ldots,M$) where $M$ can be chosen in order to bound the reproduction error in a listening area of radius $\rho$ by selecting $M \geq \lceil e \frac{\omega}{c} \frac{\rho}{2} \rceil$~\cite{zotkin2009plane}. Then according to Shannon's theorem, we can correctly reproduce the soundfield without additional errors, except for the ones due to the discretization, by using $N \geq2M+1$ plane waves.

The filter corresponding to the $l$-th loudspeaker and the $n$-th plane-wave component, can then be defined as~\cite{bianchi2016model}
\begin{equation}
    h_l(\theta_n, \omega) = \frac{4}{jL} \sum_{m=-M}^{M} \frac{e^{jm(\theta_l-\theta_n)}}{H_{m}^{(2)}(\frac{\omega}{c} \rho_l)}.
\end{equation}
The driving signal corresponding to the secondary source $l$ rendering all the $N$ plane-wave components is~\cite{bianchi2016model}
\begin{equation}
    d_{\mathrm{mr},l}(\omega) =\frac{1}{N}\sum_{n=1}^{N}\varphi(\theta_n,\omega)h_l(\theta_n, \omega).
\end{equation}
Finally, the soundfield at $\mathbf{r}_a$ is
\begin{equation}
\label{eq:mr_soundfield}
    \hat{p}_{\mathrm{mr}}(\mathbf{r}_a,\omega) = 
    \sum_{l=1}^{L}d_{\mathrm{mr},l}( \omega) g(\mathbf{r}_a|\mathbf{r}_l,\omega).  
\end{equation}

\subsubsection{Implementation with linear arrays}
Let us now consider an array of secondary sources deployed on a line segment such that $\mathbf{r}_l=[x_0,-y_0\leq y_0]^T$. In this case the allowed values for the reproduced plane wave directions belong to a subset of $[0, 2\pi)$ and specifically the allowed range is $\theta \in {\mathbf{R}|\theta_\text{min}\leq \theta \leq \theta_\text{max}}$, where
$\theta_\text{min}=\arctan(-y_0,x_0)$ and $\theta_\text{max}=\arctan(y_0,x_0)$. This angular interval is sampled using $N$ components. This limitation is due to the  geometrical constraints posed by the configuration of the array and of the listening region. Reproduction is performed towards the half-plane given by $x < x_0$~\cite{ahrens2010sound} and the linear array is not able to accomodate all the plane wave directions surrounding the listening region, as in the circular array case.
Since no closed-form solutions are known for arrays that are not circular~\cite{bianchi2016model}, the filter to be applied to the loudspeakers signals are estimated by minimizing the error due to the approximation of plane wave soundfield through secondary sources, that is~\cite{bianchi2016model}
\begin{equation}
\begin{split}
    \label{eq:driving_sig_linear}
    h_l(\theta_n,\omega) = \argmin_{h_l}|\sum_{i=1}^{I} e^{j\frac{\omega}{c}<\mathbf{r}_i,\hat{\mathbf{k}}(\theta_n)>}-\\h_l(\theta_n, \omega)g_l(\mathbf{r}_i|\mathbf{r}_l,\omega)|^2,
    \end{split}
\end{equation}
which yields~\cite{bianchi2016model}
\begin{equation}
    \label{eq:filters_linear_mr}
    h_l(\theta_n, \omega) = \left(\mathbf{G}_{\mathrm{cp}}^{H}(\omega)\mathbf{G}_{\mathrm{cp}}(\omega) + \lambda \mathbf{I}_L\right)^{-1}\mathbf{p}_{\mathrm{cp},\mathrm{pwd}}(\omega,\theta_n),
\end{equation}
where $\mathbf{p}_{\mathrm{cp},\mathrm{pwd}}(\theta_n,\omega)=[e^{j\frac{\omega}{c}<\mathbf{r}_i,\hat{\mathbf{k}}(\theta_n)>},\ldots, e^{j\frac{\omega}{c}<\mathbf{r}_I,\hat{\mathbf{k}}(\theta_n)>}]^T$ is a vector containing the pressure soundfield at the control points, due to a plane wave with direction $\theta_n$.

We can then derive the driving signals in the case of the linear array as~\cite{bianchi2016model}
\begin{equation}
    \mathbf{d}_{\mathrm{mr},l}(\omega) = \frac{\theta_\text{max}-\theta_\text{min}}{2\pi N}\sum_{n=1}^{N}\varphi(\theta_n, \omega)h_l(\theta_n, \omega),
\end{equation}
and then the desired soundfield can be obtained by inserting the derived driving signals into \eqref{eq:mr_soundfield}.

\subsection{Adaptive Wavefield Synthesis}

WaveField Synthesis (WFS)~\cite{berkhout1993acoustic} is a soundfield reproduction technique which assumes free-field reproduction and whose driving signals are derived from the Kirchoff-Helmholtz integral theorem.

Let us consider a 2D free-field environment. The WFS driving signals needed to reproduce a source placed in $\mathbf{r}_s$
can be derived as~\cite{gauthier2006adaptive}
\begin{equation}
\begin{split}
        d_\mathrm{WFS}(\mathbf{r}_l,\omega) =& \frac{4\pi}{\omega\rho} A(\omega)j\sqrt{\frac{jk}{2\pi}}\cos \Psi \frac{e^{jk||\mathbf{r}_s - \mathbf{r}_l||}}{\sqrt{||\mathbf{r}_s -\mathbf{r}_l||}} \\ &\times \sqrt{\frac{||\mathbf{r}_o - \mathbf{r}_l||} {||\mathbf{r}_o -\mathbf{r}_l||+||\mathbf{r}_s -\mathbf{r}_l||}} \Delta_l, 
\end{split}
\end{equation}
where $\rho$ denotes the air density, $\Psi$ the angle between $\mathbf{r}_s$ and the normal to the reproduction line (i.e. contour comprising the loudspeaker array) at the secondary source $\mathbf{r}_l$, $\mathbf{r}_o$ denotes a point on the \textit{reference line}, along which the amplitude error should theoretically be zero~\cite{verheijen1997sound} and finally, $\Delta_l = ||\mathbf{r}_l - \mathbf{r}_{l+1}||$ denotes the spacing between consecutive loudspeakers.

In order to solve the reproduction inaccuracies due to the WFS free-field assumption, in~\cite{gauthier2006adaptive} it was proposed a compensation technique for WFS driving signals, denoted Adaptive Wave Field Synthesis ($\mathrm{AWFS}$). Let us consider the soundfield $\mathbf{p}_\mathrm{cp,wfs}(\omega)$ obtained by reproducing at control points through the WFS driving signals and  $\mathbf{e}_\mathrm{cp}(\omega) = \mathbf{p}_\mathrm{cp}(\omega) - \mathbf{p}_\mathrm{cp,wfs}(\omega)$ as the reproduction error, then the $\mathbf{d}_\mathrm{awfs}(\omega) \in \mathbb{C}^L$ driving signals are obtained in $\mathrm{AWFS}$ by by solving the following minimization problem~\cite{gauthier2006adaptive}
\begin{equation}
\begin{split}
    \argmin_{\mathbf{d}_\mathrm{awfs}}~ &\mathbf{e}(\omega)^{H}\mathbf{e} + \\ &\lambda(\mathbf{d}_\mathrm{awfs}(\omega)-\mathbf{d}_\mathrm{wfs}(\omega))^{H} (\mathbf{d}_\mathrm{awfs}(\omega)-\mathbf{d}_\mathrm{wfs}(\omega)),
    \end{split}
\end{equation}
where $\mathbf{e} = \mathbf{p}-\hat{\mathbf{p}}_{\mathrm{wfs}}$ is the difference between the ground truth soundfield and estimated complex soundfields, $\lambda$ is a regularization parameter.

The adapted wave-field synthesis driving signals that minimize the cost function are then found  through~\cite{gauthier2006adaptive,nelson1991active}
\begin{equation}
\label{eq:awfs_inversion}
    \mathbf{d}_\mathrm{awfs}=[\mathbf{G}_{\mathrm{cp}}^{H}\mathbf{G}_{\mathrm{cp}} + \lambda \mathbf{I}]^{-1} [\mathbf{G}_{\mathrm{cp}}^{H}\mathbf{e}_{\mathrm{cp}}] + \mathbf{d}_\mathrm{wfs},
\end{equation}
where the solution is equivalent to the WFS one for $\lambda \rightarrow \inf$ and to the optimal solution in a least-mean-square sense for $\lambda \rightarrow 0$.

%% file: iii-method.tex
\section{Driving-signals compensation through complex-valued convolutional neural networks}
\label{sec:iii-method}
\begin{figure}[!t]
\centering
\subfloat[]{\resizebox{.35\columnwidth}{!}{%
\input{figures/setup/setup_circular.tex}}%
\label{subfig:setup_circular}}
\hfil
\subfloat[]{\resizebox{.35\columnwidth}{!}{%
\input{figures/setup/setup_circular_irregular.tex}}%
\label{subfig:setup_circular_irregular}}
\vfil
\subfloat[]{\resizebox{.35\columnwidth}{!}{%
\input{figures/setup/setup_linear.tex}}%
\label{subfig:setup_linear}}
\hfil
\subfloat[]{\resizebox{.35\columnwidth}{!}{%
\input{figures/setup/setup_linear_irregular.tex}}%
\label{subfig:setup_linear_irregular}}
\caption{Examples of regular circular~\protect\subref{subfig:setup_circular} and linear~\protect\subref{subfig:setup_linear} array setups, examples of irregular circular~\protect\subref{subfig:setup_circular_irregular} and linear~\protect\subref{subfig:setup_linear_irregular} array setups. }
\label{fig:setup}
\end{figure}

In this section we present the proposed technique for soundfield synthesis through complex-valued CNNs using irregular loudspeaker arrays. We first formalize the problem as the compensation of the filters obtained through the MR technique, then we describe the general pipeline of the method and the proposed network architecture.
\subsection{Problem Formulation}
Let us consider a circular or linear array of secondary sources as shown in Fig.~\ref{fig:setup}\protect\subref{subfig:setup_circular} and Fig.~\ref{fig:setup}\protect\subref{subfig:setup_linear}, respectively. An irregular loudspeaker array setup is obtained by removing some secondary sources from the setup, as shown in Fig.~\ref{fig:setup}\protect\subref{subfig:setup_circular_irregular} and Fig.~\ref{fig:setup}\protect\subref{subfig:setup_linear_irregular}. More formally, we can define an irregular loudspeaker array as an array where the spacing between the secondary sources is not constant.
\begin{figure}[!t]
\centering
\subfloat[]{%
\includegraphics[width=.45\columnwidth]{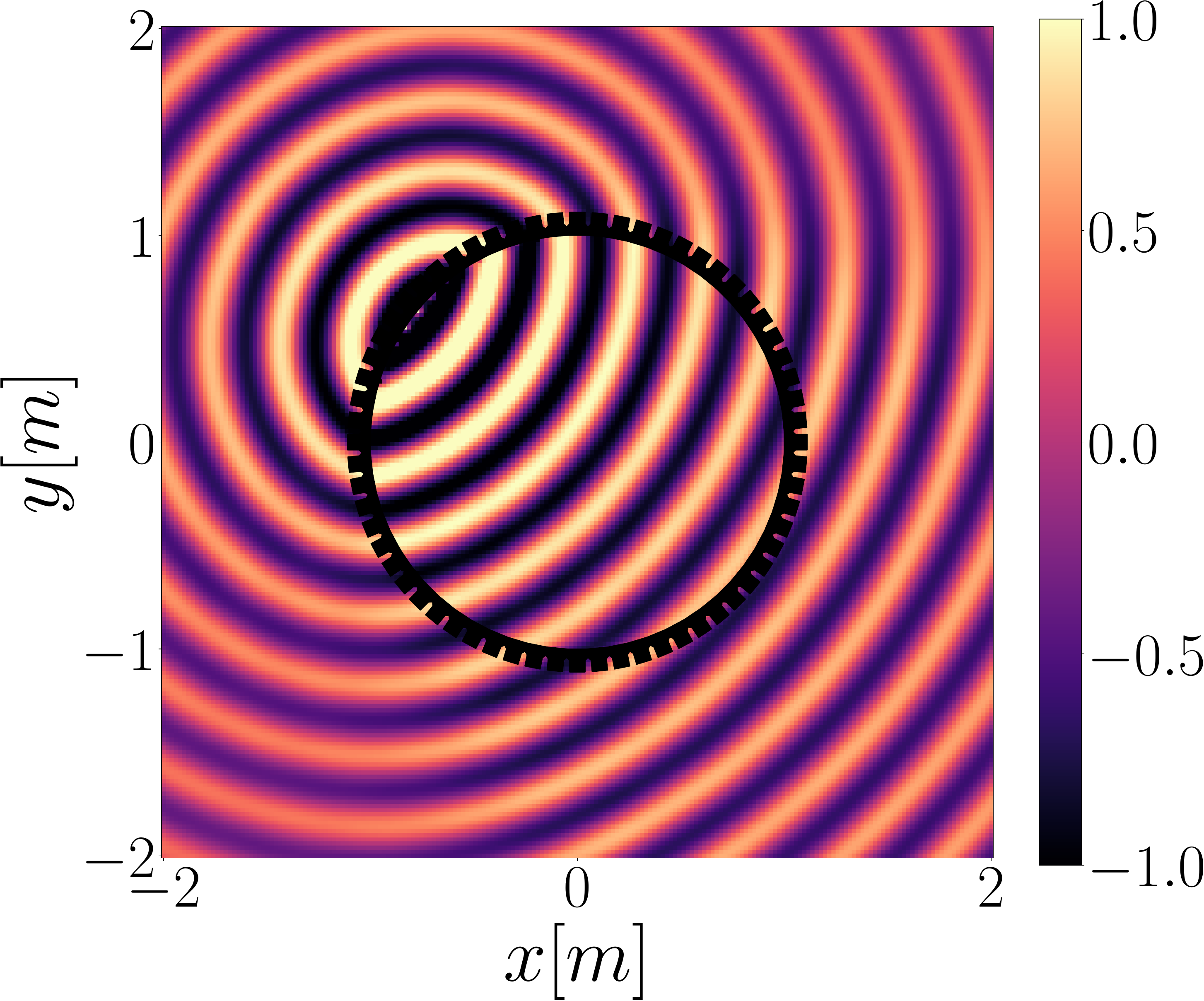}%
\label{subfig:soundfield_regular}}
\subfloat[]{%
\includegraphics[width=.45\columnwidth]{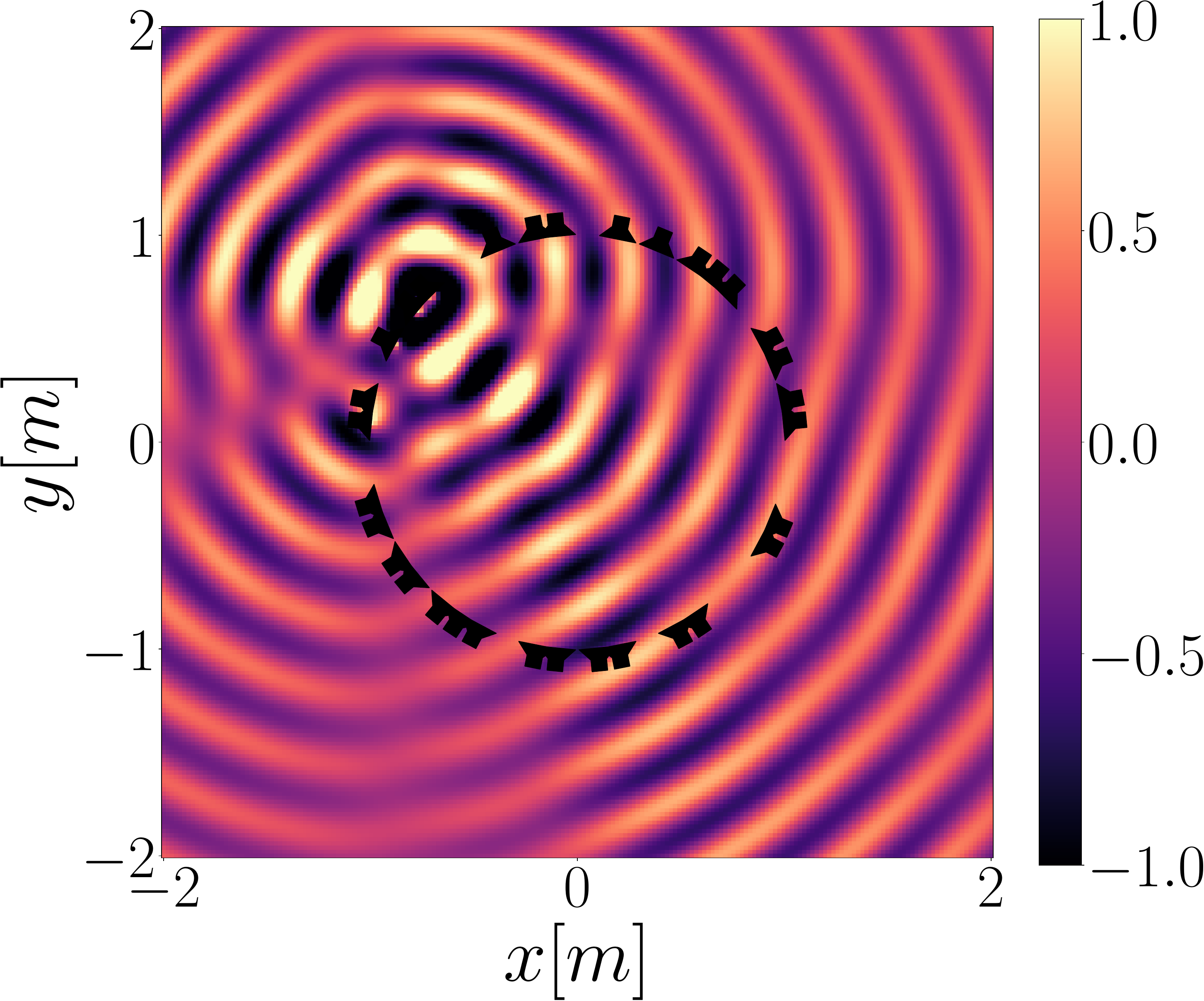}%
\label{subfig:soundfield_irregular}}
\caption{Amplitude (real) part of the soundfield for a source placed in $\mathbf{r}=[-1.2~\mathrm{m},  0.96~\mathrm{m}, 0~\mathrm{m}]$ at $f=1007~\mathrm{Hz}$ obtained using PWD through a regular \protect\subref{subfig:soundfield_regular} and irregular \protect\subref{subfig:soundfield_irregular} array of secondary sources. Black loudspeakers represent the geometry of the chosen array.}
\label{fig:soundfield_example}
\end{figure}

Given the MR soundfield synthesis technique presented in Sec.~\ref{sec:model_based_sfs}, it is possible to obtain driving signals enabling a correct reproduction of the soundfield, as shown using a circular array in Fig.~\ref{fig:soundfield_example}\protect\subref{subfig:soundfield_regular}. However, if we remove secondary sources and we do not take any countermeasure, the quality of the reproduced soundfield degrades considerably, as shown in Fig.~\ref{fig:soundfield_example}\protect\subref{subfig:soundfield_irregular}. If we consider the driving signals $\mathbf{d}_{\mathrm{mr}} \in \mathbb{C}^{L\times K}$, being $K$ the number of frequencies, obtained, either using a linear or circular array, through the MR technique, our objective is then to retrieve the function $\mathcal{U}(\cdot)$ such that 
\begin{equation}
    \mathbf{d}_{\mathrm{cnn}}(\omega_k) = \mathcal{U}(\mathbf{d}_\mathrm{mr}(\omega_k)),
\end{equation}
where $\omega_k, k=1,\ldots,K$ are the discrete angular frequencies and the driving signals $\mathbf{d}_{\mathrm{cnn}}(\omega_k) \in \mathbb{C}^{L}$ are the compensated version of $\mathbf{d}_{\mathrm{mr}}(\omega_k)$, obtained by
minimizing the following optimization problem 
\begin{equation}
    \label{eq:minim_problem_cnn}
    \mathbf{d}_{\mathrm{cnn}}(\omega_k)= \argmin_{\mathbf{d}_{\mathrm{cnn}}(\omega_k)}|p(\mathbf{r}_i, \omega_k)-\sum_{l=1}^{L}d_{\mathrm{cnn},l}(\omega_k)g(\mathbf{r}_i|\mathbf{r}_l, \omega_k)|^2,
\end{equation} 
that is, corresponding to the minimization of the reproduction error at control points $\mathbf{r}_i$.
\begin{figure*}[!t]
    \centering
    \includegraphics[width=.9\textwidth]{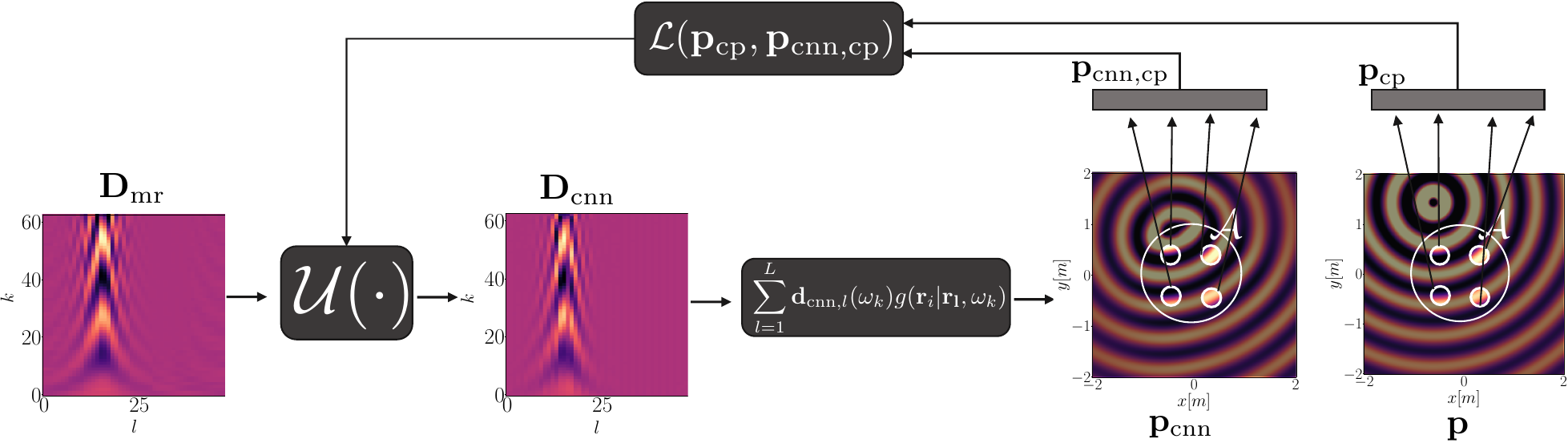}
    \caption{Schematic representation of the training procedure. Note that for simplicity, the images of $\mathbf{p}_{\mathrm{cnn}}$ and $\mathbf{p}$ correspond only the real part of the amplitude pressure soundfield obtained at a frequency $f=562~\mathrm{Hz}$ and due to a source positioned in $\mathbf{r}=[-0.61~\mathrm{m}, 1.42~\mathrm{m}]^T$.}
    \label{fig:train_scheme}
\end{figure*}
\subsection{Data representation}
Due to the adoption of complex-valued neural networks, we can directly feed the complex driving signals as input to the proposed model.
More specifically, if we consider the discrete set of $K$ frequencies and we can define the network input by stacking the driving signals into a $\mathbf{D}_{\mathrm{mr}} \in \mathbb{C}^{L \times K}$ matrix as follows
\begin{equation}
\mathbf{D}_{\mathrm{mr}}=
\begin{bmatrix}
    d_{\mathrm{mr},1}(\omega_1) & d_{\mathrm{mr},1}(\omega_2)  & \dots  & d_{\mathrm{mr},1}(\omega_K)  \\
    d_{\mathrm{mr},2}(\omega_1) & d_{\mathrm{mr},2}(\omega_2)  & \dots  & d_{\mathrm{mr},2}(\omega_K)  \\
    \vdots & \vdots & \ddots & \vdots \\
    d_{\mathrm{mr},L}(\omega_1) & d_{\mathrm{mr},L}(\omega_2)  & \dots  & d_{\mathrm{mr},L}(\omega_K)  \\
\end{bmatrix},  
\end{equation}
\subsection{Pipeline}
The pipeline of the proposed method is depicted in Fig~\ref{fig:train_scheme}.

In order to train the network we consider a set of simulated data. More specifically, we consider a set of point sources positioned at locations $\mathbf{r}_{s}$ outside the listening region. For each source we compute the corresponding driving signal matrix $\mathbf{D}_{\mathrm{mr}}$ and, by applying \eqref{eq:green_function}, the corresponding ground-truth pressure soundfield at control points $\mathbf{p}_{\mathrm{cp}}$.

The matrix $\mathbf{D}_{\mathrm{mr}}$ is fed as input to the network $\mathcal{U}(\cdot)$, whose output is the matrix containing the compensated filters $\mathbf{D}_{\mathrm{cnn}}$.

The prediction of the soundfield due to $\mathbf{r}_{s}$ at the selected control points $\mathbf{r}_{i}$ at frequency $\omega_k$ is given by the convolution in the frequency domain between the estimated filters and the point-to-point Green's function, i.e.
\begin{equation}
    {p}_{\mathrm{cnn},\mathrm{cp},i}(\omega_k) = \sum_{l=1}^{L}d_{\mathrm{cnn},l}(\omega)g(\mathbf{r_i}|\mathbf{r_l}, \omega_k).
\end{equation}

The parameters of the network $\mathcal{U}(\cdot)$ are optimized through the loss function

\begin{equation}
\label{eq:loss_function}
\begin{split}
\mathcal{L}(\mathbf{p}_{\mathrm{cnn},\mathrm{cp}},\mathbf{p}_{\mathrm{cp}}) =
    &\sum_{k=1}^{K}  (|\mathbf{p}_{\mathrm{cp}}(\omega_k)-\mathbf{p}_{\mathrm{cnn},\mathrm{cp},i}(\omega_k)|)
\end{split}
\end{equation}

The loss in \eqref{eq:loss_function} is defined for a single source in $\mathbf{r}_s$. However, it is on a batch of sources. The batch index is here omitted for the sake of compactness. 

\subsection{Network Architecture}
In order to estimate the compensated driving signals from the ones obtained using the $\mathrm{MR}$ method using an irregular loudspeaker array, we make use of a complex-valued 2D convolutional architecture denoted as $\mathcal{U}(\cdot)$. 
Since the main novelty contained in this manuscript stands in the application of deep learning to soundfield synthesis and not on the proposed deep learning techniques, we designed the network architectures by selecting standard design choices from the literature and adapting them to the particular considered scenario.

The network takes as input  $\mathbf{D}_\mathrm{mr}$ and outputs the matrix $\mathbf{D}_\mathrm{cnn}$. While the proposed architecture is made to work with an odd size, for what concerns the frequency number $K$, and a number of loudspeakers $L$ being a power of two, only minor adjustments would be needed in order to adapt it to different scenarios.

The proposed network is composed of the following layers:
\begin{enumerate}[label=\roman*),noitemsep]
\item A complex convolutional layer, with $128$ filters, which  outputs a $(L/2)-1\times (K-1)/2\times 128$ feature map.
\item A complex convolutional layer, with $256$ filters, which  outputs a $(L/4)-1 \times (K-3)/4\times 128$ feature map.
\item A complex convolutional layer, with $512$ filters, which  outputs a $(L-8)/8\times (K-7)/8\times 512$ feature map.

\item A transposed complex convolutional layer, with $256$ filters, which  outputs a $(L/4)-1  \times (K-3)/4 \times 256$ feature map.
\item A transposed complex convolutional layer, with $128$ filters, which  outputs a $(L/2-1)\times  (K-1)/2 \times 128$ feature map.
\item A transposed complex convolutional layer, with $128$ filters, which  outputs a $(L*2)\times K\times 128$ feature map.
\item A transposed complex convolutional layer, with $1$ filter, which  outputs a $(L*2)\times K\times 1$ feature map.
\end{enumerate}
The input is processed by subsequently compressing it along the 2D dimensions and increasing the number of filters, since this procedure helps in learning higher-level features hierarchically~\cite{lecun2015deep} at different scales. The chosen number of filters is similar to the ones commonly used in the literature, such as in VGG16~\cite{Simonyan15}. Since the proposed model is compensating the input driving signals, it is necessary that the output has the same dimensions as the input. For this reason the architecture has a mirrored structure that first compresses the input data using 2D convolutional layers and then expands them through 2D transposed convolutional layers to generate the compensated driving signals.

All layers have a $(3 \times 3)$ kernel, which is a common choice among CNN-based architectures~\cite{songgong2022acoustic}, with the exception of layer v) having a $4 \times 3$ kernel.  This choice is made to account for the fact that in the considered scenario the number of frequencies is not a power of two. No padding is applied, stride value is equal to $2 \times 2$ and the chosen activation is the 
Complex Parametric Rectified Linear Unit CPReLU, which has been proposed and used for audio-related applications ~\cite{pandey2019exploring} and it is extremely powerful due to the high number of parameters contained in the activation. Similarly to the CReLU activation~\cite{kuroe2003activation,trabelsi2018deep} , CPReLU applies separate PReLUs~\cite{he2015delving} on the real and imaginary part of a neuron. More specifically it is defined as
\begin{equation}
    \mathrm{CPReLU}(z) = \mathrm{PReLU}(\Re(z)) + j\mathrm{PReLU}(\Im(z)),
\end{equation}
where $z \in \mathcal{C}$ represents the value of a neuron and $\Re$ and $\Im$ denote the operators extracting the real and imaginary parts, respectively, out of a complex number.

In the layer vii), zero-padding is applied, stride is equal to $1 \times 1$ and a linear activation is used.
We introduce a skip connection, which has been proven to be able to speed up training~\cite{Kaiming2016} by feeding as input to layer v) the addition of the outputs of layer iv) and ii).
All convolutional layers, with the exception of vii) are followed by dropout, in order to prevent overfitting~\cite{srivastava2014dropout}.
The complex-valued layers of the network were implemented by means of the CVNN~\cite{j_agustin_barrachina_2022_7303587} library using Tensorflow as backend. 

\begin{figure}[!t]
\centering
\subfloat[]{%
\includegraphics[width=.45\columnwidth]{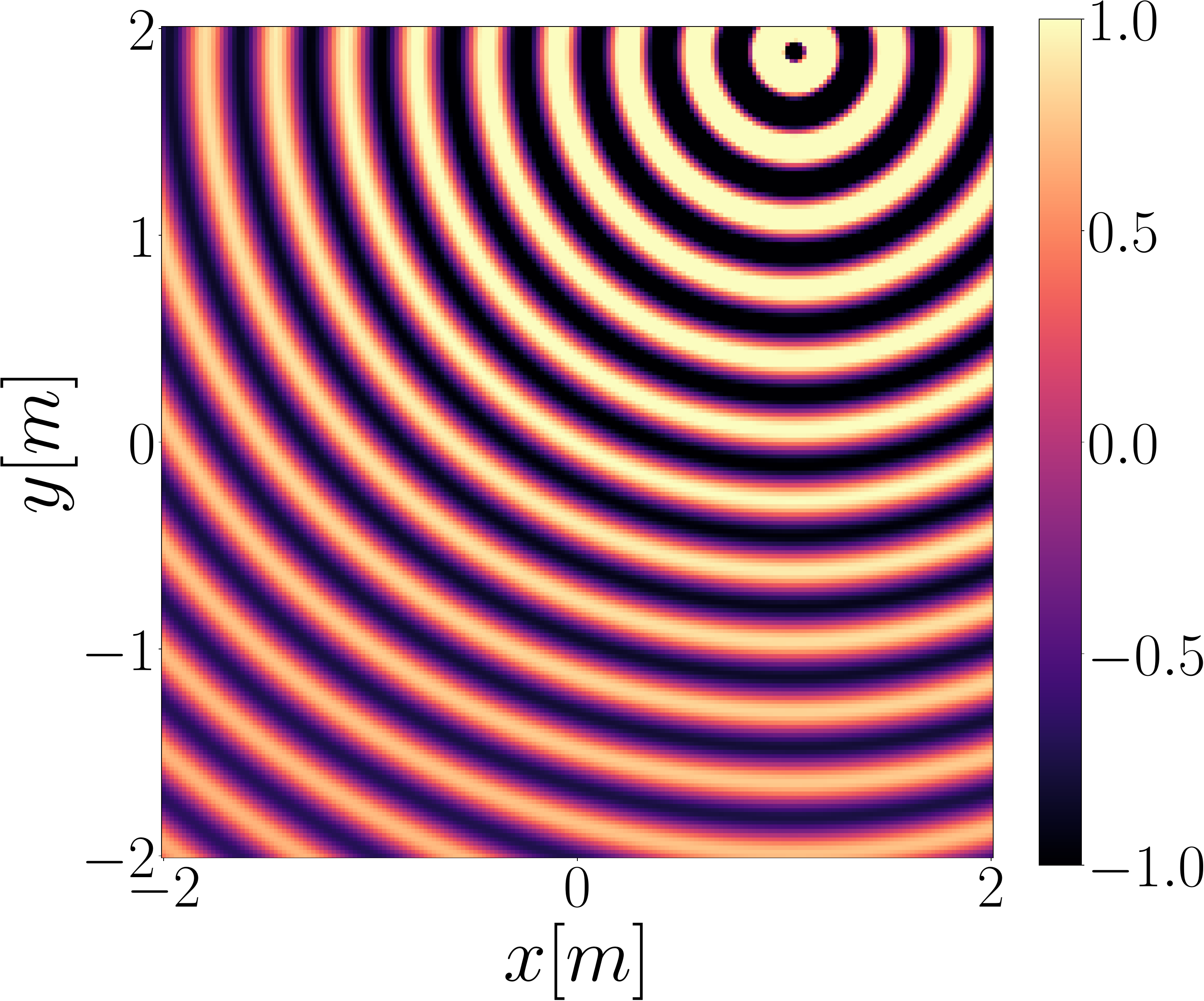}%
\label{subfig:sf_linear_real_gt}}
\subfloat[]{%
\includegraphics[width=.45\columnwidth]{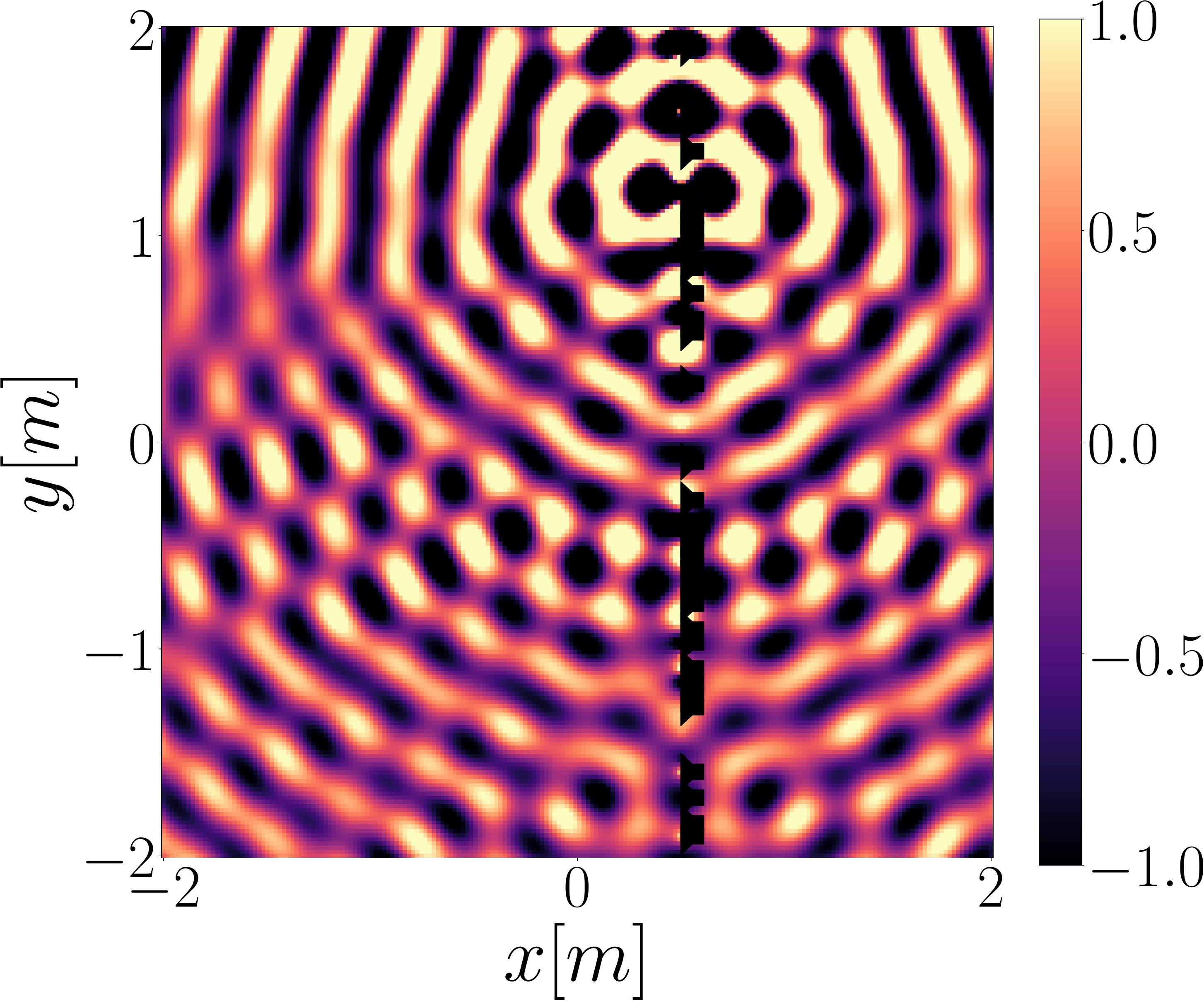}%
\label{subfig:sf_linear_real_pwd}}
\vspace{-1em}
\subfloat[]{%
\includegraphics[width=.45\columnwidth]{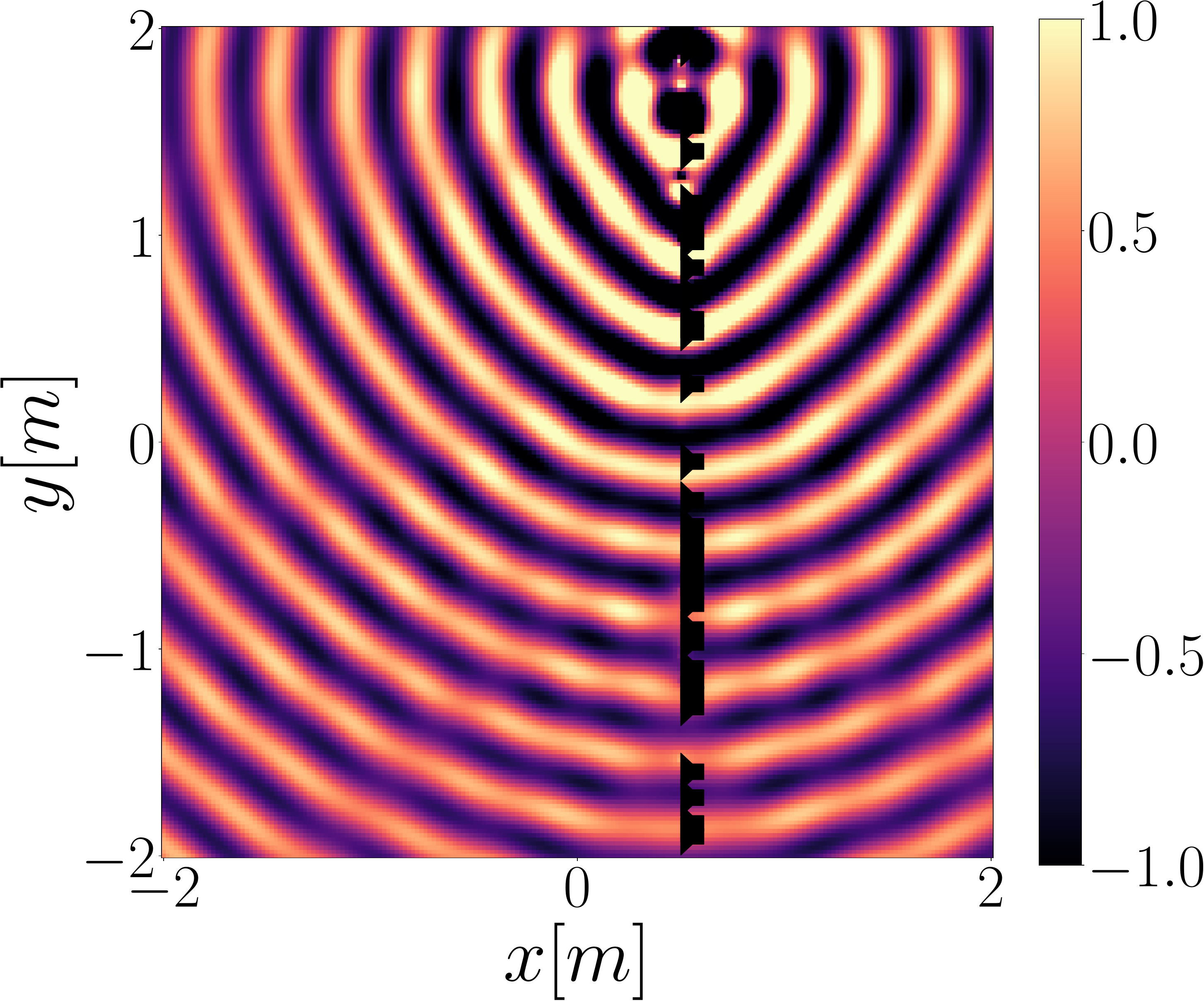}%
\label{subfig:sf_linear_real_pwd_cnn}}
\subfloat[]{%
\includegraphics[width=.45\columnwidth]{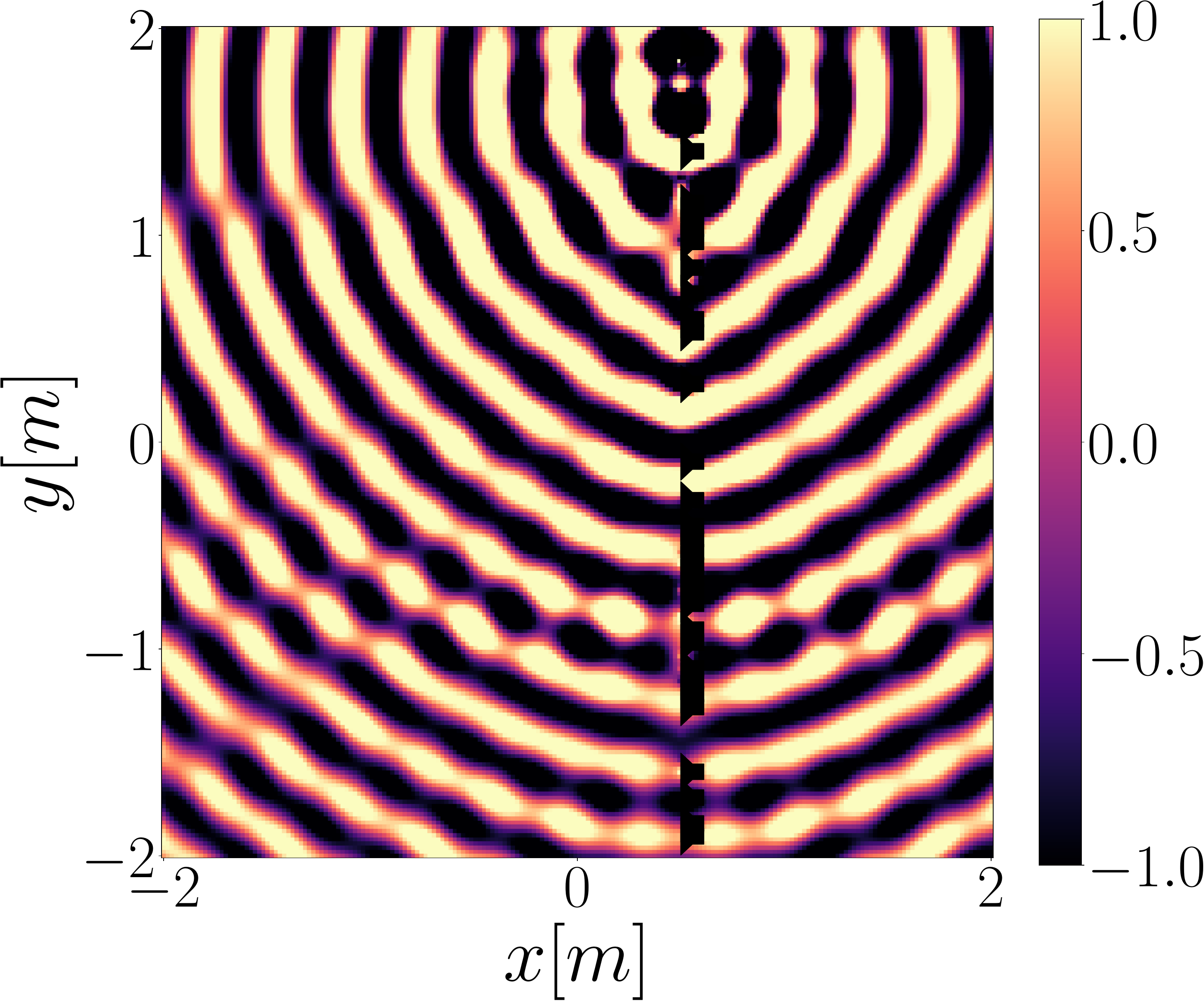}%
\label{subfig:sf_linear_real_pm}}
\vspace{-1em}
\subfloat[]{%
\includegraphics[width=.45\columnwidth]{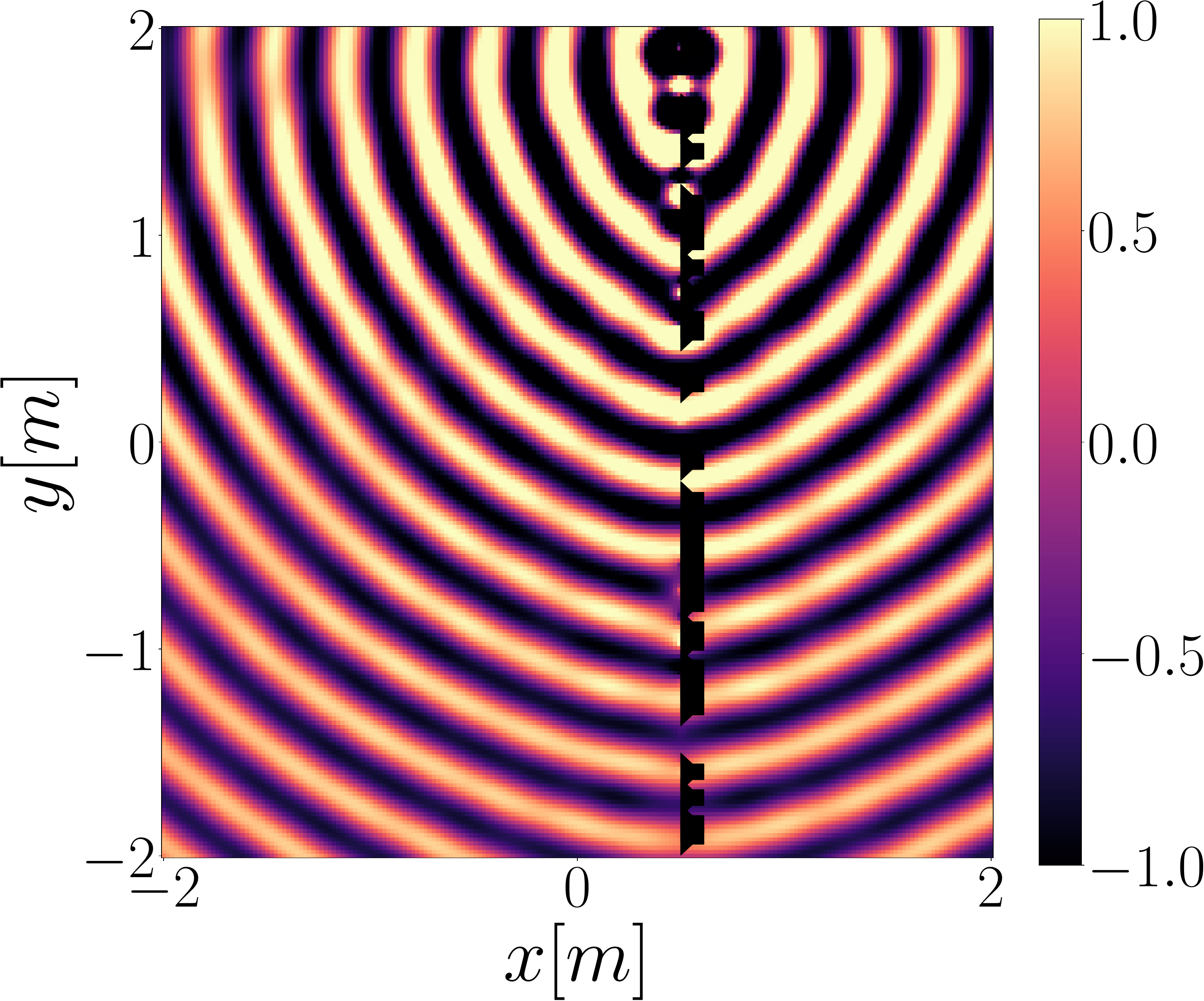}%
\label{subfig:sf_linear_real_awfs}}
\subfloat[]{%
\includegraphics[width=.45\columnwidth]{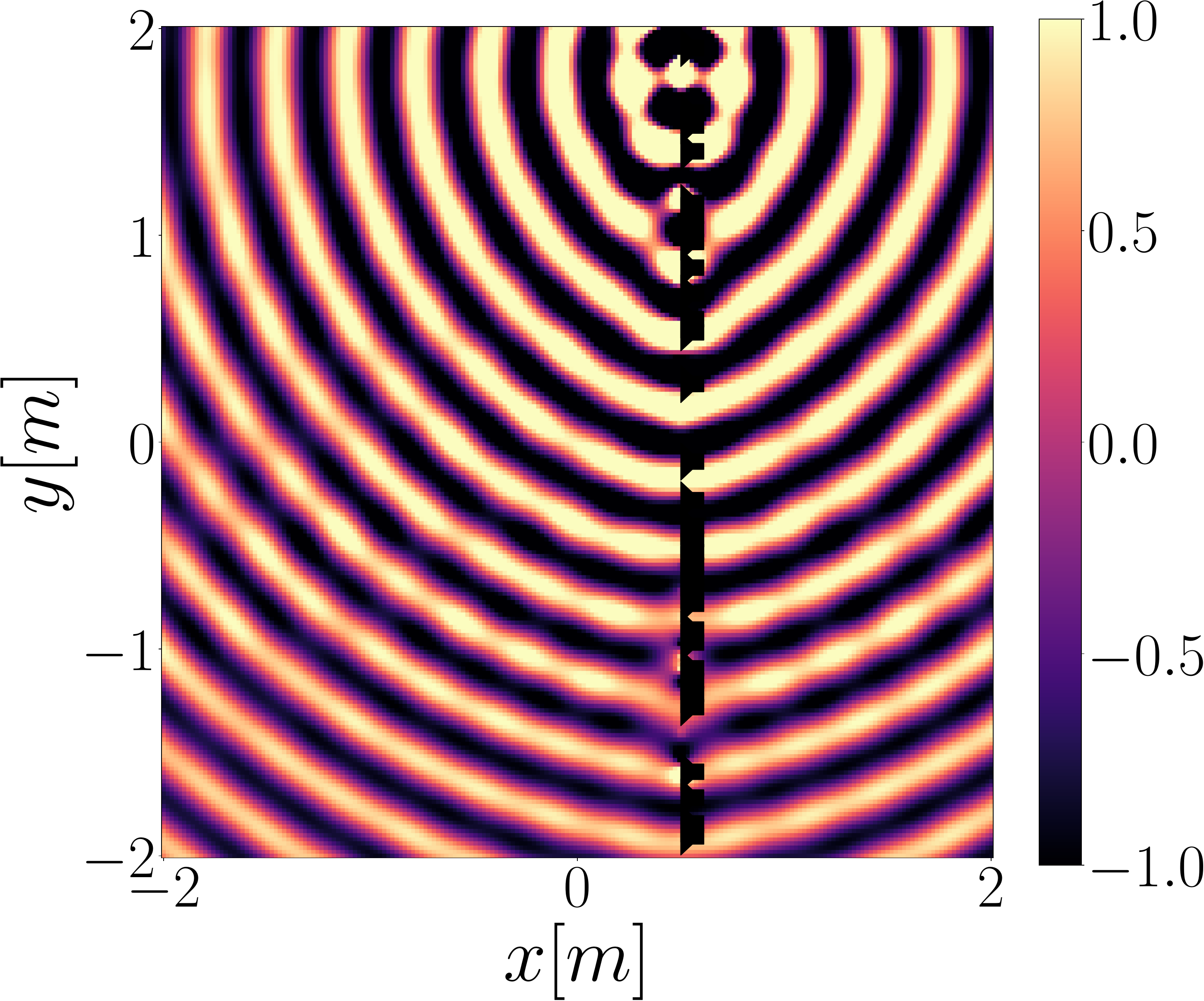}%
\label{subfig:sf_linear_real_pwd_apwd}}
\caption{Amplitude (real part) of the soundfield for a source placed in $\mathbf{r}=[1.05~\mathrm{m}, 1.88~\mathrm{m}, 0~\mathrm{m}]^T$ at $f=1007~\mathrm{Hz}$ , ground truth is shown in \protect\subref{subfig:sf_linear_real_gt}. Reproduction through an irregular linear array of $L=32$ loudspeakers using $\mathrm{MR}$~\protect\subref{subfig:sf_linear_real_pwd},$\mathrm{CNN}$ ~\protect\subref{subfig:sf_linear_real_pwd_cnn}, $\mathrm{PM}$~\protect\subref{subfig:sf_linear_real_pm}, $\mathrm{AWFS}$~\protect\subref{subfig:sf_linear_real_awfs} and $\mathrm{AMR}$~\protect\subref{subfig:sf_linear_real_pwd_apwd}. }
\label{fig:soundfield_example_real_linear}
\end{figure}

%% file: figures/setup/setup_circular.tex
\input{figures/setup/speaker_pic}
\begin{tikzpicture}

\foreach \Angle [count=\xi] in 
{0,30,60,90,120,150,180,210,240,270,300,330,360}
  \pic[rotate=\Angle-90,scale=0.35] (sp\xi) at (\Angle:1cm) {Speaker};
\draw[thick,->](180:1.75cm) -- (0:1.75cm) node[anchor=north west] {\large $x$};
\draw[thick,->]  (270:1.75cm) -- (90:1.75cm) node[anchor=north east] {\large $y$};

\end{tikzpicture}

%% file: figures/setup/setup_circular_irregular.tex

\input{figures/setup/speaker_pic}
\begin{tikzpicture}

\foreach \Angle [count=\xi] in 
{0,30,90,120,180,210,270,300,330}
  \pic[rotate=\Angle-90,scale=0.35] (sp\xi) at (\Angle:1cm) {Speaker};
\draw[thick,->](180:1.75cm) -- (0:1.75cm) node[anchor=north west] {\large $x$};
\draw[thick,->]  (270:1.75cm) -- (90:1.75cm) node[anchor=north east] {\large $y$};

\end{tikzpicture}

%% file: figures/setup/setup_linear.tex
\input{figures/setup/speaker_pic}
\begin{tikzpicture}
foreach
\foreach \ycoord [count=\xi] in 
{-1.5,-1,-0.5,0,0.5,1,1.5}
\pic[rotate=-90,scale=0.35] (sp\xi) at (.65,\ycoord) {Speaker};
\draw[thick,->](180:1.75cm) -- (0:1.75cm) node[anchor=north west] {\large $x$};
\draw[thick,->]  (270:1.75cm) -- (90:1.75cm) node[anchor=north east] {\large $y$};

\end{tikzpicture}

%% file: figures/setup/setup_linear_irregular.tex
\input{figures/setup/speaker_pic}
\begin{tikzpicture}
foreach
\foreach \ycoord [count=\xi] in 
{-1.5,,-0.5,0,1,}
\pic[rotate=-90,scale=0.35] (sp\xi) at (.65,\ycoord) {Speaker};
\draw[thick,->](180:1.75cm) -- (0:1.75cm) node[anchor=north west] {\large $x$};
\draw[thick,->]  (270:1.75cm) -- (90:1.75cm) node[anchor=north east] {\large $y$};

\end{tikzpicture}

%% file: iv-results.tex
\section{Results}
\label{sec:iv-results}

\begin{figure*}[!ht]
\centering
\subfloat[]{%
\includegraphics[width=.4\columnwidth]{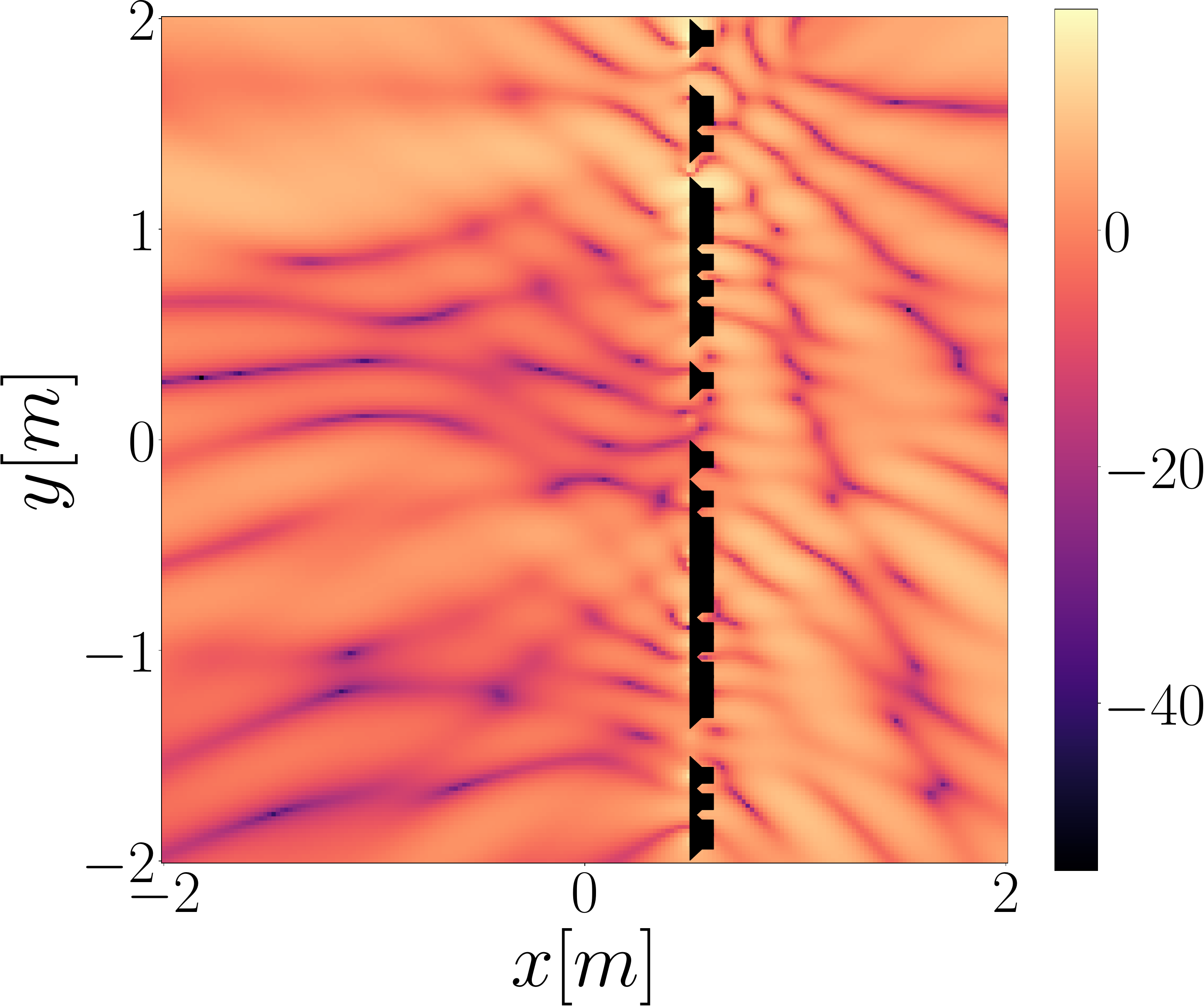}%
\label{subfig:nmse_linear_real_pwd}}
\hspace{-0.4em}
\subfloat[]{%
\includegraphics[width=.4\columnwidth]{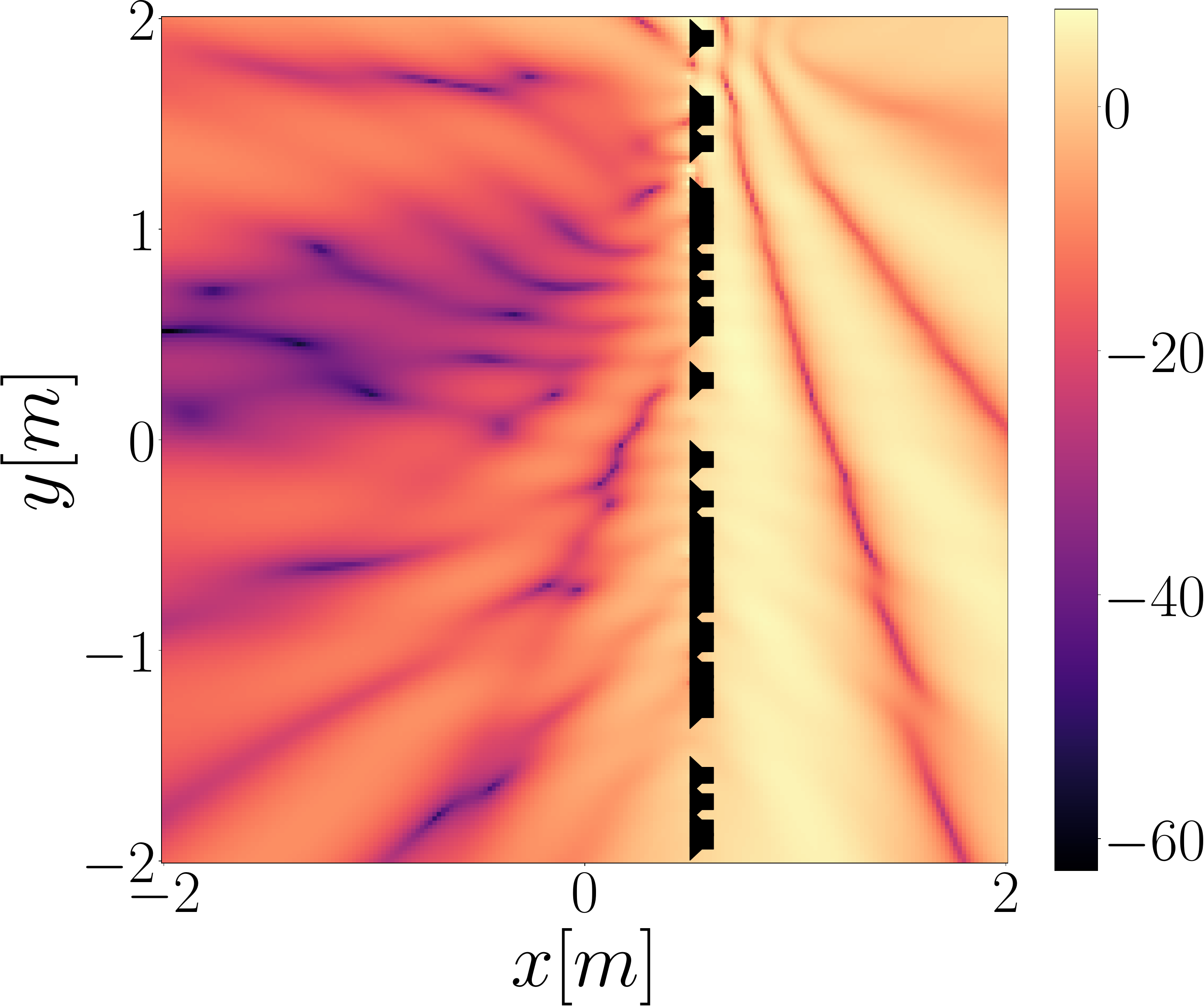}
\label{subfig:nmse_linear_real_pwd_cnn}}
\hspace{-0.4em}
\subfloat[]{%
\includegraphics[width=.4\columnwidth]{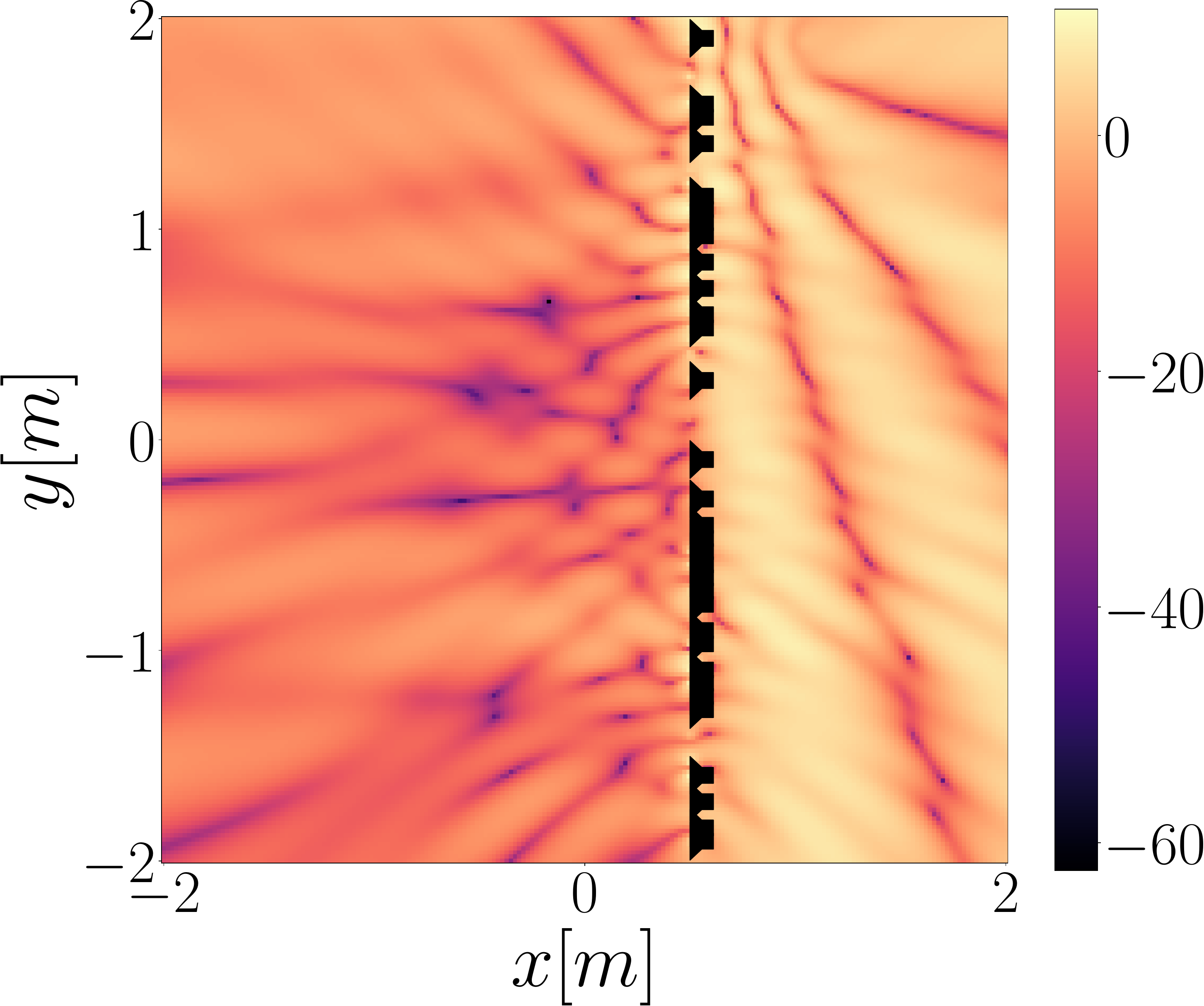}%
\label{subfig:nmse_linear_real_pm}}
\hspace{-0.4em}
\subfloat[]{%
\includegraphics[width=.4\columnwidth]{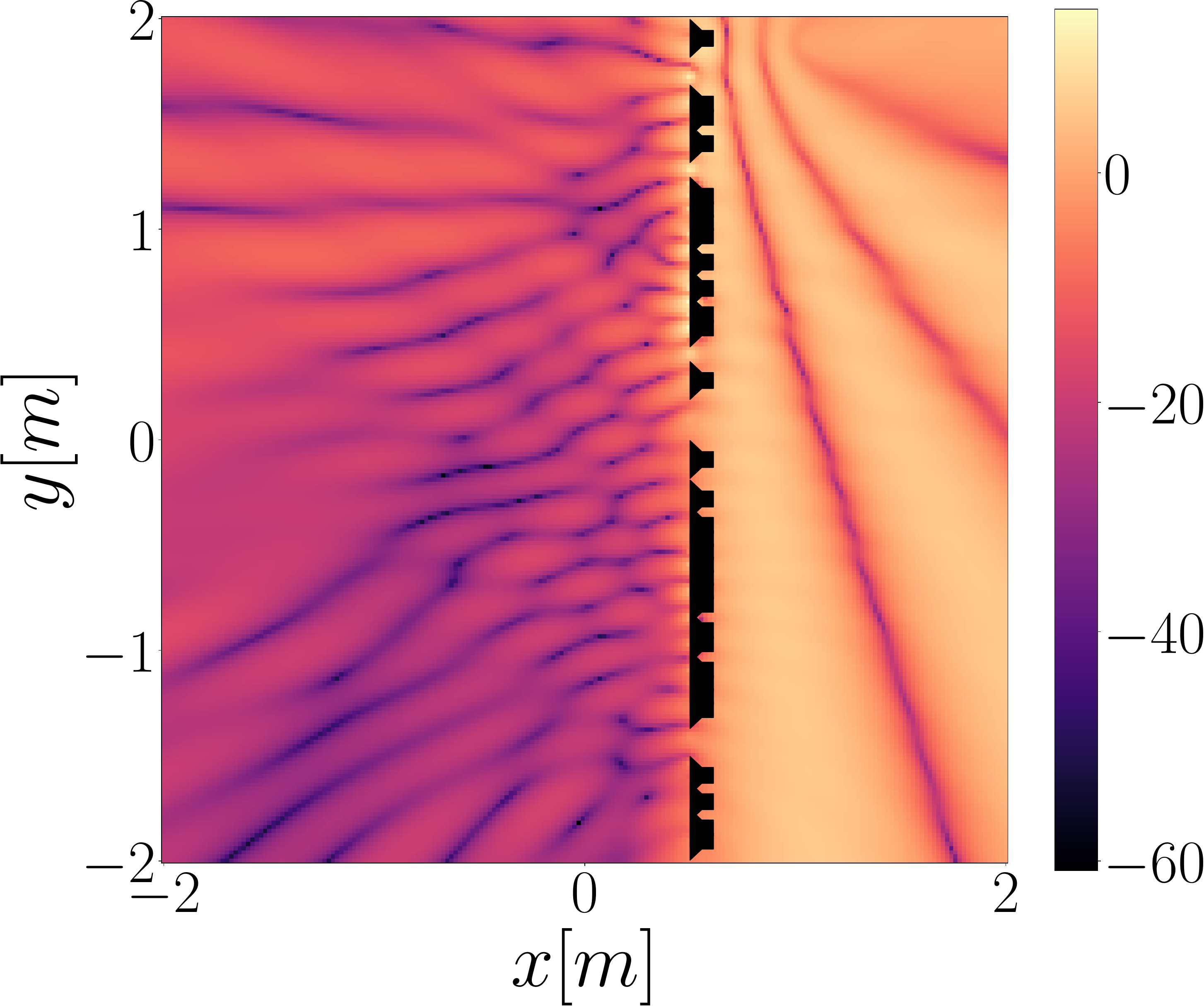}%
\label{subfig:nmse_linear_real_awfs}}
\hspace{-0.4em}
\subfloat[]{%
\includegraphics[width=.4\columnwidth]{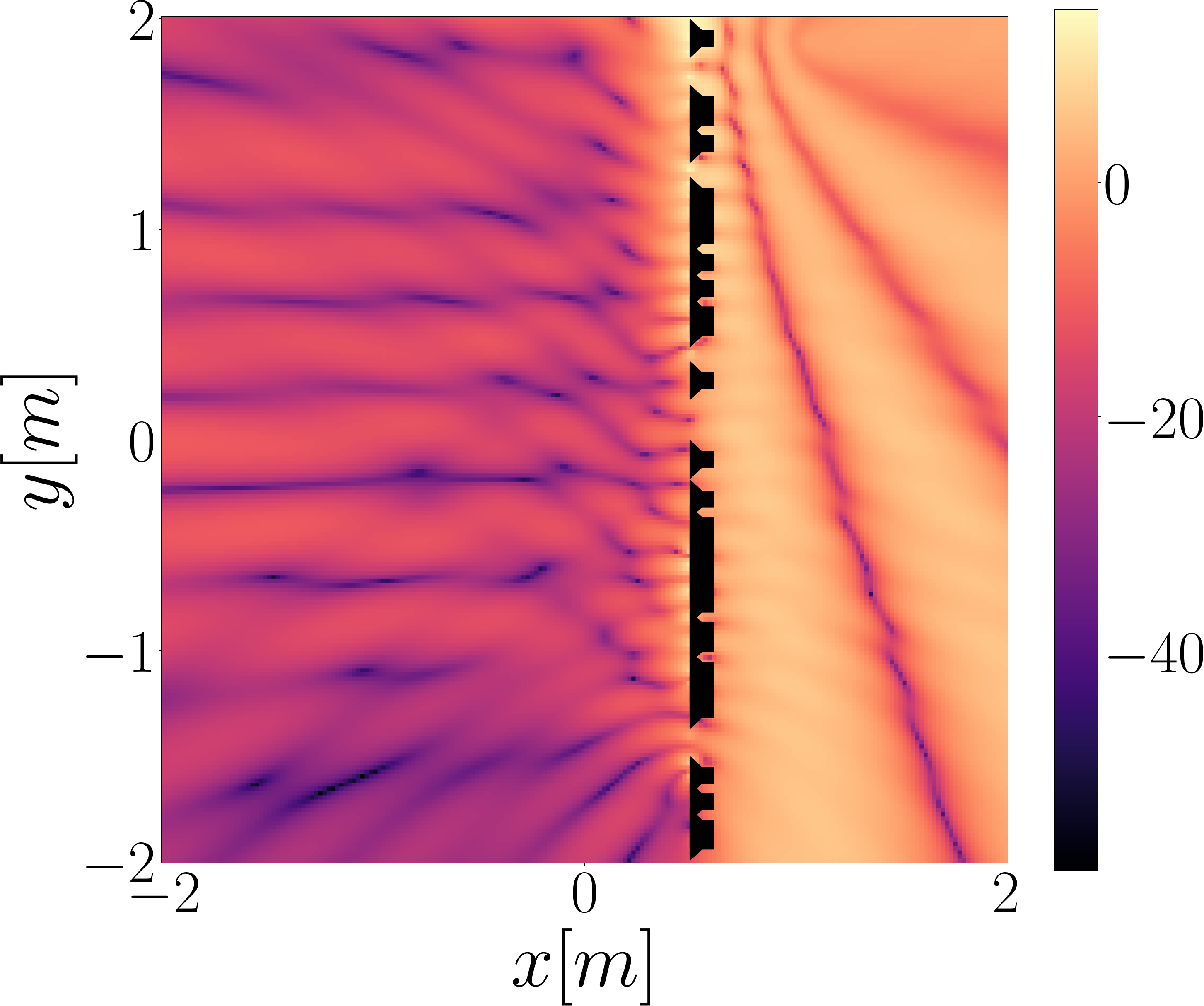}%
\label{subfig:nmse_linear_real_pwd_apwd}}

\caption{Normalized Reproduction Error (NRE) distribution in $\mathrm{dB}$ for a source placed in $\mathbf{r}=[1.05~\mathrm{m}, 1.88~\mathrm{m}, 0~\mathrm{m}]^T$ at $f=1007~\mathrm{Hz}$ when using $\mathrm{MR}$~\protect\subref{subfig:nmse_linear_real_pwd}, $\mathrm{CNN}$~\protect\subref{subfig:nmse_linear_real_pwd_cnn}, $\mathrm{PM}$~\protect\subref{subfig:nmse_linear_real_pm}, $\mathrm{AWFS}$~\protect\subref{subfig:nmse_linear_real_awfs} and $\mathrm{AMR}$~\protect\subref{subfig:nmse_linear_real_pwd_apwd}. Black loudspeakers represent the geometry of the chosen array.}
\label{fig:soundfield_example_nre_linear}
\end{figure*}
In this section we present simulation and experimental results aimed at estimating the accuracy of the soundfield synthesised with the proposed method, referred in the following as $\mathrm{CNN}$, with respect to the techniques presented in Sec.~\ref{sec:ii-background}, namely the model-based soundfield rendering technique~\cite{bianchi2016model} ($\mathrm{MR}$), the Pressure Matching technique~\cite{nelson1994active}($\mathrm{PM}$) and the Adaptive WaveField Synthesis ($\mathrm{AWFS}$). We also consider an adaptive version of the $\mathrm{MR}$ technique by applying the $\mathrm{AWFS}$ procedure defined in \eqref{eq:awfs_inversion} to the driving signals obtained via the model-based technique. We will refer to this method as $\mathrm{AMR}$ in the following.

The $\mathrm{MR}$ technique assumes setups where loudspeakers are regularly spaced, therefore its performances are expected to be non-optimal when it is applied to an irregular array, as in the case of this manuscript. Moreover, since the $\mathrm{CNN}$ technique compensates the driving signals extracted via $\mathrm{MR}$, the synthesis accuracy obtained through the latter can be considered as the higher bound with respect to the reproduction error. 

We consider also the $\mathrm{PM}$ method since, similarly to $\mathrm{CNN}$, it does not pose any constraint with respect to the configuration of the loudspeaker array.

We avoid a comparison with a mode matching technique, even if it is suitable to work with irregular setups, due to the inherently different optimization procedure. While the $\mathrm{PM}$ and $\mathrm{CNN}$ approaches minimize the pressure obtained at a series of control points with no need of feedback measurements, the mode matching technique, instead, minimizes the expansion of the soundfield obtained using spherical wavefunctions, whose coefficients must be estimated through microphone measurements~\cite{koyama2021sound}. 

The simulation results refer to circular and linear speakers deployments, while the experimental ones to a circular array setup only. We first present aspects of the setup that are in common between the configurations. We then discuss separately the different scenarios.
The code used in order to generate the data, train the model as well as the setup and additional results can be found at \url{https://polimi-ispl.github.io/deep_learning_soundfield_synthesis_irregular_array/}.The $\mathrm{WFS}$ driving functions needed to apply $\mathrm{AWFS}$ were computed using the Sound Field Synthesis (SFS) Toolbox for Python~\cite{wierstorf2012sound}
\subsection{Model parameters}
In order to train the network we simulate a set of point sources $\mathcal{S}$, which is then separated into three sets $\mathcal{S}_\mathrm{train}$, $\mathcal{S}_\mathrm{val}$, $\mathcal{S}_\mathrm{test}$ used for the training, validation and testing phases, respectively. These datasets are independent from each other, meaning more formally that
\begin{equation}
    \mathcal{S}_\mathrm{train}\cap \mathcal{S}_\mathrm{val} = \mathcal{S}_\mathrm{train} \cap \mathcal{S}_\mathrm{test} = \mathcal{S}_\mathrm{test} \cap \mathcal{S}_\mathrm{val} = \emptyset.
\end{equation}
The network is trained using the Adam optimizer~\cite{kingma2014adam} with a learning rate $\mathrm{lr}=10^{-4}$. We set the maximum number of epochs to $5000$ and saved only the model corresponding to the best validation loss value. We apply early stopping by ending the training after $10$ epochs of no improvement in terms of validation loss. The networks usually needed $100$ to $200$ epochs before reaching convergence. 

The regularization constant $\lambda$ used to regularize the least squares solution in $\mathrm{PM}$ (see \eqref{eq:pm_formulation} and $\mathrm{MR}$ (see \eqref{eq:filters_linear_mr}),$\mathrm{AMR}$ and $\mathrm{AWFS}$ (see \eqref{eq:awfs_inversion}) was set to $10^{-3} \sigma_\mathrm{max}$, where $\sigma_\mathrm{max}$ is the maximum singular value of $\mathbf{G}_\mathrm{cp}^H\mathbf{G}_\mathrm{cp}$, similarly to~\cite{ueno2019three}.

\subsection{Evaluation metrics}
\begin{figure*}[!ht]
\centering
\subfloat[]{\input{figures/NRE_16_linear.tex}%
\label{subfig:nre_16_missing_linear}}
\hfil
\subfloat[]{\input{figures/SSIM_16_linear.tex}%
\label{subfig:ssim_16_missing_linear}}
\vfil
\subfloat[]{\input{figures/NRE_32_linear.tex}%
\label{subfig:nre_32_missing_linear}}
\hfil
\subfloat[]{\input{figures/SSIM_32_linear.tex}%
\label{subfig:ssim_32_missing_linear}}
\vfil
\subfloat[]{\input{figures/NRE_48_linear.tex}%
\label{subfig:nre_48_missing_linear}}
\hfil
\subfloat[]{\input{figures/SSIM_48_linear.tex}%
\label{subfig:ssim_48_missing_linear}}
\caption{Irregular linear array soundfield synthesis performances with respect to frequency: \protect\subref{subfig:nre_16_missing_linear} NRE when $L=48$, \protect\subref{subfig:nre_32_missing_linear} NRE when $L=32$, \protect\subref{subfig:nre_48_missing_linear} NRE $L=16$, \protect\subref{subfig:ssim_16_missing_linear}. SSIM  when $L=48$, \protect\subref{subfig:ssim_32_missing_linear} SSIM  when $L=32$, \protect\subref{subfig:ssim_48_missing_linear} SSIM  when $L=16$.} 
\label{fig:results_linear_frequency}
\end{figure*}
In order to evaluate the performances of the proposed method, we adopt two different metrics, the Normalized Reproduction Error ($\mathrm{NRE}$)~\cite{ueno2019three} and the Structural Similarity Index Measure (SSIM)~\cite{wang2004image}. 
The $\mathrm{NRE}$ measures the reproduction accuracy and for a single emitting source $\mathbf{r}_s$ and frequency $\omega_k$ is defined as
\begin{equation}
    \mathrm{NRE}(\mathbf{r}_s,\omega_k) = 10\log_{10} \frac{\sum_{a=1}^{A}|\hat{p}(\mathbf{r}_{a},\omega_k)-p(\mathbf{r}_{a},\omega_k))|^2}{\sum_{a=1}^{A}|p(\mathbf{r}_{a},\omega_k))|^2}, 
\end{equation}
where $\hat{p}(\mathbf{r}_{a},\omega_k)$ corresponds to the pressure soundfield estimated at point $\mathbf{r}_{a}$ using either the $\mathrm{MR}$, $\mathrm{PM}$ or $\mathrm{CNN}$ techniques, while $p(\mathbf{r}_{a},\omega_k)$ is the ground-truth.

As already done in~\cite{lluis2020sound} we also evaluate the accuracy in terms of $\mathrm{SSIM}$, which enables to evaluate how the considered techniques are able to reproduce the overall shape of the pressure soundfield for each frequency point.
For a single emitting source $\mathbf{r}_s$ and frequency $\omega_k$, the $\mathrm{SSIM}$ is given by
\begin{equation}
    \mathrm{SSIM}(\mathbf{r}_s, \omega_k) = \frac{(2\mu_{\hat{\mathbf{p}}}\mu_{\mathbf{p}}+c_1)({2\sigma_{\hat{\mathbf{p}}\mathbf{p}}+c_2)}}{(\mu^2_{\hat{\mathbf{p}}}+\mu^2_{\mathbf{p}}+c_1)(\sigma^2_{\hat{\mathbf{p}}}+\sigma^2_{\mathbf{p}}+c_1)},
\end{equation} 
where $\mathbf{p} \in \mathbb{R}^{A}$ and $\hat{\mathbf{p}} \in \mathbb{R}^{A}$ correspond to absolute value of the pressure soundfield, normalized between $0$ and $1$, measured in the listening area $\mathcal{A}$ at frequency $\omega_k$ when the source $\mathbf{r}_s$ is active, in the ground truth case, and when either $\mathrm{CNN}$, $\mathrm{PM}$ or $\mathrm{MR}$ are used, respectively. 
The value $\mu_{(\cdot)}$ and $\sigma^2_{(\cdot)}$ are the average and variance of the matrix at subscript, respectively. Finally $\sigma_{(\cdot,\cdot)}$ is the covariance between the entries of the two matrices given as argument. In order to stabilize the division with a weak denominator, the $\mathrm{SSIM}$ calculation includes the two constants $c_1=(h_1R)^2$ and $c_2=(h_2R)^2$ where $R$ is the dynamic range of the entry values ($1$ in the case of normalized matrices), while $h_1=0.01$ and $h_2=0.03$, following the standard recommendation~\cite{lluis2020sound}.
\subsection{Linear Array}
\begin{figure}[!t]
\centering
\subfloat[]{%
\includegraphics[width=.45\columnwidth]{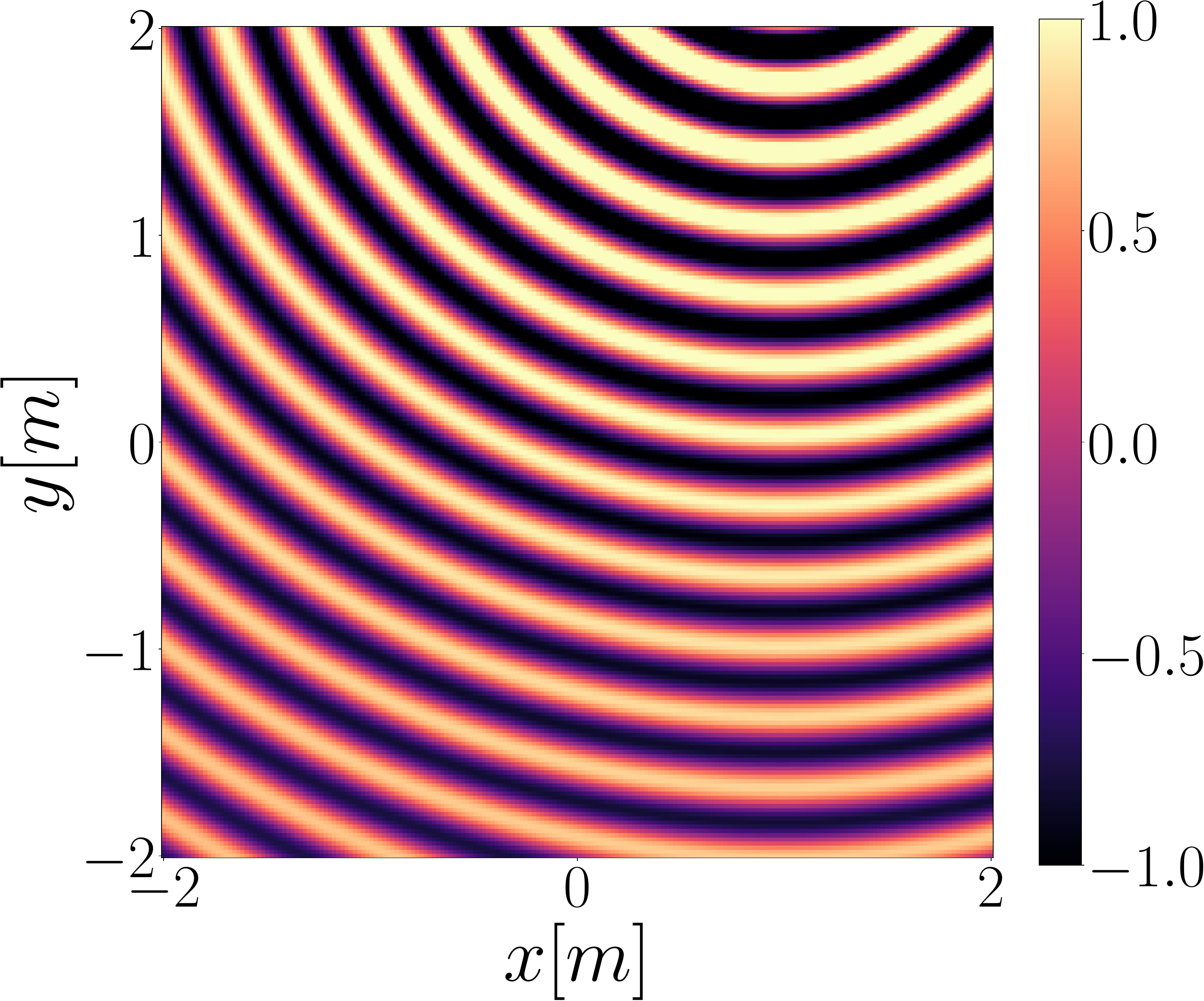}%
\label{subfig:sf_circular_real_gt}}
\subfloat[]{%
\includegraphics[width=.45\columnwidth]{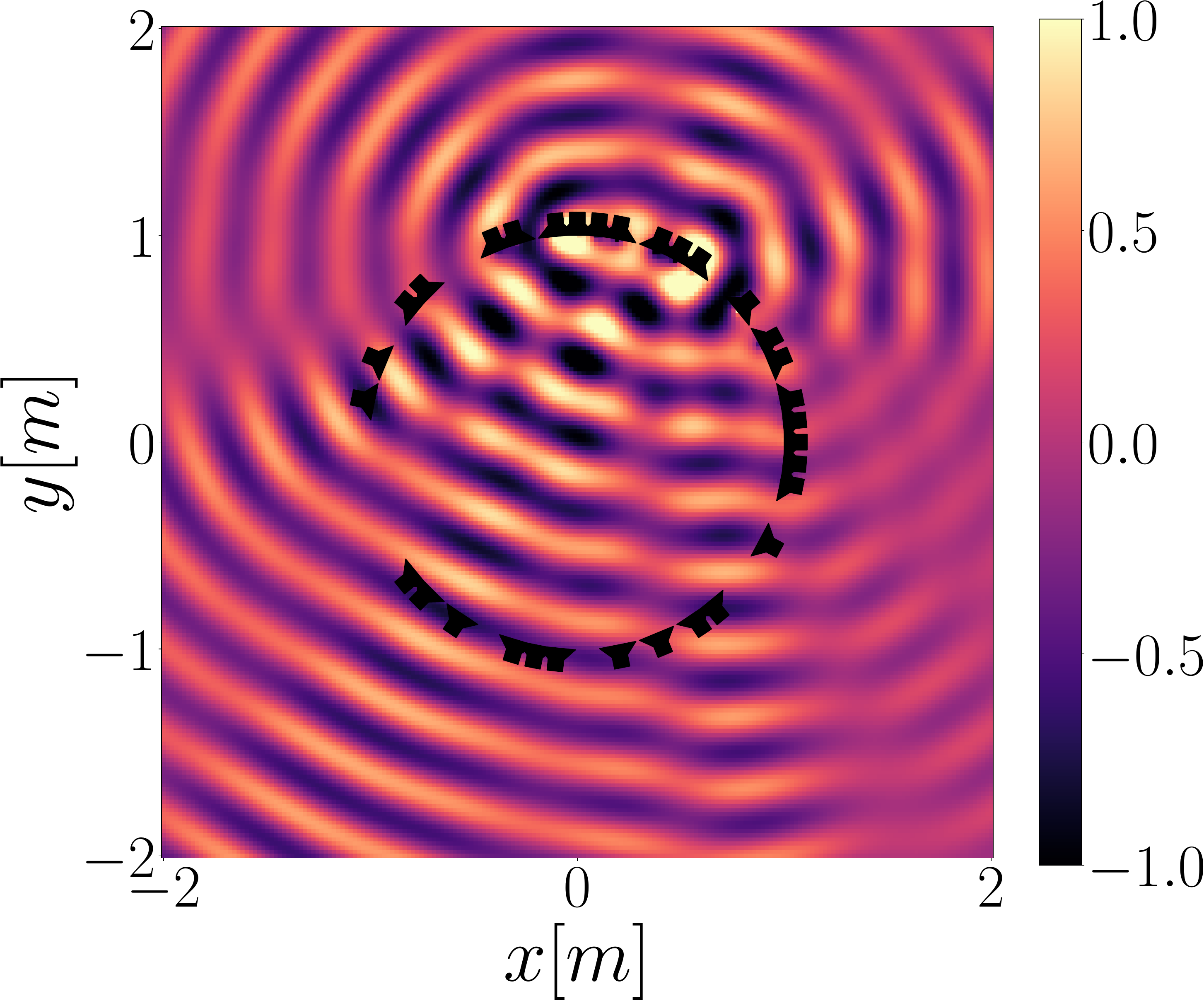}%
\label{subfig:sf_circular_real_pwd}}
\vspace{-1em}
\subfloat[]{%
\includegraphics[width=.45\columnwidth]{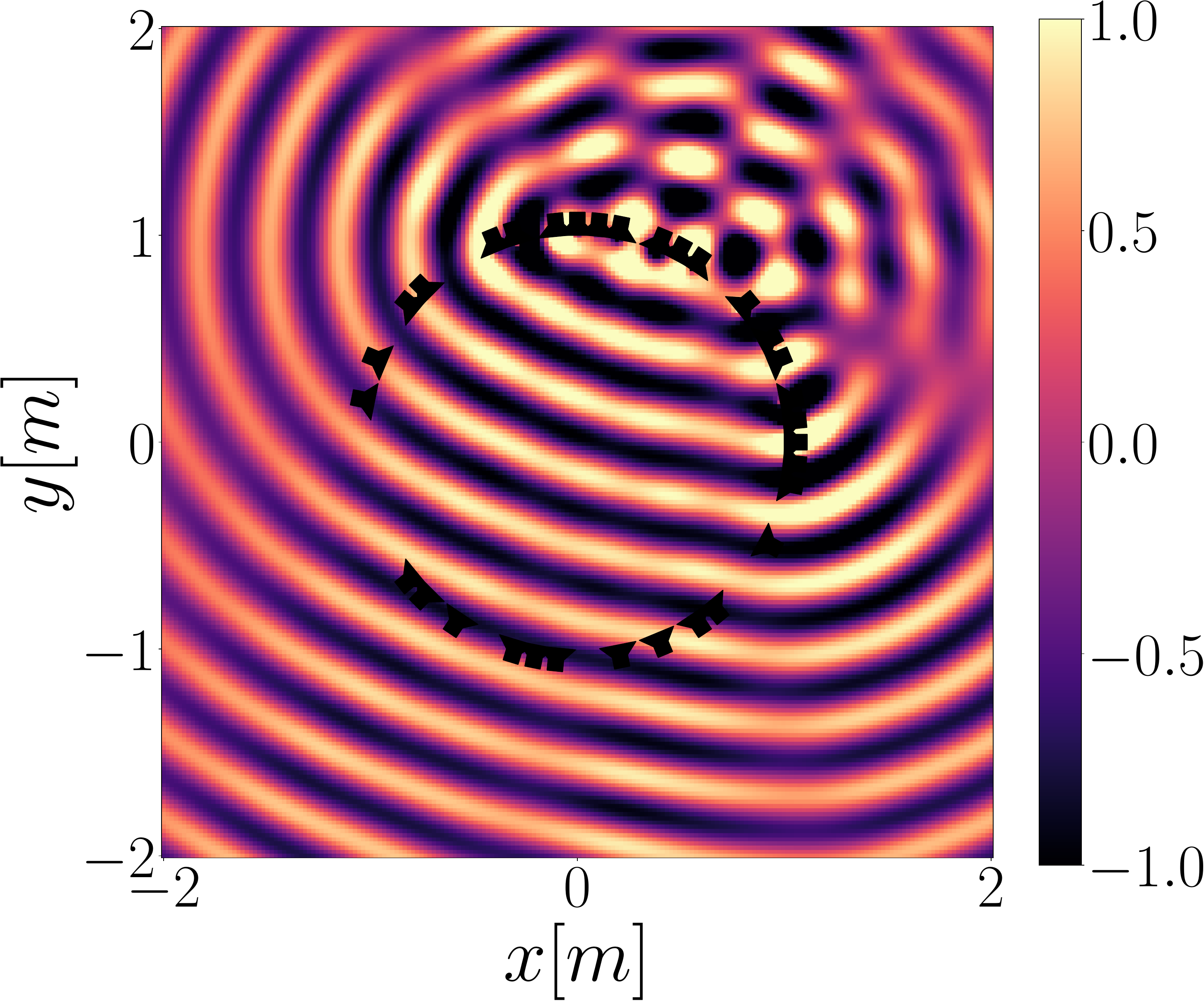}%
\label{subfig:sf_circular_real_pwd_cnn}}
\subfloat[]{%
\includegraphics[width=.45\columnwidth]{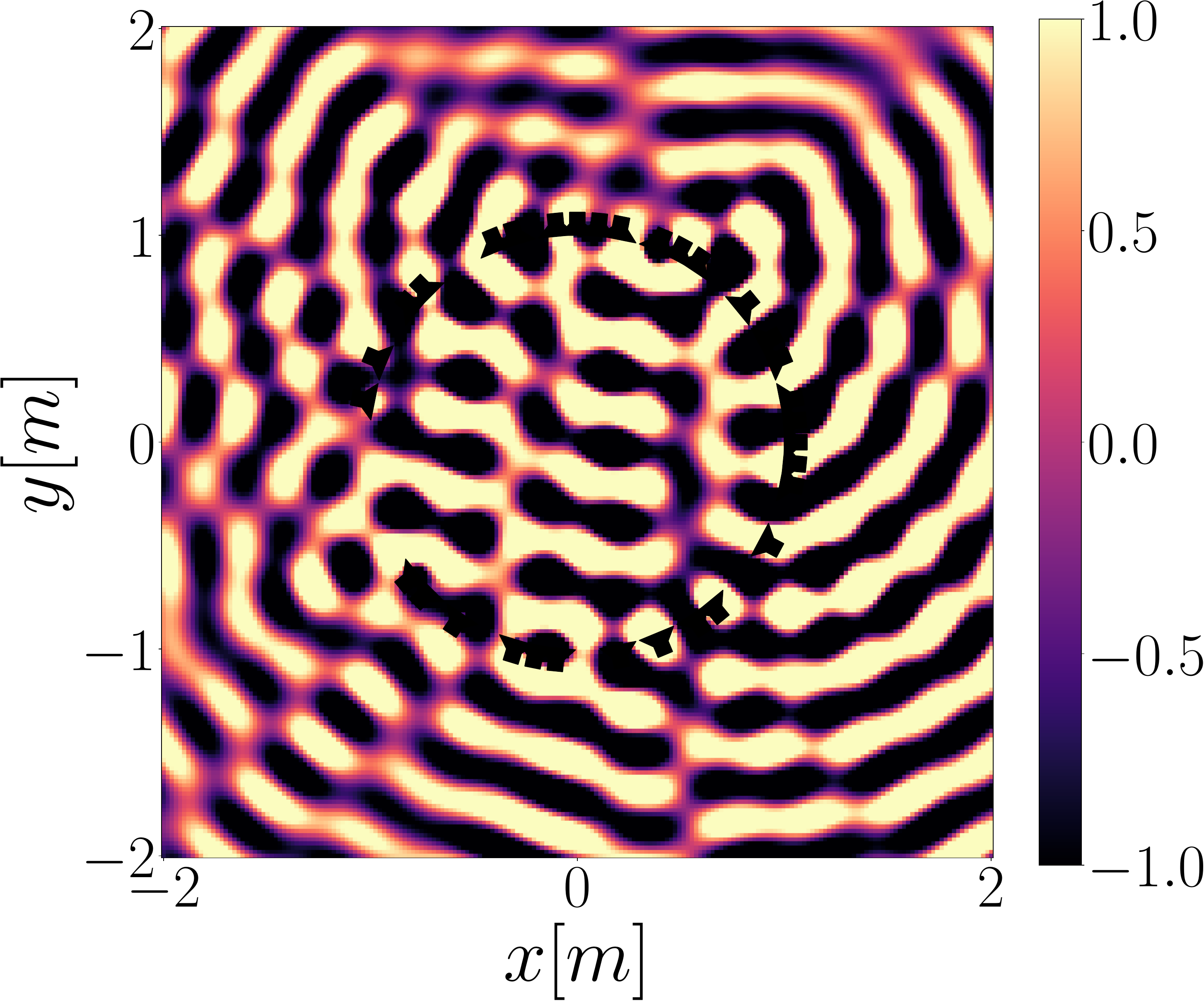}%
\label{subfig:sf_circular_real_pm}}
\vspace{-1em}
\subfloat[]{%
\includegraphics[width=.45\columnwidth]{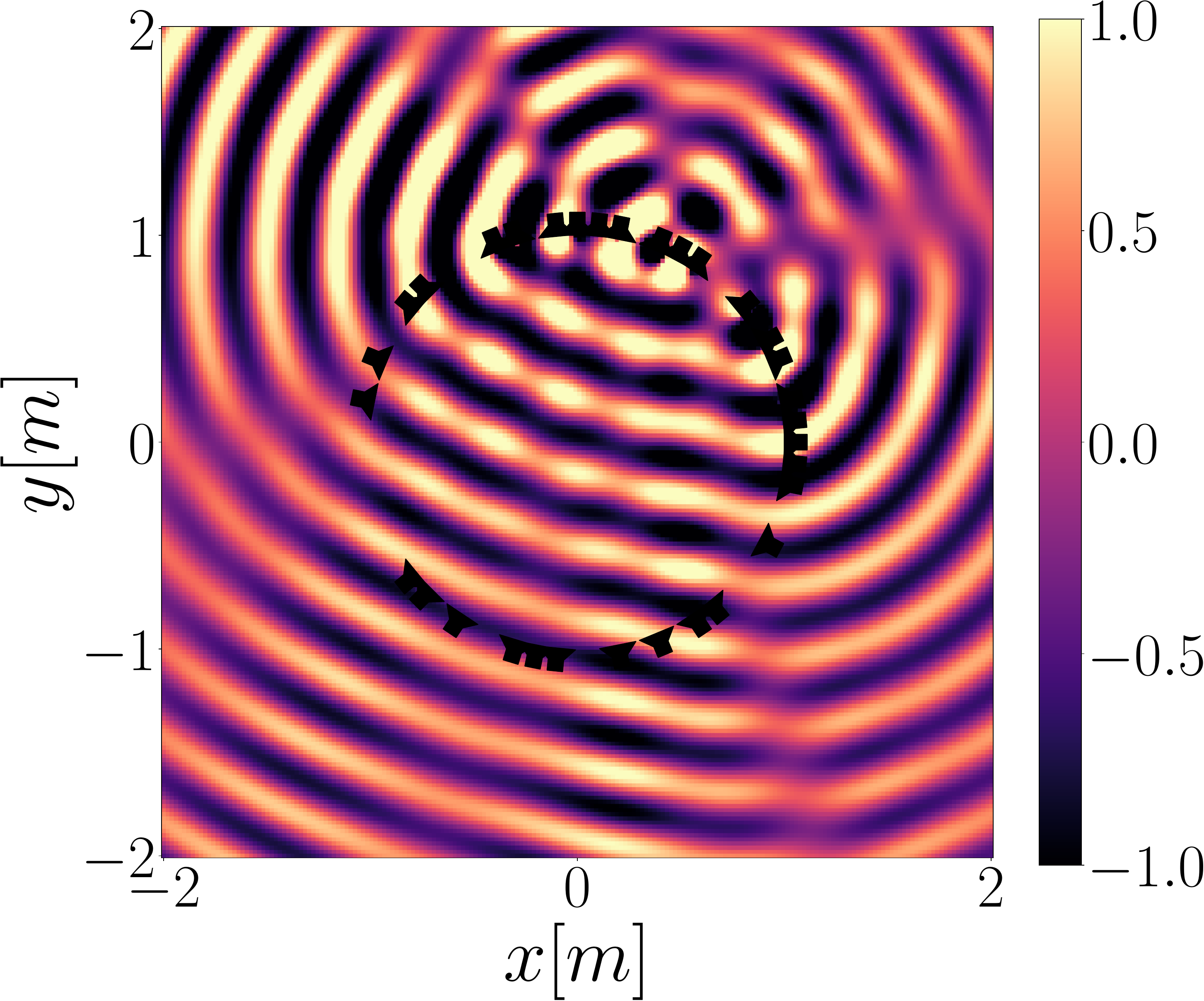}%
\label{subfig:sf_circular_real_awfs}}
\subfloat[]{%
\includegraphics[width=.45\columnwidth]{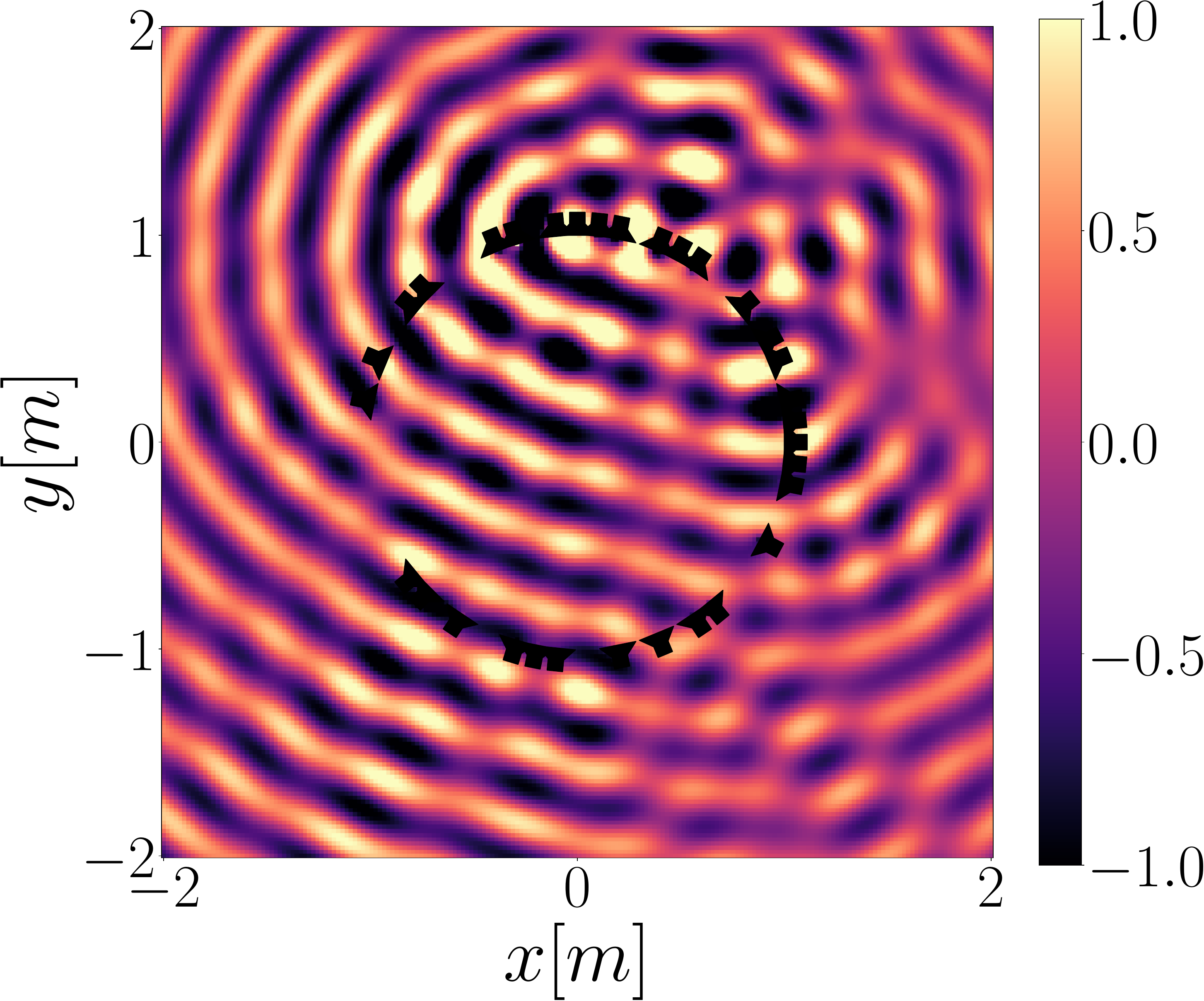}%
\label{subfig:sf_circular_real_pwd_apwd}}

\caption{Real part of the soundfield for a source placed in $\mathbf{r}=[0.99, ~\mathrm{m}, 2.88~\mathrm{m}, 0~\mathrm{m}]^T$ at $f=1007~\mathrm{Hz}$ , ground truth is shown in \protect\subref{subfig:sf_circular_real_gt}. Reproduction performances using the irregular circular array of $L=32$ loudspeakers are shown using $\mathrm{MR}$~\protect\subref{subfig:sf_circular_real_pwd}, $\mathrm{CNN}$ ~\protect\subref{subfig:sf_circular_real_pwd_cnn},$\mathrm{PM}$~\protect\subref{subfig:sf_circular_real_pm},$\mathrm{AWFS}$ 
~\protect\subref{subfig:sf_circular_real_awfs}, and $\mathrm{AMR}$ ~\protect\subref{subfig:sf_circular_real_pwd_apwd}. Black loudspeakers represent the geometry of the chosen array.}
\label{fig:soundfield_example_real_circular}
\end{figure}

\begin{figure*}[!t]
\centering
\subfloat[]{%
\includegraphics[width=.4\columnwidth]{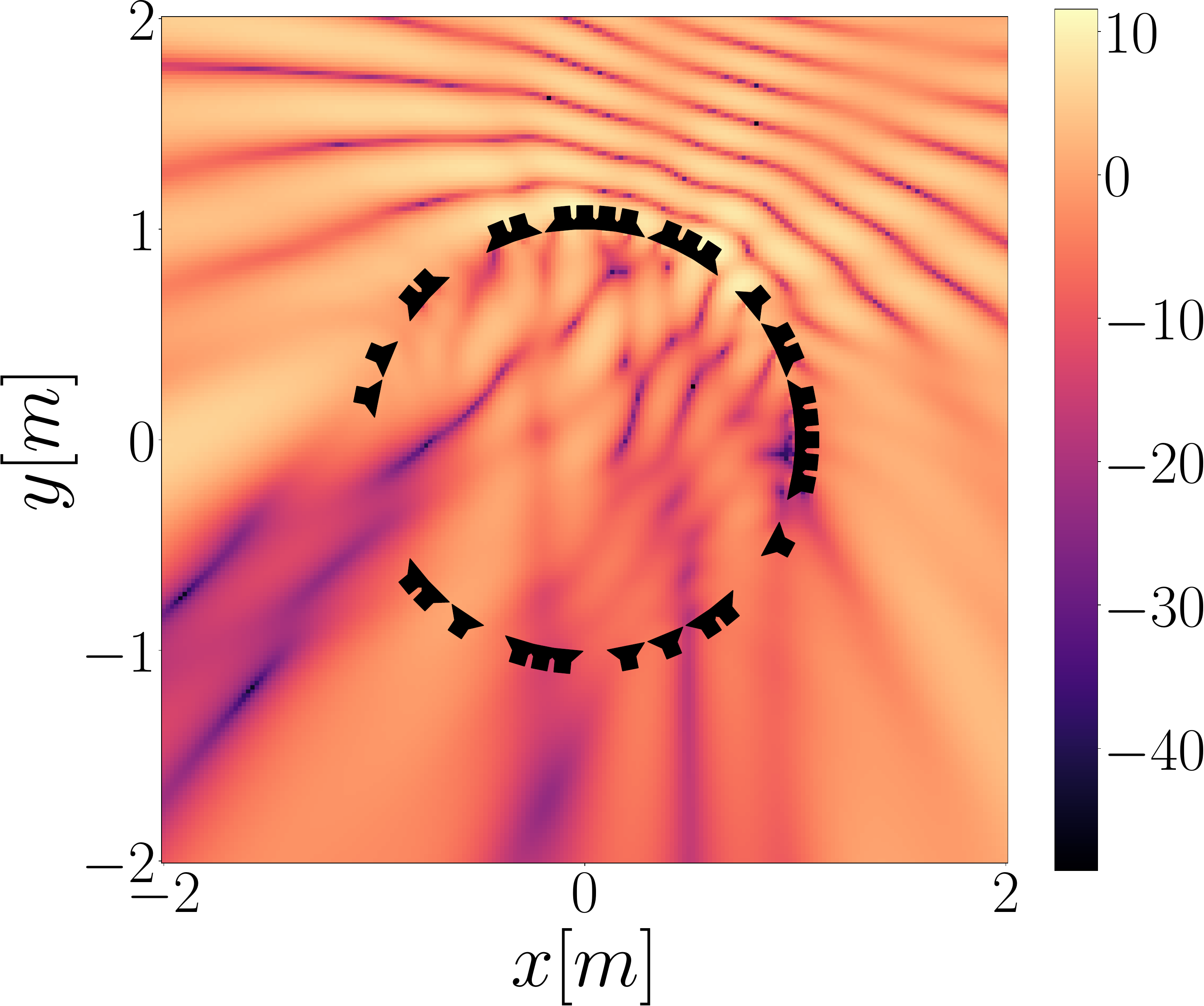}%
\label{subfig:nmse_circular_real_pwd}}
\hspace{-0.4em}
\subfloat[]{%
\includegraphics[width=.4\columnwidth]{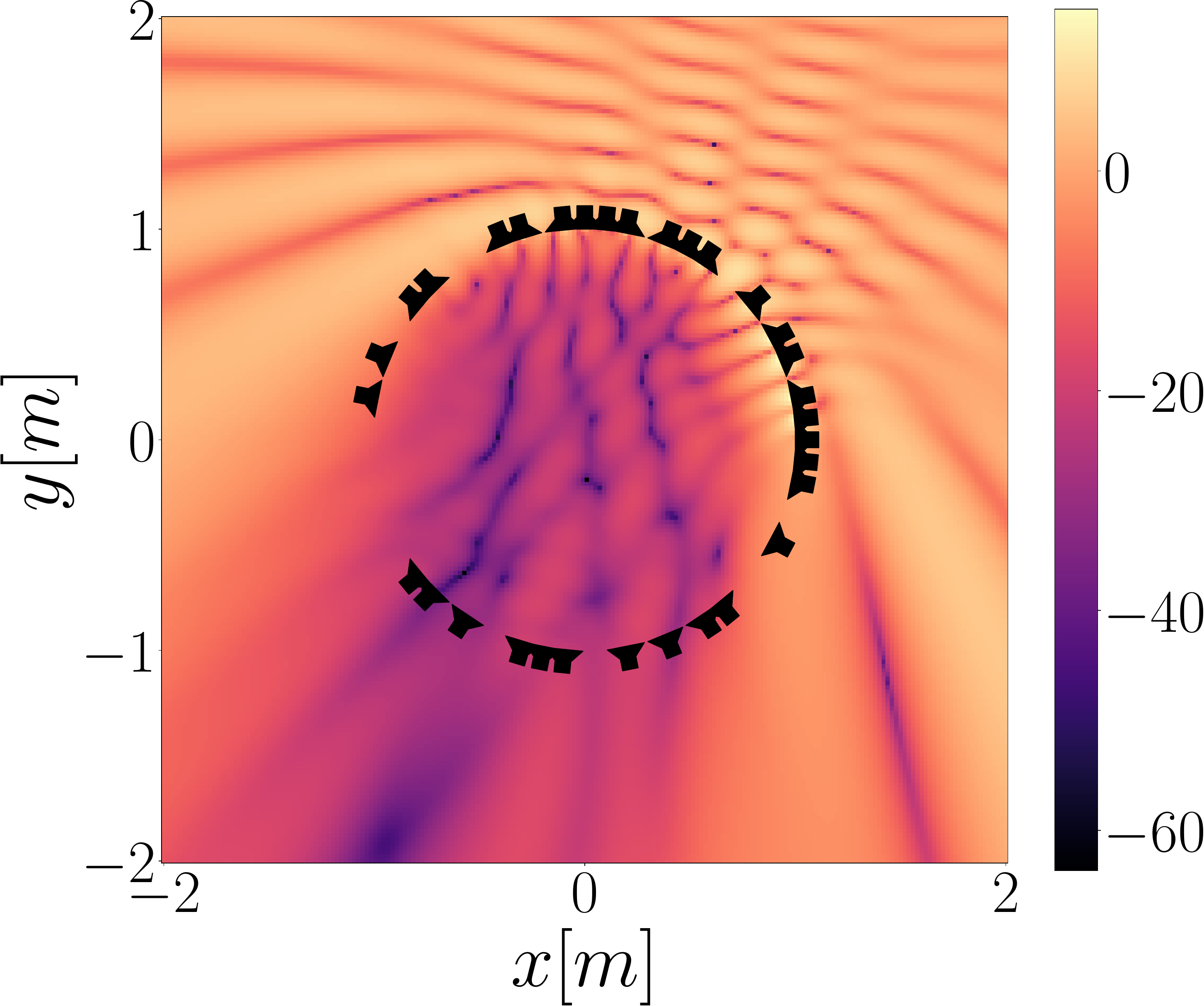}%
\label{subfig:nmse_circular_real_pwd_cnn}}
\hspace{-0.4em}
\subfloat[]{%
\includegraphics[width=.4\columnwidth]{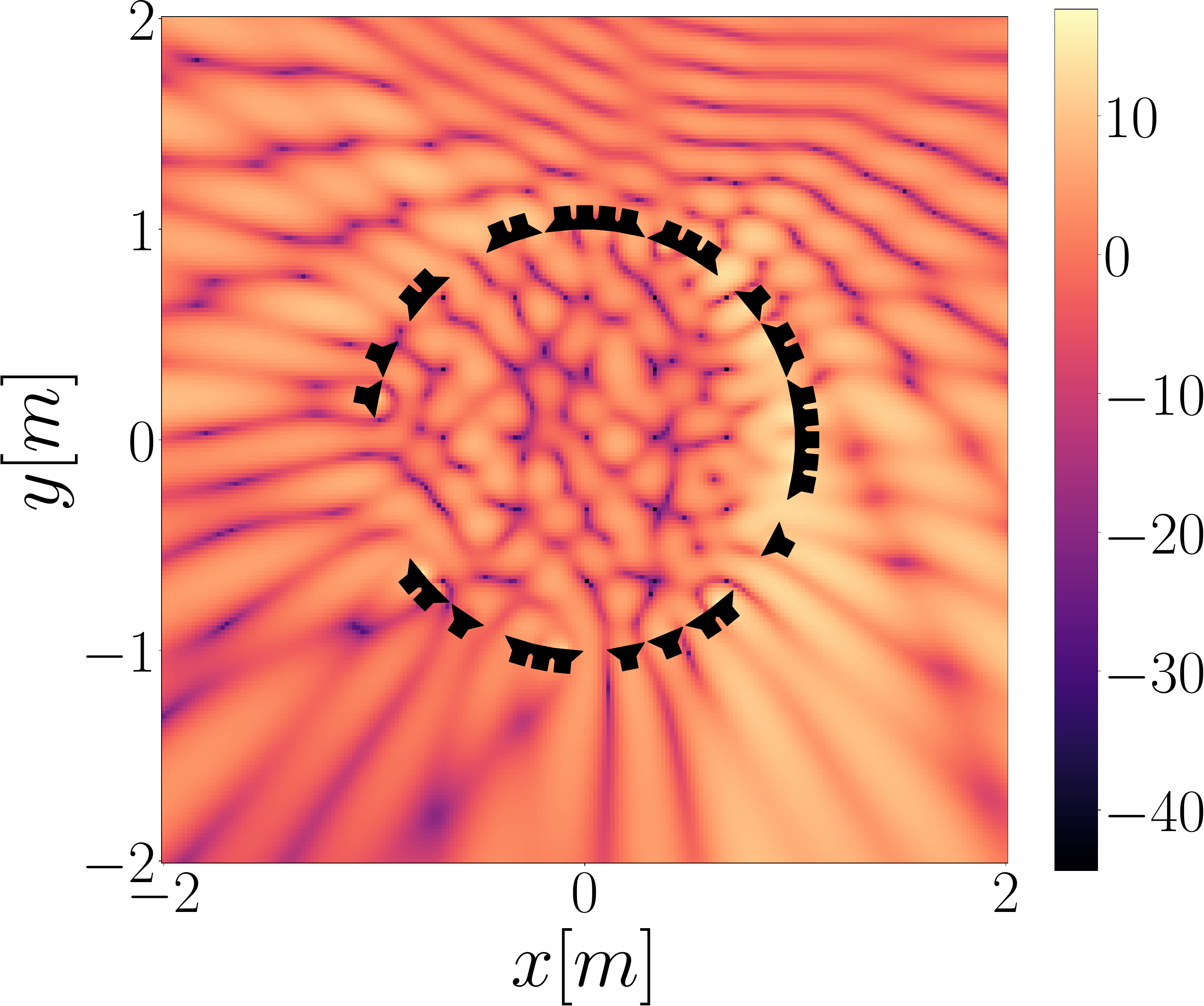}%
\label{subfig:nmse_circular_real_pm}}
\hspace{-0.4em}
\subfloat[]{%
\includegraphics[width=.4\columnwidth]{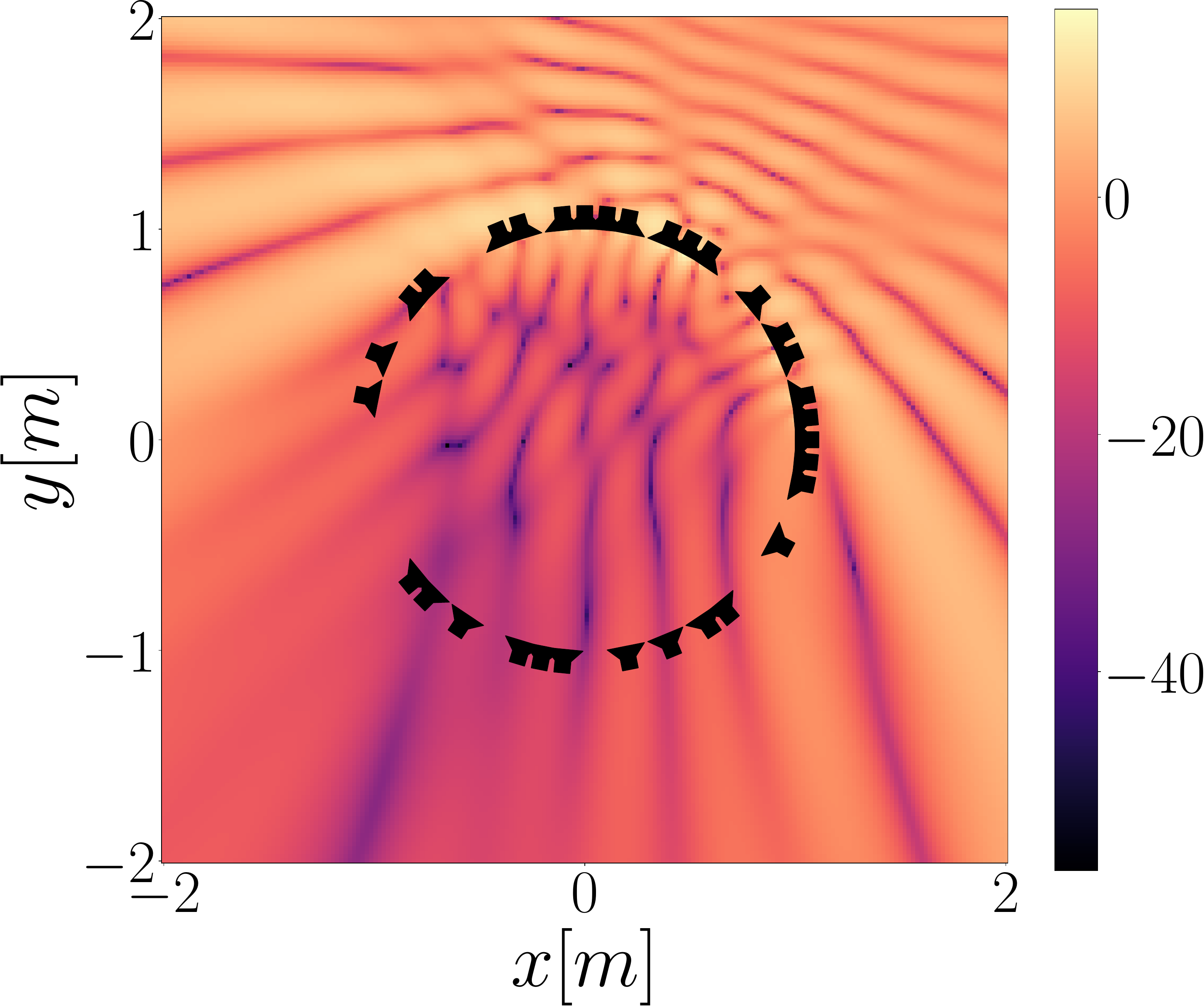}%
\label{subfig:nmse_circular_real_awfs}}
\hspace{-0.4em}
\subfloat[]{%
\includegraphics[width=.4\columnwidth]{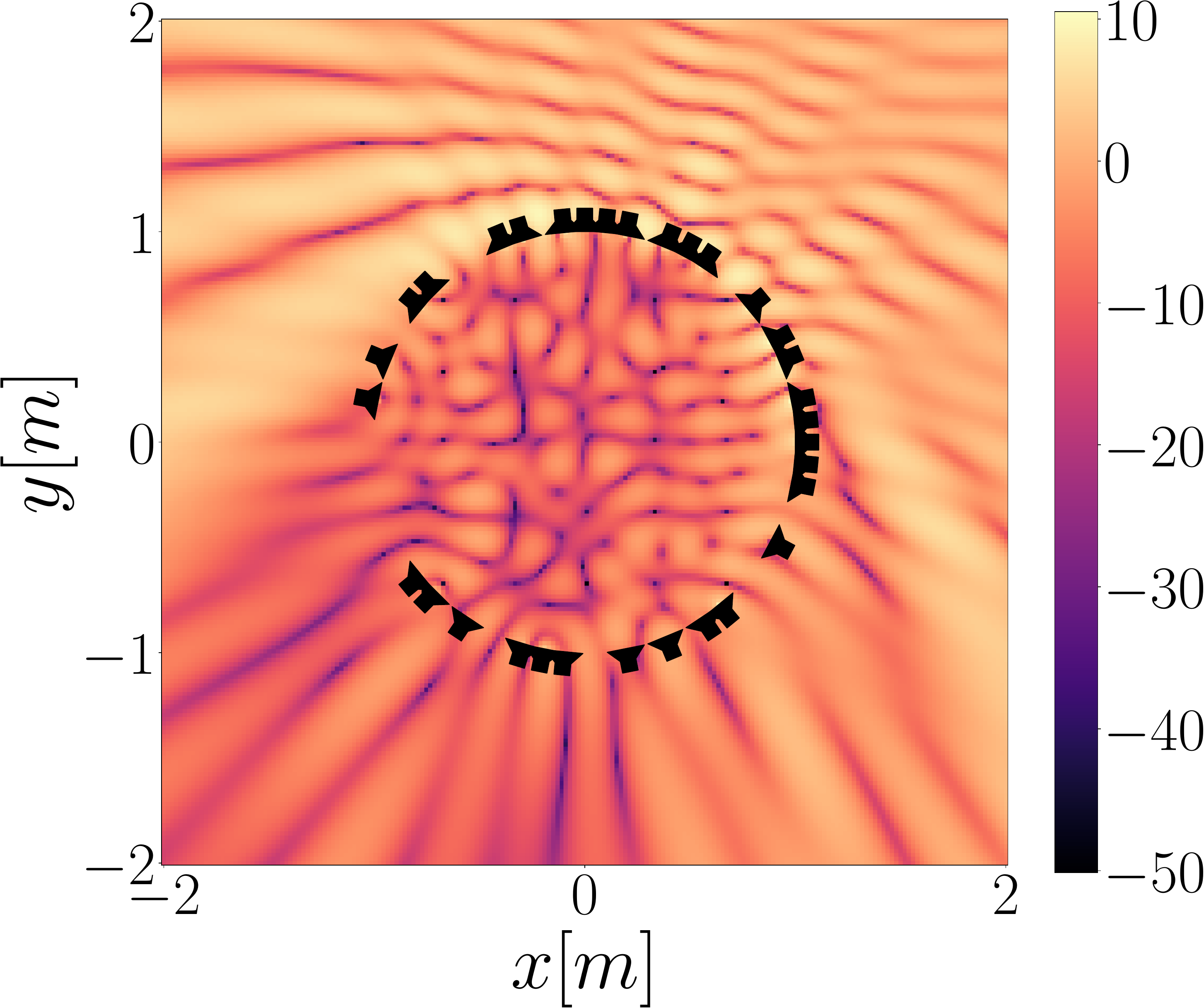}%
\label{subfig:nmse_circular_real_pwd_apwd}}
\caption{Normalized Reproduction Error (NRE) distribution in $\mathrm{dB}$ for a source placed in $\mathbf{r}=[0.99~\mathrm{m}, 2.88~\mathrm{m}, 0~\mathrm{m}]^T$ at $f=1007~\mathrm{Hz}$ when using: $\mathrm{MR}$~\protect\subref{subfig:nmse_circular_real_pwd}, $\mathrm{CNN}$ ~\protect\subref{subfig:nmse_circular_real_pwd_cnn}, $\mathrm{PM}$~\protect\subref{subfig:nmse_circular_real_pm}, $\mathrm{AWFS}$~\protect\subref{subfig:nmse_circular_real_awfs} and $\mathrm{AMR}$~\protect\subref{subfig:nmse_circular_real_pwd_apwd}. Black loudspeakers represent the geometry of the chosen array}
\label{fig:soundfield_example_nre_circular}
\end{figure*}
In this section we present results related to soundfield synthesis when considering a linear array setup.
\subsubsection{Setup}
We considered a regular linear array consisting of $L=64$ secondary sources with a spacing of $0.0625~\mathrm{m}$. From this configuration, we generated three irregular array setups by randomly removing $16$, $32$ or $48$ loudspeakers, resulting in three irregular arrays with $L=48$, $L=32$ and $L=16$ secondary sources, respectively.
The listening area $\mathcal{A}$ considered for reproduction was a $2~\mathrm{m} \times 2~\mathrm{m}$ surface located on the half plane on the left of the array, sampled using $A=25000$ points with a spacing of $0.02~\mathrm{m}$. We used $I=60$ control points placed on a grid inside $\mathcal{A}$ both for computing the losses during the training of $\mathrm{CNN}$ model and for calculating the driving signals through $\mathrm{PM}$ and $\mathrm{AWFS}$ and the filters needed to compute $\mathrm{MR}$ through~\eqref{eq:filters_linear_mr}  and $\mathrm{AMR}$.

In order to train the network, we considered the cardinality of $\mathcal{S}_\mathrm{train}$, $\mathcal{S}_\mathrm{val}$ and $\mathcal{S}_\mathrm{test}$ equal to $3920$, $980$, and $2500$, respectively. In particular, the sources in $\mathcal{S}_\mathrm{test}$ are 
generated by shifting the sources contained in $\mathcal{S}_\mathrm{train} \cup \mathcal{S}_\mathrm{val}$ by $0.08 \mathrm{m}$. We considered sources emitting a signal with spectrum $A(\omega_k)=1$ at $K=63$ frequencies spaced by $23~\mathrm{Hz}$, in the range between $46~\mathrm{Hz}$ and $1500~\mathrm{Hz}$.
\subsubsection{Results}
In Fig.~\ref{fig:soundfield_example_real_linear} we show the real part of the reproduced sound pressure distribution at frequency $f=1007~\mathrm{Hz}$ for a point source located in~$\mathbf{r}=[1.05~[\mathrm{m}], 1.88~[\mathrm{m}], 0~[\mathrm{m}]]^T$, synthesized using $L=32$ loudspeakers. More specifically, Fig.~\ref{fig:soundfield_example_real_linear}\protect\subref{subfig:sf_linear_real_gt} refers to the ground truth soundfield, while the fields for $\mathrm{MR}$, $\mathrm{CNN}$ $\mathrm{PM}${,$\mathrm{AWFS}$ and $\mathrm{AMR}$} are shown in  Fig.~\ref{fig:soundfield_example_real_linear}\protect\subref{subfig:sf_linear_real_pwd}, Fig.~\ref{fig:soundfield_example_real_linear}\protect\subref{subfig:sf_linear_real_pwd_cnn}, Fig.~\ref{fig:soundfield_example_real_linear}\protect\subref{subfig:sf_linear_real_pm}, Fig.~\ref{fig:soundfield_example_real_linear}\protect\subref{subfig:sf_linear_real_awfs} and Fig.~\ref{fig:soundfield_example_real_linear}\protect\subref{subfig:sf_linear_real_pwd_apwd}, respectively. It is apparent the fact that the $\mathrm{CNN}$ model obtains the best results, by reducing the number of irregularities in the wavefront, both with respect to the $\mathrm{MR}$ technique, whose driving signals are the input to the $\mathrm{CNN}$ model, and to the $\mathrm{PM}$ technique. While the differences in performances with respect to the $\mathrm{AWFS}$ and $\mathrm{AMR}$ techniques are less evident, the $\mathrm{CNN}$ model is still able to perform best. These considerations are also confirmed by inspecting the $\mathrm{NRE}$ for the same scenario, as shown in Fig.~\ref{fig:soundfield_example_nre_linear}.

In Fig~\ref{fig:results_linear_frequency}\protect\subref{subfig:nre_16_missing_linear}-\protect\subref{subfig:nre_32_missing_linear}-\protect\subref{subfig:nre_48_missing_linear} we present results showing the $\mathrm{NRE}$ averaged over all $|\mathcal{S}_\mathrm{test}|$ sources, when considering an irregular array of $L=48,32$ and $16$ secondary sources. The $\mathrm{CNN}$ achieves the best $\mathrm{NRE}$ over the whole range of considered frequencies in all cases, both with respect to the $\mathrm{MR}$ and $\mathrm{PM}$ techniques, where the latter shows also a higher irregularity. When comparing the $\mathrm{cnn}$ with respect to the linear optimizers-based $\mathrm{AWFS}$ and $\mathrm{AMR}$ methods, the former still obtains better performances in all scenarios, however the gap in performances diminishes together with the number of active loudspeakers, being almost indistinguishable for $L=16$. As expected, fewer are the active secondary sources, higher is the error.

In Fig~\ref{fig:results_linear_frequency}\protect\subref{subfig:ssim_16_missing_linear}-\protect\subref{subfig:ssim_32_missing_linear}-\protect\subref{subfig:ssim_48_missing_linear} we present results showing the $\mathrm{SSIM}$ averaged over all $|\mathcal{S}_\mathrm{test}|$ sources, when considering an irregular array of $L=48,~32$ and $16$ sources, respectively. 
For $L=48$ the results are more or less similar for all methods, $\mathrm{CNN}$ is worse at the lowest frequencies, while slightly better at the higher ones.
In the case of $L=32$ the $\mathrm{SSIM}$  curves are similar for most methods except for $\mathrm{CNN}$ which obtains slightly lower results below $600~\mathrm{Hz}$, but performs better than the other methods for higher frequency values. 
Finally, In the case of $L=16$ the $\mathrm{SSIM}$ is comparable for all considered methods, with $\mathrm{CNN}$ obtaining slightly better results over $600~\mathrm{Hz}$.
\subsection{Circular Array}
\begin{figure*}[!t]
\centering
\subfloat[]{\input{figures/NRE_16_circular.tex}%
\label{subfig:nre_16_missing_circular}}
\hfil
\subfloat[]{\input{figures/SSIM_16_circular.tex}%
\label{subfig:ssim_16_missing_circular}}
\vfil
\subfloat[]{\input{figures/NRE_32_circular.tex}%
\label{subfig:nre_32_missing_circular}}
\hfil
\subfloat[]{\input{figures/SSIM_32_circular.tex}%
\label{subfig:ssim_32_missing_circular}}
\vfil
\subfloat[]{\input{figures/NRE_48_circular.tex}%
\label{subfig:nre_48_missing_circular}}
\hfil
\subfloat[]{\input{figures/SSIM_48_circular.tex}%
\label{subfig:ssim_48_missing_circular}}
\caption{Irregular circular array soundfield synthesis performances with respect to frequency: \protect\subref{subfig:nre_16_missing_circular} NRE when $L=48$, \protect\subref{subfig:nre_32_missing_circular} NRE when $L=32$, \protect\subref{subfig:nre_48_missing_circular} NRE when $L=16$. \protect\subref{subfig:ssim_16_missing_circular} SSIM when $L=48$, \protect\subref{subfig:ssim_32_missing_circular} SSIM when $L=32$, \protect\subref{subfig:ssim_48_missing_circular} SSIM when $L=16$. }
\label{fig:results_circular_frequency}
\end{figure*}
\begin{figure*}[!t]
\centering
\subfloat[]{\input{figures/NRE_16_circular_radius.tex}%
\label{subfig:nre_16_missing_circular_radius}}
\hfil
\subfloat[]{\input{figures/SSIM_16_circular_radius.tex}%
\label{subfig:ssim_16_missing_circular_radius}}
\vfil
\subfloat[]{\input{figures/NRE_32_circular_radius.tex}%
\label{subfig:nre_32_missing_circular_radius}}
\hfil
\subfloat[]{\input{figures/SSIM_32_circular_radius.tex}%
\label{subfig:ssim_32_missing_circular_radius}}
\vfil
\subfloat[]{\input{figures/NRE_48_circular_radius.tex}%
\label{subfig:nre_48_missing_circular_radius}}
\hfil
\subfloat[]{\input{figures/SSIM_48_circular_radius.tex}%
\label{subfig:ssim_48_missing_circular_radius}}
\caption{Irregular circular array soundfield synthesis performances with respect to distance from the center of the reproduction area at frequency $f=1007~\mathrm{Hz}$: \protect\subref{subfig:nre_16_missing_circular_radius} NRE when $L=48$, \protect\subref{subfig:nre_32_missing_circular_radius} NRE when $L=32$, \protect\subref{subfig:nre_48_missing_circular_radius} NRE when $L=16$, \protect\subref{subfig:ssim_16_missing_circular_radius} SSIM when $L=48$, \protect\subref{subfig:ssim_32_missing_circular_radius} SSIM when $L=32$ \protect\subref{subfig:ssim_48_missing_circular_radius} SSIM when $L=16$.} 
\label{fig:results_circular_radius}
\end{figure*}
In this section we present results related to soundfield synthesis when considering a circular array setup.
\subsubsection{Setup}
We considered a regular circular array consisting of $L=64$ secondary sources with a radius of $1~\mathrm{m}$.
The listening area considered for reproduction was the area surrounded by the speakers, which amounts to $3.14 \mathrm{m}^2$, sampled using $A=7770$ listening points, with a spacing of $0.02~\mathrm{m}$. We used $I=276$ control points placed in a grid inside $\mathcal{A}$ to compute the losses during the training of $\mathrm{CNN}$ model and to calculate the driving signals through $\mathrm{PM}$ ,$\mathrm{AWFS}$ and $\mathrm{AMR}$.

In order to train the network we used $|\mathcal{S}_\mathrm{train}|=4096$ and $|\mathcal{S}_\mathrm{val}|=1024$, respectively.
The $\mathcal{S}_\mathrm{train}$ and $\mathcal{S}_\mathrm{val}$ sets were generated by sampling uniformly with $256$ points $20$ circumferences whose radius was uniformly distributed in the range $[1.5 \mathrm{m}, 3.5 \mathrm{m}]$ from the center of the array. Finally, $|\mathcal{S}_\mathrm{test}|=2560$ sources were used to test and the dataset was created by shifting the sources contained in $\mathcal{S}_\mathrm{train} \cup \mathcal{S}_\mathrm{val}$ by $0.05 \mathrm{m}$, but sampling the $20$ circumferences with $128$ uniformly distributed points. We considered sources emitting a signal with spectrum $A(\omega_k)=1$ at $K=63$ frequencies spaced by $23~\mathrm{Hz}$, in the range between $46~\mathrm{Hz}$ and $1500~\mathrm{Hz}$.

\subsubsection{Results}
In Fig.~\ref{fig:soundfield_example_real_circular}\protect\subref{subfig:sf_circular_real_gt} we show the real part of the ground truth sound pressure distribution for an emitting point source placed in~$\mathbf{r}=[0.99~\mathrm{m}, 2.88~\mathrm{m}, 0~\mathrm{m}]^T$. In Fig.~\ref{fig:soundfield_example_real_circular}\protect\subref{subfig:sf_circular_real_pwd}, Fig.~\ref{fig:soundfield_example_real_circular}\protect\subref{subfig:sf_circular_real_pwd_cnn}, Fig.~\ref{fig:soundfield_example_real_circular}\protect\subref{subfig:sf_circular_real_pm}, Fig.~\ref{fig:soundfield_example_real_circular}\protect\subref{subfig:sf_circular_real_awfs} and Fig.~\ref{fig:soundfield_example_real_circular}\protect\subref{subfig:sf_circular_real_pwd_apwd}, the real part of the sound pressure obtained through $\mathrm{MR}$, $\mathrm{CNN}$, $\mathrm{PM}$, $\mathrm{AWFS}$ and $\mathrm{AMR}$ is shown, respectively when 32 speakers are active. It is clear how the $\mathrm{CNN}$ model performs best, by reducing the number of irregularities in the wavefront, with respect to the $\mathrm{MR}$, $\mathrm{AWFS}$, $\mathrm{AMR}$ techniques and especially with respect to the $\mathrm{PM}$ technique, whose reproduced soundfield is extremely irregular. 
These considerations are also confirmed by inspecting the $\mathrm{NRE}$ obtained for the same scenario, shown in Fig.~\ref{fig:soundfield_example_nre_circular}, where the $\mathrm{NRE}$ in the case of $\mathrm{CNN}$, shown in Fig.~\ref{fig:soundfield_example_nre_circular}\protect\subref{subfig:nmse_circular_real_pwd_cnn}, is sensibly lower in the listening area $\mathcal{A}$ with respect to the ones obtained through $\mathrm{MR}$ and $\mathrm{PM}$, shown in Fig.~\ref{fig:soundfield_example_nre_circular}\protect\subref{subfig:nmse_circular_real_pwd}, Fig.~\ref{fig:soundfield_example_nre_circular}\protect\subref{subfig:nmse_circular_real_pm},~\ref{fig:soundfield_example_nre_circular}\protect\subref{subfig:nmse_circular_real_awfs} and ~\ref{fig:soundfield_example_nre_circular}\protect\subref{subfig:nmse_circular_real_pwd_apwd}, respectively.

In Fig~\ref{fig:results_circular_frequency}\protect\subref{subfig:nre_16_missing_circular}-\protect\subref{subfig:nre_32_missing_circular}-\protect\subref{subfig:nre_48_missing_circular} we present results showing the $\mathrm{NRE}$ averaged over all $|\mathcal{S}_\mathrm{test}|$ sources, when considering an irregular array of $L=48,~32$ and $16$ secondary sources. 
Similarly to the linear array case, the $\mathrm{CNN}$ achieves $\mathrm{NRE}$ results that are on par or better than the other considered techniques. This is more evident when the number of secondary sources is lower. While $\mathrm{MR}$ is approximately constant in the considered frequency range, the error of $\mathrm{CNN}$ tends to increase with the frequency, even if it remains lower than the one of $\mathrm{MR}$. Analogously, $\mathrm{PM}$ exhibits an error that increases with the frequency, becoming extremely irregular for the upper frequency range and more sparse setups.
$\mathrm{AMR}$ shows a behavior similar to $\mathrm{CNN}$ but reaching higher $\mathrm{NRE}$ values. When considering the $\mathrm{AWFS}$ technique, the $\mathrm{CNN}$ technique performs better both in the $L=48$ and $L=32$ cases, while performances when using an array with $L=16$ loudspeaker are practically on par.

In Fig~\ref{fig:results_circular_frequency}\protect\subref{subfig:ssim_16_missing_circular}-\protect\subref{subfig:ssim_32_missing_circular}-\protect\subref{subfig:ssim_48_missing_circular} we present the $\mathrm{SSIM}$ metric averaged over all $|\mathcal{S}_\mathrm{test}|$ sources, when considering an irregular array of $L=48,32$ and $16$ sources, respectively. Differently from the linear array case, the $\mathrm{SSIM}$ obtained through $\mathrm{CNN}$ is similar or better than the other considered methods, especially for higher frequency values. This is probably due to both the smaller listening area considered, allowing for a smaller number of irregularities in the reproduced wavefront, and the fact that the array surrounds the listening area enabling reproduction from a higher number of directions.

In the case of the circular array, we also computed the $\mathrm{NRE}$ and $\mathrm{SSIM}$ when varying the location of the emitting source, in particular when it moves farther from the center of the array in the range $1.5~\mathrm{m}<\rho<3.5~\mathrm{m}$, while keeping the frequency fixed at $1007~\mathrm{Hz}$. The results of the $\mathrm{NRE}$ metric are shown in Fig~\ref{fig:results_circular_radius}\protect\subref{subfig:nre_16_missing_circular_radius}-\protect\subref{subfig:nre_32_missing_circular_radius}-\protect\subref{subfig:nre_48_missing_circular_radius} for the arrays with $48$, $32$ an $16$ secondary sources, respectively. All methods present a mostly constant behavior with respect to the whole considered radius range, with $\mathrm{CNN}$ and $\mathrm{PM}$ the most and less accurate, respectively. As expected the $\mathrm{NRE}$ worsens when decreasing the number of active secondary sources. Coherently with the $\mathrm{NRE}$ results, for $L=48$ the $\mathrm{CNN}$ and $\mathrm{AWFS}$ performances are extemely similar.
The results for the $\mathrm{SSIM}$ metric are shown in Fig~\ref{fig:results_circular_radius}\protect\subref{subfig:ssim_16_missing_circular_radius}-\protect\subref{subfig:ssim_32_missing_circular_radius}-\protect\subref{subfig:ssim_48_missing_circular_radius} for the arrays with $48$, $32$ an $16$ secondary sources, respectively. In this case, the accuracy slightly worsens as the distance of the sources increases. While $\mathrm{CNN}$, $\mathrm{MR}$ and $\mathrm{AWFS}$ are close to each other, 
$\mathrm{AWR}$ and $\mathrm{PM}$ turns out to be the worse.

\subsection{Real Data}
In this section we present results related to soundfield
synthesis when considering a circular array setup and data obtained from Room Impulse Responses (RIRs) measurements contained in the  dataset from~\cite{Zhao2022room}.

\subsubsection{Setup}
RIRs were measured in an anechoic room of size $4.90~\mathrm{m}\times 7.22~\mathrm{m}\times 5.29~\mathrm{m}$ with an average reverberation time of $0.045~\mathrm{s}$ using an array of $L=60$ loudspeakers (Genelec 8010A) with radius of $1.5~\mathrm{m}$, the spacing between each loudspeaker is of approximately $ 0.157~\mathrm{m}$. From this configuration,  three irregular array setups were generated by randomly removing $12$, $28$ or $44$ loudspeakers, resulting in three irregular configurations with $L=48$, $L=32$ and $L=16$ secondary sources, respectively. The RIRs related to the reproduction zone are measured by considering the square microphone (DPA 4060) array configuration, specifically related to the Zone E in~\cite{Zhao2022room}, consisting of $64$ microphones sampling with a spacing of $0.04~\mathrm{m}$ a square of size $0.28~\mathrm{m} \times 0.28~\mathrm{m}$ placed in the center of the area comprised by the microphone array. Both microphones and loudspeakers were placed at the same height of $1.45~\mathrm{m}$ from the floor. A total of $16$ control points inside the reproduction area  were considered in order to compute the losses using the $~\mathrm{CNN}$ model and the driving signals through the $\mathrm{PM}$, $\mathrm{AWFS}$ and $\mathrm{AMR}$ techniques. The considered sampling frequency is of $Fs=48000~\mathrm{Hz}$~\cite{Zhao2022room}.

In order to generate the dataset, we simulated through Pyroomacoustics~\cite{scheibler2018pyroomacoustics} a total of $4264$ point sources placed in a $8~\mathrm{m} \times 8~\mathrm{m}$ grid surrounding the loudspeaker array. The sources were split into $|\mathcal{S}_\mathrm{train}|=1705$, $|\mathcal{S}_\mathrm{val}|=427$ and $|\mathcal{S}_\mathrm{test}|=2132$ to create the training, validation and test sets, respectively. We considered sources emitting a signal with spectrum $A(\omega_k)=1$ at $K=63$ frequencies spaced by $23~\mathrm{Hz}$, in the range between $50~\mathrm{Hz}$ and $1500~\mathrm{Hz}$.

\subsubsection{Results}
\begin{figure}[!t]
\centering
\subfloat[]{%
\includegraphics[width=.45\columnwidth]{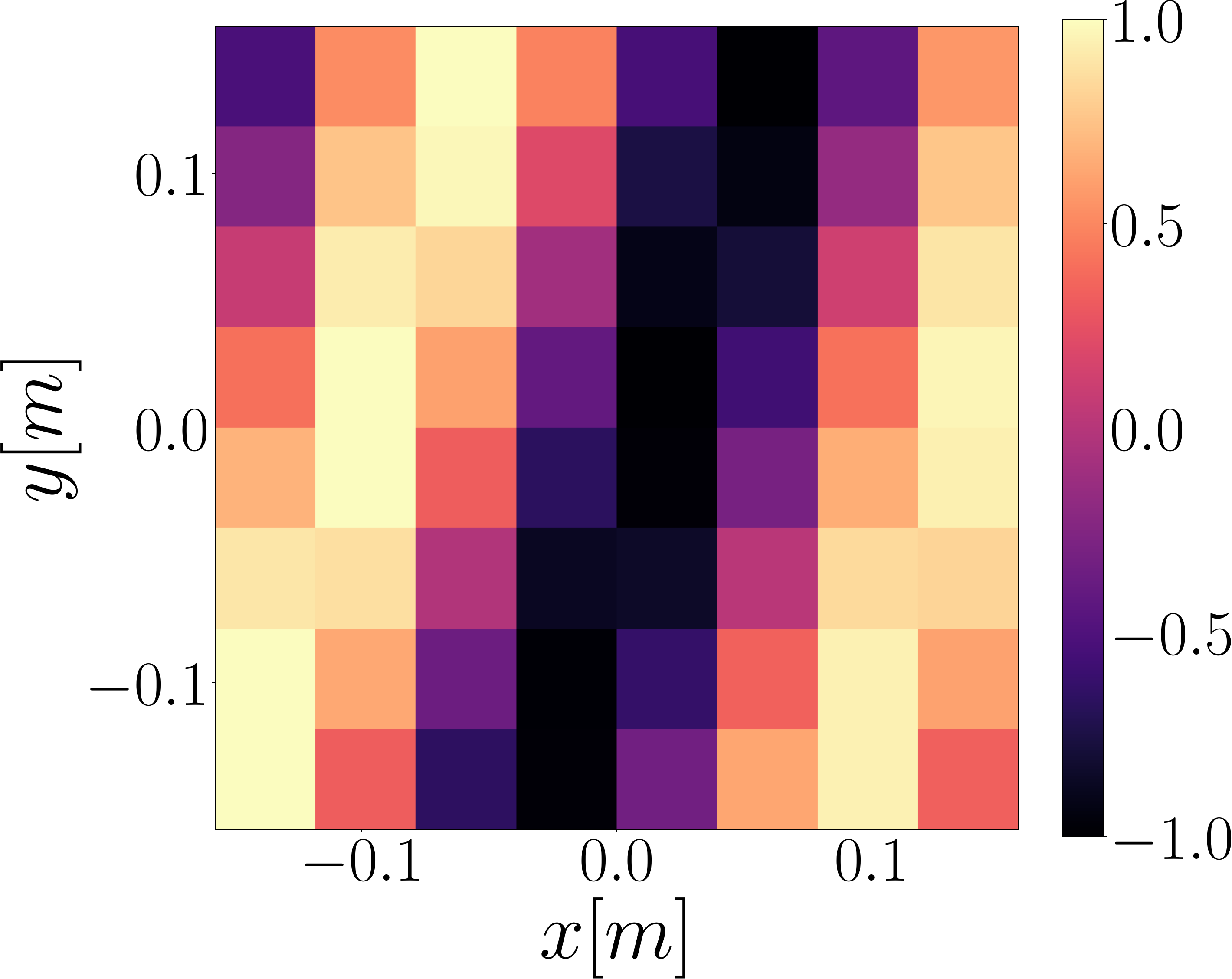}%
\label{subfig:sf_circular_real_gt_EXP}}
\subfloat[]{%
\includegraphics[width=.45\columnwidth]{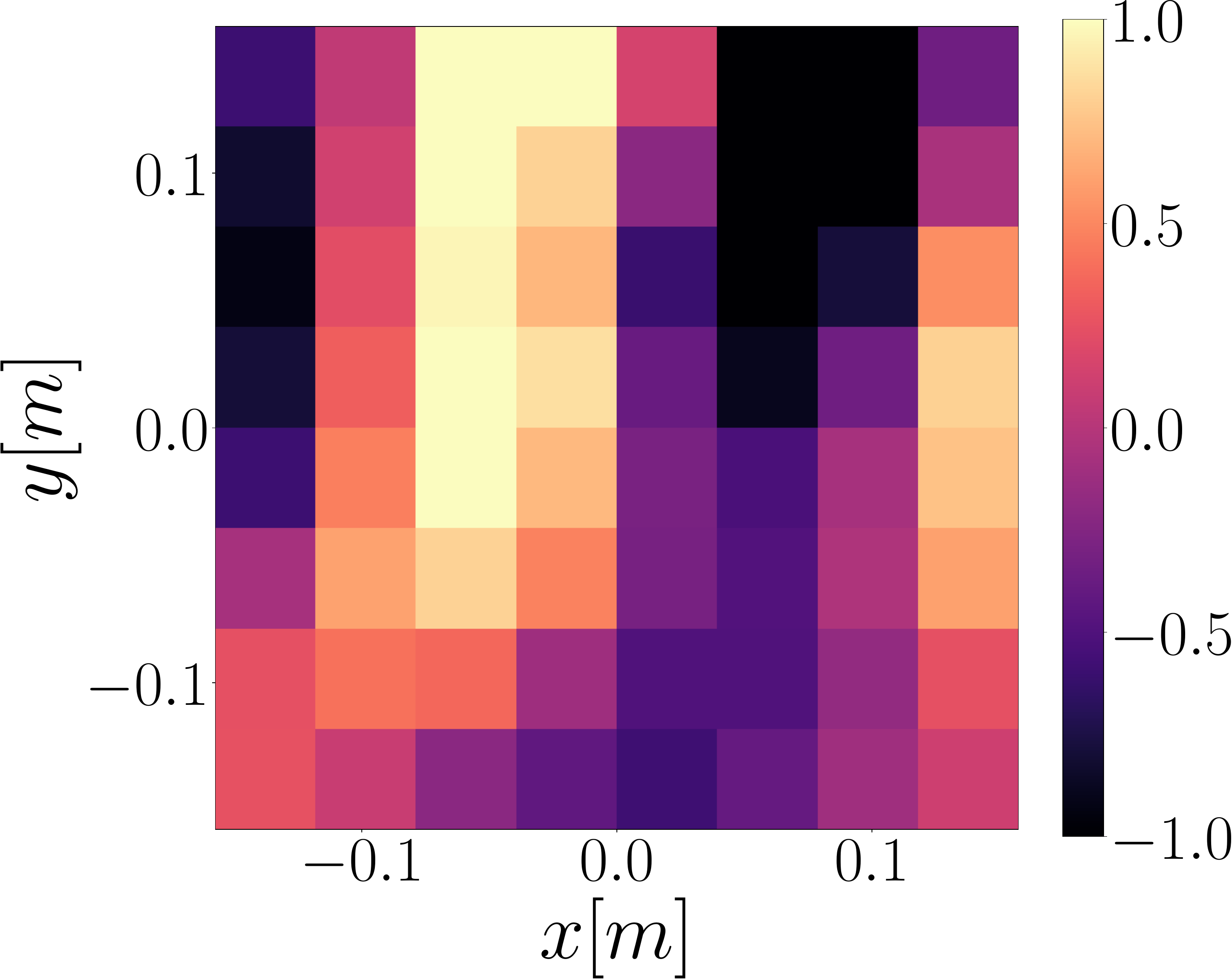}%
\label{subfig:sf_circular_real_pwd_EXP}}
\vspace{-1em}
\subfloat[]{%
\includegraphics[width=.45\columnwidth]{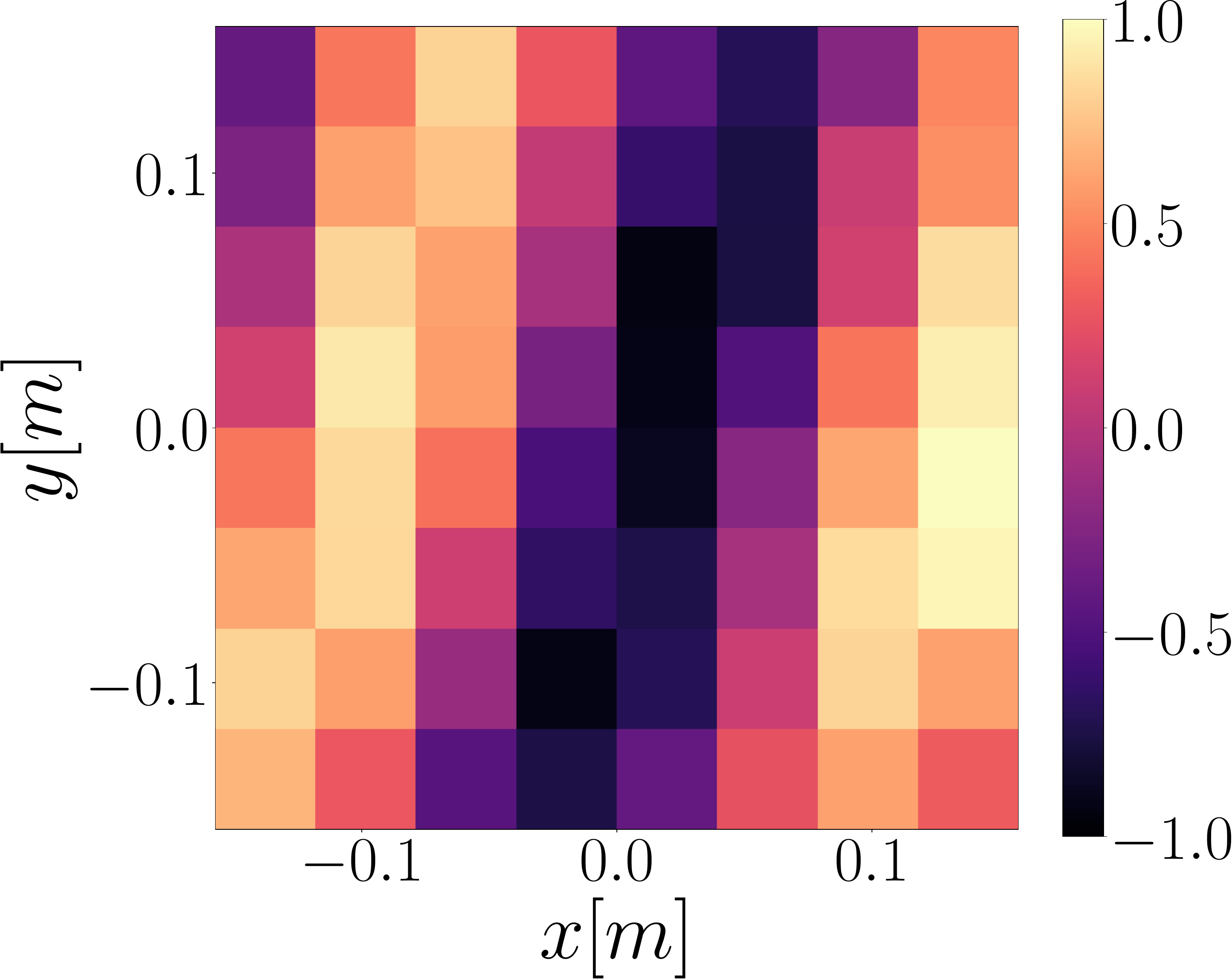}%
\label{subfig:sf_circular_real_pwd_cnn_EXP}}
\subfloat[]{%
\includegraphics[width=.45\columnwidth]{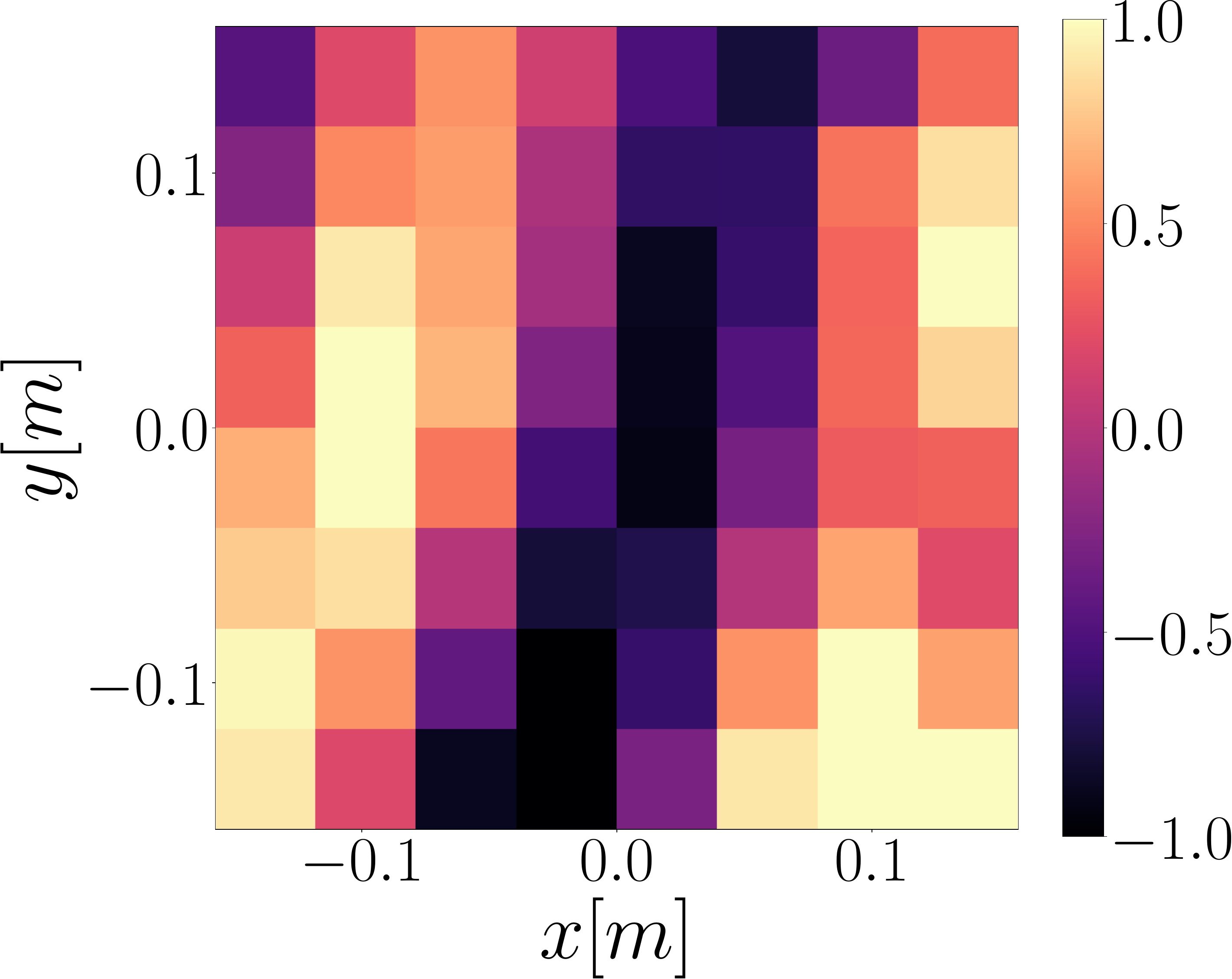}%
\label{subfig:sf_circular_real_pm_EXP}}
\vspace{-1em}
\subfloat[]{%
\includegraphics[width=.45\columnwidth]{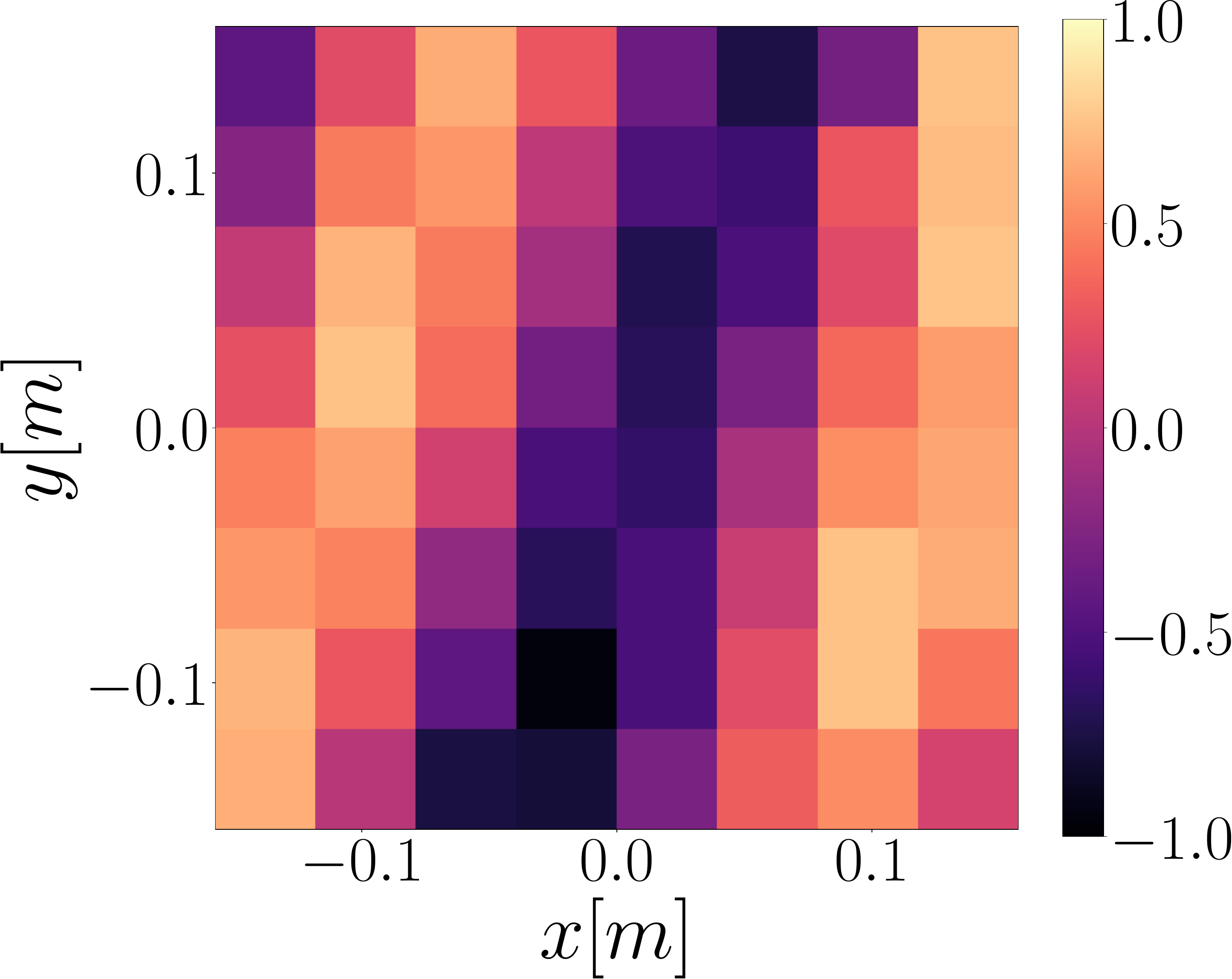}%
\label{subfig:sf_circular_real_awfs_EXP}}
\subfloat[]{%
\includegraphics[width=.45\columnwidth]{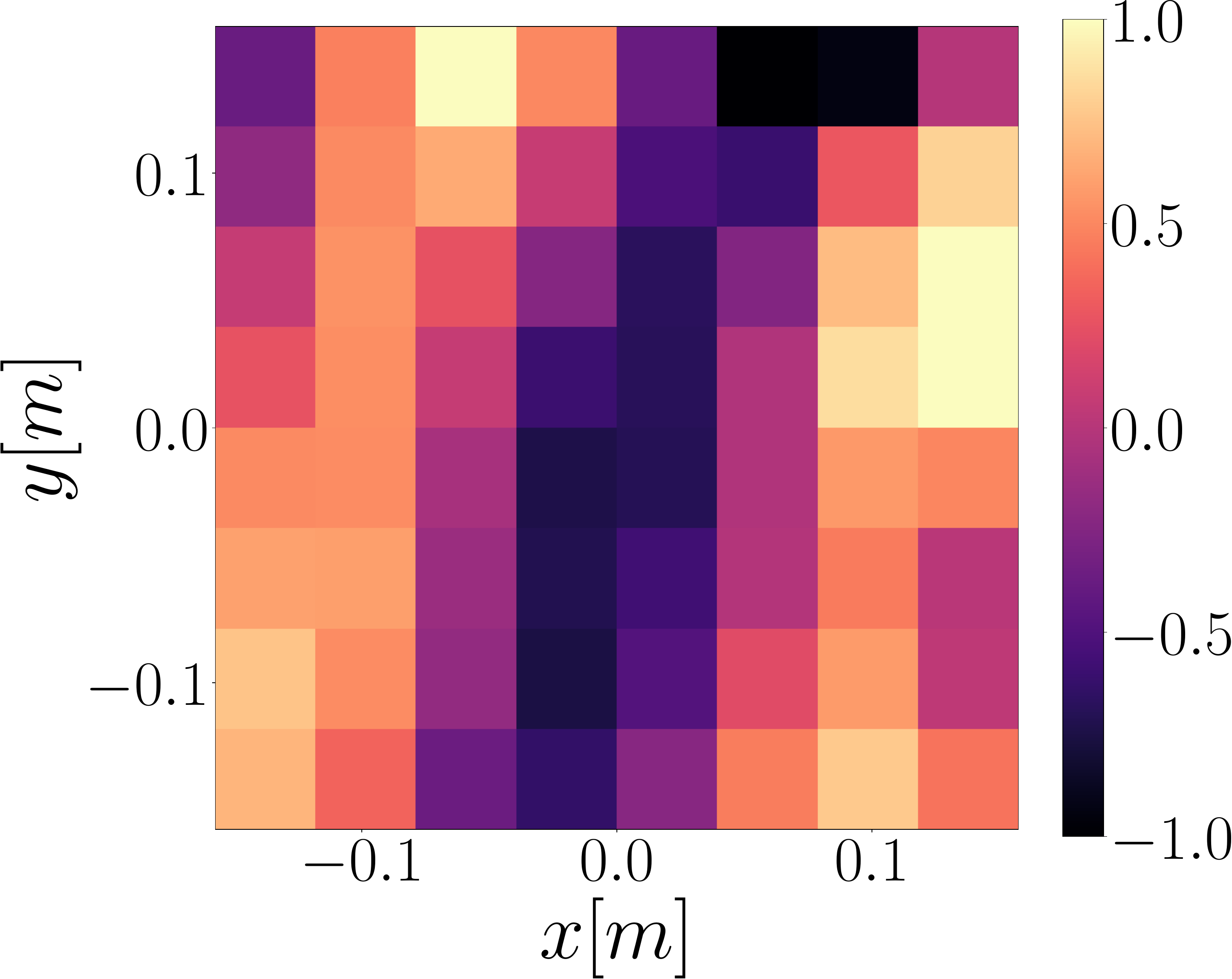}%
\label{subfig:sf_circular_real_pwd_apwd_EXP}}
\caption{Amplitude (real part) of the soundfield for a source placed in $\mathbf{r}=[-3.76, ~\mathrm{m}, -1.14~\mathrm{m}, 0~\mathrm{m}]^T$ at $f=1500~\mathrm{Hz}$ , ground truth is shown in \protect\subref{subfig:sf_circular_real_gt_EXP}. Reproduction performances using the irregular circular array of $L=32$ loudspeakers are shown using $\mathrm{MR}$~\protect\subref{subfig:sf_circular_real_pwd_EXP}, ~\protect\subref{subfig:sf_circular_real_pwd_cnn_EXP}$\mathrm{CNN}$,$\mathrm{PM}$~\protect\subref{subfig:sf_circular_real_pm_EXP},~\protect\subref{subfig:sf_circular_real_awfs_EXP}$\mathrm{AWFS}$ and~\protect\subref{subfig:sf_circular_real_pwd_apwd_EXP}$\mathrm{APWD}$. Black loudspeakers represent the geometry of the chosen array.}
\label{fig:soundfield_example_real_circular_EXP}
\end{figure}
\begin{figure*}[!t]
\centering
\subfloat[]{%
\includegraphics[width=.40\columnwidth]{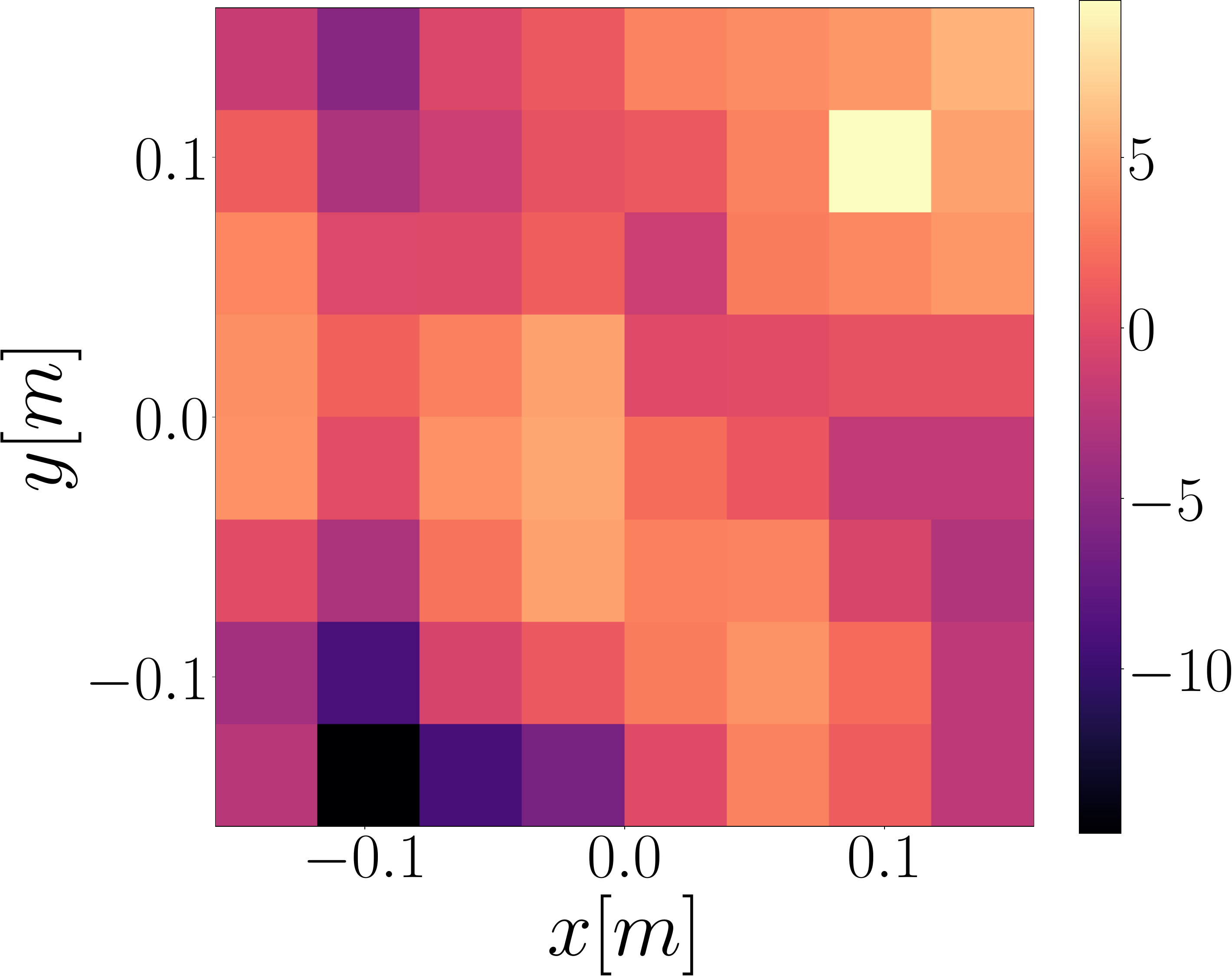}%
\label{subfig:nmse_circular_real_pwd_EXP}}
\hspace{-0.4em}
\subfloat[]{%
\includegraphics[width=.40\columnwidth]{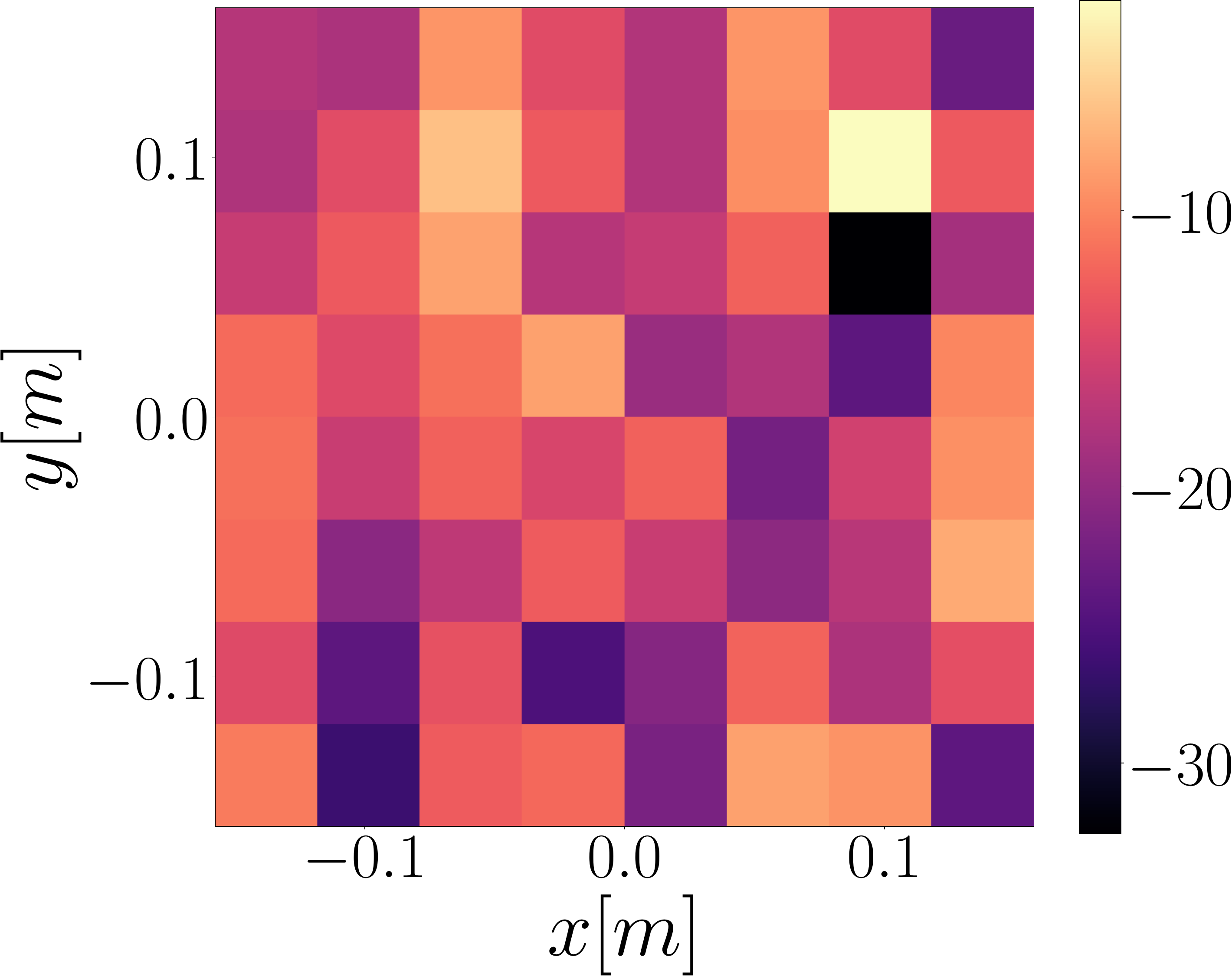}%
\label{subfig:nmse_circular_real_pwd_cnn_EXP}}
\hspace{-0.4em}
\subfloat[]{%
\includegraphics[width=.40\columnwidth]{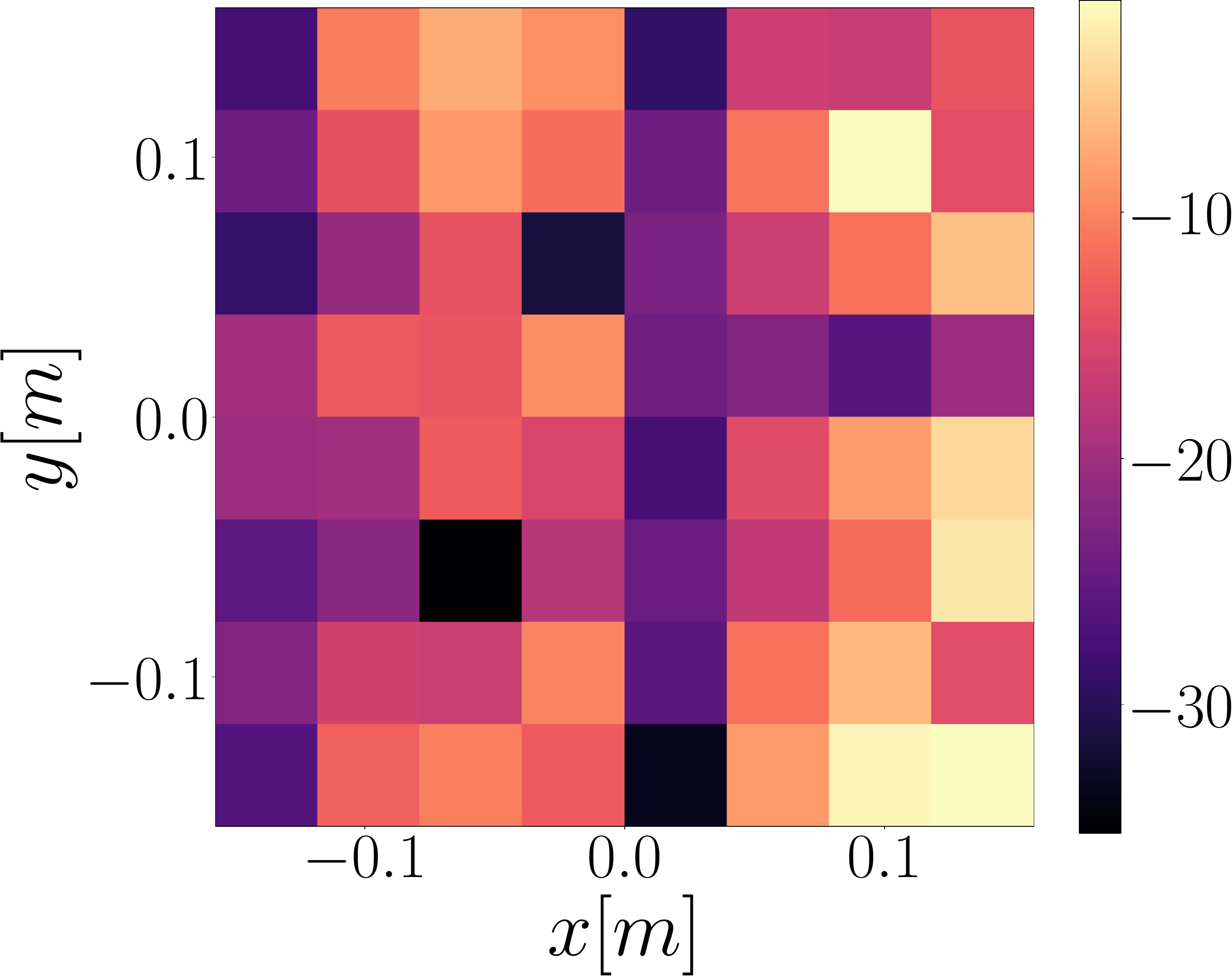}%
\label{subfig:nmse_circular_real_pm_EXP}}
\subfloat[]{%
\includegraphics[width=.40\columnwidth]{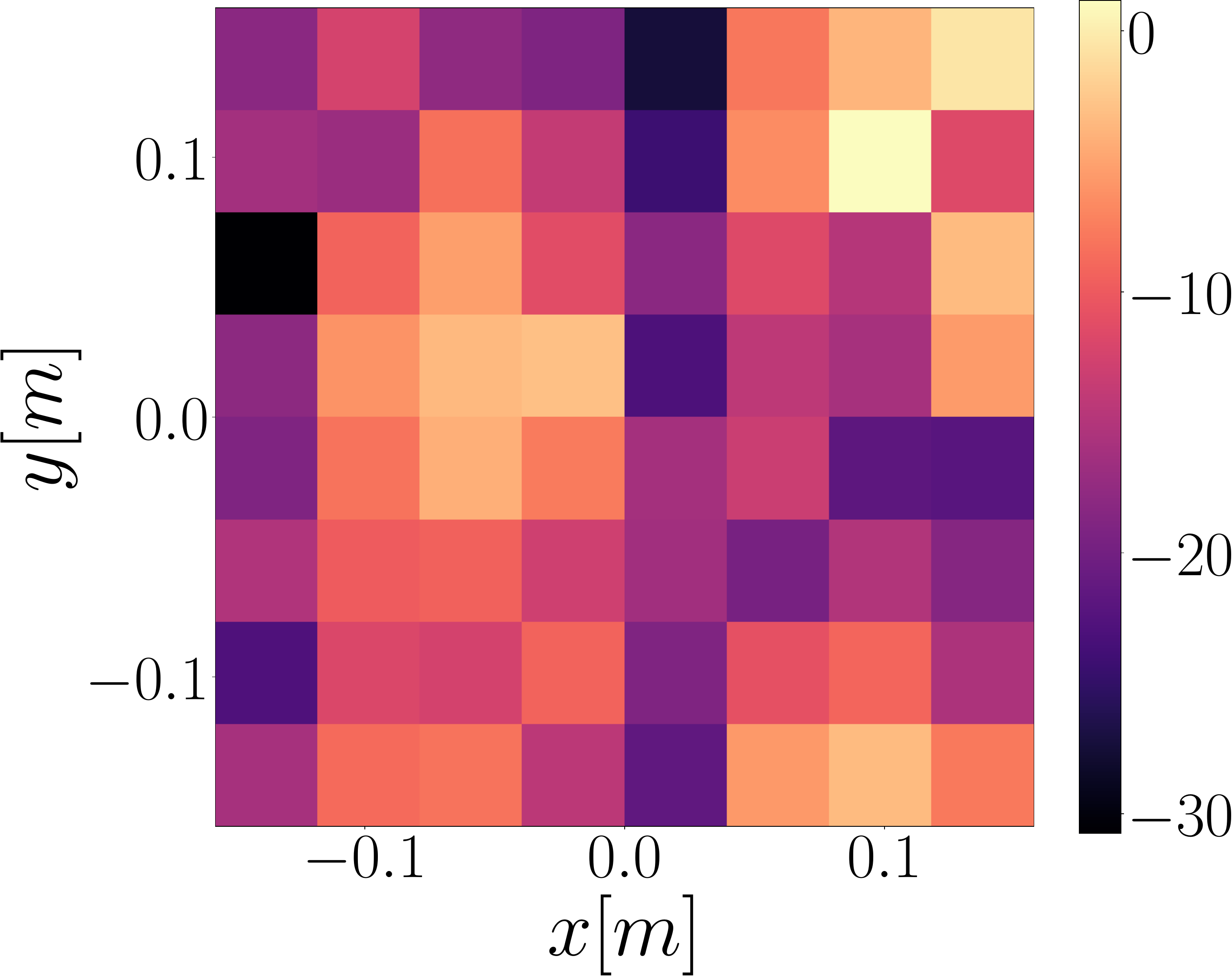}%
\label{subfig:nmse_circular_real_awfs_EXP}}
\hspace{-0.4em}
\subfloat[]{%
\includegraphics[width=.40\columnwidth]{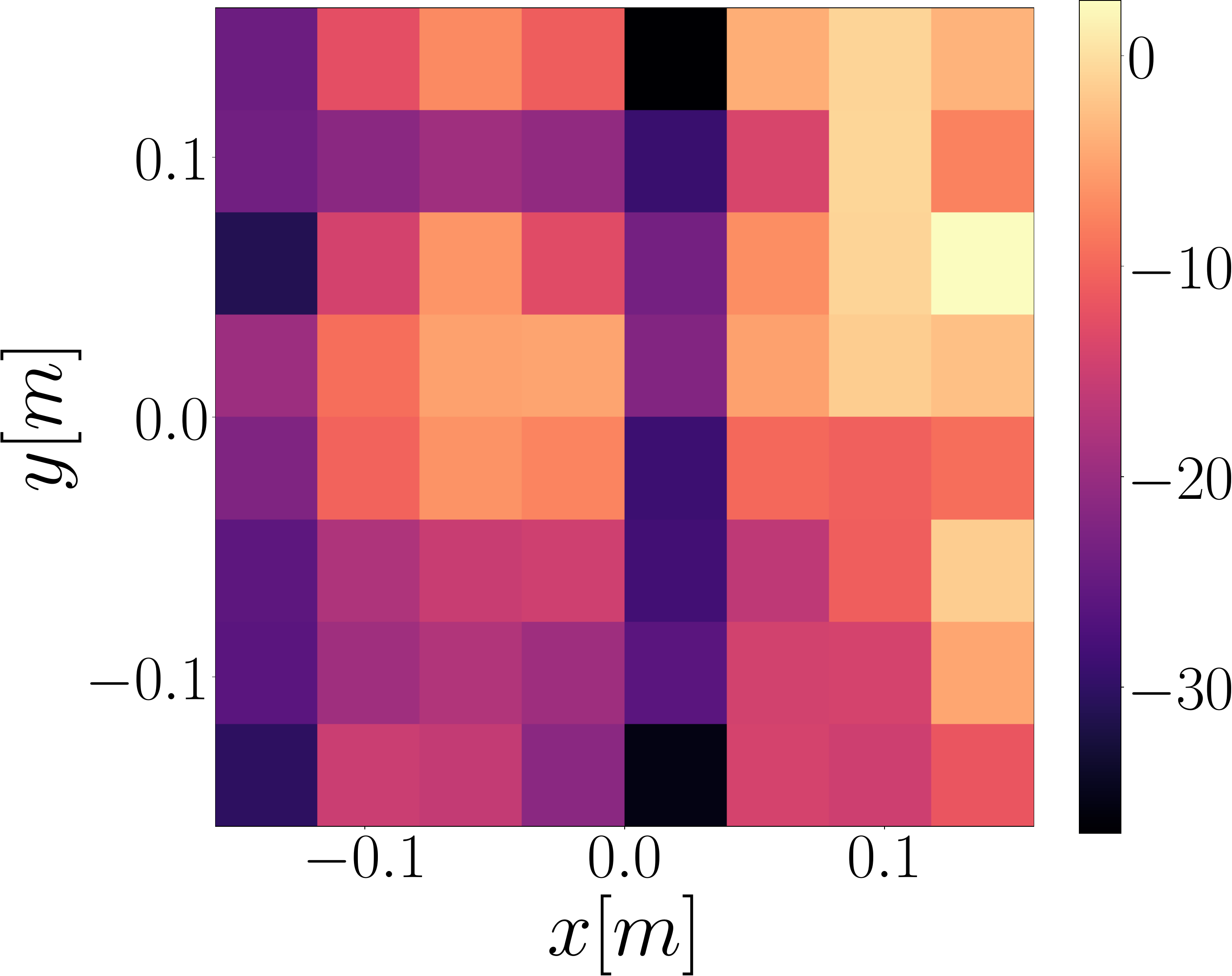}%
\label{subfig:nmse_circular_real_pwd_apwd_EXP}}
\caption{Normalized Reproduction Error (NRE) distribution in $\mathrm{dB}$ for a source placed in $\mathbf{r}=[-3.76, ~\mathrm{m}, -1.14~\mathrm{m}, 0~\mathrm{m}]^T$ at $f=1500~\mathrm{Hz}$ when using: $\mathrm{MR}$~\protect\subref{subfig:nmse_circular_real_pwd_EXP}, $\mathrm{CNN}$ ~\protect\subref{subfig:nmse_circular_real_pwd_cnn_EXP},~$\mathrm{PM}$~\protect\subref{subfig:nmse_circular_real_pm_EXP}, $\mathrm{AWFS}$~\protect\subref{subfig:nmse_circular_real_awfs_EXP} and $\mathrm{AMR}$~\protect\subref{subfig:nmse_circular_real_pwd_apwd_EXP}.
\label{fig:soundfield_example_nre_circular_EXP}}
\end{figure*}
In Fig.~\ref{fig:soundfield_example_real_circular_EXP}\protect\subref{subfig:sf_circular_real_gt_EXP} we show the real part of the ground truth sound pressure distribution for a point source placed in~$\mathbf{r}=[-3.76~\mathrm{m}, -1.14~\mathrm{m}, 0~\mathrm{m}]^T$ at $f=1500~\mathrm{Hz}$. In Fig.~\ref{fig:soundfield_example_real_circular_EXP}\protect\subref{subfig:sf_circular_real_pwd_EXP}, Fig.~\ref{fig:soundfield_example_real_circular_EXP}\protect\subref{subfig:sf_circular_real_pwd_cnn_EXP}, Fig.~\ref{fig:soundfield_example_real_circular_EXP}\protect\subref{subfig:sf_circular_real_pm_EXP}, Fig.~\ref{fig:soundfield_example_real_circular_EXP}\protect\subref{subfig:sf_circular_real_awfs_EXP} and Fig.~\ref{fig:soundfield_example_real_circular_EXP}\protect\subref{subfig:sf_circular_real_pwd_apwd_EXP}, the real part of the sound pressure obtained through $\mathrm{MR}$, $\mathrm{CNN}$, $\mathrm{PM}$,  $\mathrm{AWFS}$ and $\mathrm{AMR}$ is shown, respectively, when 32 speakers are active. We can see that the $\mathrm{CNN}$ technique is the one that is able to better reproduce the soundfield, closely followed by the $\mathrm{AWFS}$, $\mathrm{AMR}$ and $\mathrm{PM}$, the $\mathrm{MR}$ technique is the one that seems to perform worst at generating the desired ground truth soundfield.
Similar considerations can be drawn by inspecting the $\mathrm{NRE}$ obtained for the same scenario, shown in Fig.~\ref{fig:soundfield_example_nre_circular_EXP}, where the $\mathrm{NRE}$ for the  listening area $\mathcal{A}$ in the case of $\mathrm{CNN}$, Fig.~\ref{fig:soundfield_example_nre_circular_EXP}\protect\subref{subfig:nmse_circular_real_pwd_cnn_EXP}, $\mathrm{MR}$ Fig.~\ref{fig:soundfield_example_nre_circular_EXP}\protect\subref{subfig:nmse_circular_real_pwd_EXP}, $\mathrm{PM}$ Fig.~\ref{fig:soundfield_example_nre_circular_EXP}\protect\subref{subfig:nmse_circular_real_pm_EXP}, $\mathrm{AWFS}$ Fig.~\ref{fig:soundfield_example_nre_circular_EXP}\protect\subref{subfig:nmse_circular_real_awfs_EXP} and $\mathrm{AMR}$ ~\ref{fig:soundfield_example_nre_circular_EXP}\protect\subref{subfig:nmse_circular_real_pwd_apwd_EXP}. 

\begin{figure}[!t]
\centering
\subfloat[]{\input{figures/NRE_12_real}%
\label{subfig:nre_12_missing_real}}

\centering
\subfloat[]{\input{figures/NRE_28_real}%
\label{subfig:nre_28_missing_real}}

\centering
\subfloat[]{\input{figures/NRE_44_real}%
\label{subfig:nre_44_missing_real}}

\caption{Irregular circular array soundfield synthesis performances (real measurements) with respect to frequency, NRE when: \protect\subref{subfig:nre_12_missing_real}  $L=48$, \protect\subref{subfig:nre_28_missing_real}  $L=32$, \protect\subref{subfig:nre_44_missing_real} NRE when $L=16$.}
\label{fig:results_real_frequency}
\end{figure}

In Fig~\ref{fig:results_real_frequency}\protect\subref{subfig:nre_12_missing_real}-\protect\subref{subfig:nre_28_missing_real}-\protect\subref{subfig:nre_44_missing_real} we present results showing the $\mathrm{NRE}$ averaged over all $|\mathcal{S}_\mathrm{test}|$ sources, when considering an irregular array of $L=48,~32$ and $16$ secondary sources. 
In the case of $L=48$ $\mathrm{CNN}$, $\mathrm{AMR}$ and  $\mathrm{PM}$ performances are similar under $700~\mathrm{Hz}$, while over this value $\mathrm{CNN}$ is the method that minimizes the $\mathrm{NRE}$ the most. No major difference can be observed for  $L=32$. Finally for what concerns the $L=16$ scenario $~\mathrm{CNN}$ performances are on par with $\mathrm{AMR}$ under $800~\mathrm{Hz}$, for higher values the error obtained with the latter strongly increases. On the other way around, while $\mathrm{CNN}$ performs better than $\mathrm{AWFS}$ under $600~\mathrm{Hz}$, the two methods perform similarly over $800~\mathrm{Hz}$ frequency values, with the latter obtaining slightly better results. The $\mathrm{MR}$ method is the one working worst in all cases. 

We avoid showing the SSIM results due to the fact that being it strongly dependent on the variance of the data it is not representative of the quality of the generated data in this specific case, since the ground truth soundfields are simulated, while the RIRs used for reproduction are measured, causing the data to have significantly different distributions.

%% file: figures/NRE_16_linear.tex
\pgfplotstableread{figures/linear_array/nmse_pwd_freq_db_missing_16.txt}\PWD%
\pgfplotstableread{figures/linear_array/nmse_pwd_cnn_freq_db_missing_16.txt}\PWDCNN%
\pgfplotstableread{figures/linear_array/nmse_pm_freq_db_missing_16.txt}\PM%
\pgfplotstableread{figures/linear_array/nmse_wfs_freq_db_missing_16.txt}\WFS%
\pgfplotstableread{figures/linear_array/nmse_awfs_freq_db_missing_16.txt}\AWFS%
\pgfplotstableread{figures/linear_array/nmse_pwd_apwd_freq_db_missing_16.txt}\APWD%

\begin{tikzpicture}
	\begin{axis}[
	    enlargelimits=false,
    	legend columns=3,
        legend style={at={(1,1)},anchor=south east},
	  	width=\columnwidth,
        height=.45\columnwidth,
	  	xlabel={Frequency},
	    ylabel={NRE},
	    y unit=\decibel,
	    x unit =\hertz,
	    grid=major,
	    legend entries={\scriptsize{$\mathrm{MR}$},\scriptsize{$\mathrm{CNN}$},
	    \scriptsize{$\mathrm{PM}$}, 
     \scriptsize{$\mathrm{AWFS}$}, \scriptsize{$\mathrm{AMR}$}},
	    ]
 	
    \newcommand\marksize{2}
    \newcommand\markrepeat{4}
    \addplot[color=black,mark=*,mark size=\marksize pt,mark options={solid,fill=black,mark repeat=\markrepeat}]table[x=0,y=1] from \PWD;
    \addplot[color=black,mark=square,mark size= \marksize pt,mark options={solid,fill=gray,mark repeat=\markrepeat}]table[x=0,y=1] from \PWDCNN;
    \addplot[color=black,mark=triangle, mark size= \marksize pt,mark options={solid,fill=black,mark repeat=\markrepeat}]table[x=0,y=1] from \PM;
    \addplot[color=black,mark=pentagon, mark size= \marksize pt,mark options={solid,fill=black,mark repeat=\markrepeat}]table[x=0,y=1] from \AWFS;
    \addplot[color=black,mark=diamond, mark size= \marksize pt,mark options={solid,fill=white,mark repeat=\markrepeat}]table[x=0,y=1] from \APWD; 
    
    \end{axis}
\end{tikzpicture}

%% file: figures/SSIM_16_linear.tex
\pgfplotstableread{figures/linear_array/ssim_pwd_freq_missing_16.txt}\PWD%
\pgfplotstableread{figures/linear_array/ssim_pwd_cnn_freq_missing_16.txt}\PWDCNN%
\pgfplotstableread{figures/linear_array/ssim_pm_freq_missing_16.txt}\PM%
\pgfplotstableread{figures/linear_array/ssim_wfs_freq_missing_16.txt}\WFS%
\pgfplotstableread{figures/linear_array/ssim_awfs_freq_missing_16.txt}\AWFS%
\pgfplotstableread{figures/linear_array/ssim_pwd_apwd_freq_missing_16.txt}\APWD%

\begin{tikzpicture}
	\begin{axis}[
	    enlargelimits=false,
    	legend columns=3,
        legend style={at={(1,1)},anchor=south east},
		ymin=0, ymax=1, 
	  	width=\columnwidth,
        height=.45\columnwidth,
	  	xlabel={Frequency},
	    ylabel={SSIM},
	    x unit =\hertz,
	    grid=major,
	    legend entries={\scriptsize{$\mathrm{MR}$},\scriptsize{$\mathrm{CNN}$},
	    \scriptsize{$\mathrm{PM}$}, 
     \scriptsize{$\mathrm{AWFS}$}, \scriptsize{$\mathrm{AMR}$}},
	    ]
 	
    \newcommand\marksize{2}
    \newcommand\markrepeat{4}
    \addplot[color=black,mark=*,mark size=\marksize pt,mark options={solid,fill=black,mark repeat=\markrepeat}]table[x=0,y=1] from \PWD;
    \addplot[color=black,mark=square,mark size= \marksize pt,mark options={solid,fill=gray,mark repeat=\markrepeat}]table[x=0,y=1] from \PWDCNN;
    \addplot[color=black,mark=triangle, mark size= \marksize pt,mark options={solid,fill=black,mark repeat=\markrepeat}]table[x=0,y=1] from \PM;
    \addplot[color=black,mark=pentagon, mark size= \marksize pt,mark options={solid,fill=black,mark repeat=\markrepeat}]table[x=0,y=1] from \AWFS;
    \addplot[color=black,mark=diamond, mark size= \marksize pt,mark options={solid,fill=white,mark repeat=\markrepeat}]table[x=0,y=1] from \APWD;

    \end{axis}
\end{tikzpicture}

%% file: figures/NRE_32_linear.tex
\pgfplotstableread{figures/linear_array/nmse_pwd_freq_db_missing_32.txt}\PWD%
\pgfplotstableread{figures/linear_array/nmse_pwd_cnn_freq_db_missing_32.txt}\PWDCNN%
\pgfplotstableread{figures/linear_array/nmse_pm_freq_db_missing_32.txt}\PM%
\pgfplotstableread{figures/linear_array/nmse_wfs_freq_db_missing_32.txt}\WFS%
\pgfplotstableread{figures/linear_array/nmse_awfs_freq_db_missing_32.txt}\AWFS%
\pgfplotstableread{figures/linear_array/nmse_pwd_apwd_freq_db_missing_32.txt}\APWD%

\begin{tikzpicture}
	\begin{axis}[
	    enlargelimits=false,
    	legend columns=3,
        legend style={at={(1,1)},anchor=south east},
	  	width=\columnwidth,
        height=.45\columnwidth,
	  	xlabel={Frequency},
	    ylabel={NRE},
	    y unit=\decibel,
	    x unit =\hertz,
	    grid=major,
	    legend entries={\scriptsize{$\mathrm{MR}$},\scriptsize{$\mathrm{CNN}$},
	    \scriptsize{$\mathrm{PM}$}, 
     \scriptsize{$\mathrm{AWFS}$}, \scriptsize{$\mathrm{AMR}$}},
	    ]
 	
    \newcommand\marksize{2}
    \newcommand\markrepeat{4}
    \addplot[color=black,mark=*,mark size=\marksize pt,mark options={solid,fill=black,mark repeat=\markrepeat}]table[x=0,y=1] from \PWD;
    \addplot[color=black,mark=square,mark size= \marksize pt,mark options={solid,fill=gray,mark repeat=\markrepeat}]table[x=0,y=1] from \PWDCNN;
    \addplot[color=black,mark=triangle, mark size= \marksize pt,mark options={solid,fill=black,mark repeat=\markrepeat}]table[x=0,y=1] from \PM;
    \addplot[color=black,mark=pentagon, mark size= \marksize pt,mark options={solid,fill=black,mark repeat=\markrepeat}]table[x=0,y=1] from \AWFS;
    \addplot[color=black,mark=diamond, mark size= \marksize pt,mark options={solid,fill=white,mark repeat=\markrepeat}]table[x=0,y=1] from \APWD;   
    
    \end{axis}
\end{tikzpicture}

%% file: figures/SSIM_32_linear.tex
\pgfplotstableread{figures/linear_array/ssim_pwd_freq_missing_32.txt}\PWD%
\pgfplotstableread{figures/linear_array/ssim_pwd_cnn_freq_missing_32.txt}\PWDCNN%
\pgfplotstableread{figures/linear_array/ssim_pm_freq_missing_32.txt}\PM%
\pgfplotstableread{figures/linear_array/ssim_wfs_freq_missing_32.txt}\WFS%
\pgfplotstableread{figures/linear_array/ssim_awfs_freq_missing_32.txt}\AWFS%
\pgfplotstableread{figures/linear_array/ssim_pwd_apwd_freq_missing_32.txt}\APWD%

\begin{tikzpicture}
	\begin{axis}[
	    enlargelimits=false,
    	legend columns=3,
        legend style={at={(1,1)},anchor=south east},
		ymin=0, ymax=1, 
	  	width=\columnwidth,
        height=.45\columnwidth,
	  	xlabel={Frequency},
	    ylabel={SSIM},
	    x unit =\hertz,
	    grid=major,
	    legend entries={\scriptsize{$\mathrm{MR}$},\scriptsize{$\mathrm{CNN}$},
	    \scriptsize{$\mathrm{PM}$}, 
     \scriptsize{$\mathrm{AWFS}$}, \scriptsize{$\mathrm{AMR}$}},
	    ]
 	
    \newcommand\marksize{2}
    \newcommand\markrepeat{4}
    \addplot[color=black,mark=*,mark size=\marksize pt,mark options={solid,fill=black,mark repeat=\markrepeat}]table[x=0,y=1] from \PWD;
    \addplot[color=black,mark=square,mark size= \marksize pt,mark options={solid,fill=gray,mark repeat=\markrepeat}]table[x=0,y=1] from \PWDCNN;
    \addplot[color=black,mark=triangle, mark size= \marksize pt,mark options={solid,fill=black,mark repeat=\markrepeat}]table[x=0,y=1] from \PM;
    \addplot[color=black,mark=pentagon, mark size= \marksize pt,mark options={solid,fill=black,mark repeat=\markrepeat}]table[x=0,y=1] from \AWFS;
    \addplot[color=black,mark=diamond, mark size= \marksize pt,mark options={solid,fill=white,mark repeat=\markrepeat}]table[x=0,y=1] from \APWD;

    \end{axis}
\end{tikzpicture}

%% file: figures/NRE_48_linear.tex
\pgfplotstableread{figures/linear_array/nmse_pwd_freq_db_missing_48.txt}\PWD%
\pgfplotstableread{figures/linear_array/nmse_pwd_cnn_freq_db_missing_48.txt}\PWDCNN%
\pgfplotstableread{figures/linear_array/nmse_pm_freq_db_missing_48.txt}\PM%
\pgfplotstableread{figures/linear_array/nmse_wfs_freq_db_missing_48.txt}\WFS%
\pgfplotstableread{figures/linear_array/nmse_awfs_freq_db_missing_48.txt}\AWFS%
\pgfplotstableread{figures/linear_array/nmse_pwd_apwd_freq_db_missing_48.txt}\APWD%

\begin{tikzpicture}
	\begin{axis}[
	    enlargelimits=false,
    	legend columns=3,
        legend style={at={(1,1)},anchor=south east},
	  	width=\columnwidth,
        height=.45\columnwidth,
	  	xlabel={Frequency},
	    ylabel={NRE},
	    y unit=\decibel,
	    x unit =\hertz,
	    grid=major,
	    legend entries={\scriptsize{$\mathrm{MR}$},\scriptsize{$\mathrm{CNN}$},
	    \scriptsize{$\mathrm{PM}$}, 
     \scriptsize{$\mathrm{AWFS}$}, \scriptsize{$\mathrm{AMR}$}},
	    ]
 	
    \newcommand\marksize{2}
    \newcommand\markrepeat{4}
    \addplot[color=black,mark=*,mark size=\marksize pt,mark options={solid,fill=black,mark repeat=\markrepeat}]table[x=0,y=1] from \PWD;
    \addplot[color=black,mark=square,mark size= \marksize pt,mark options={solid,fill=gray,mark repeat=\markrepeat}]table[x=0,y=1] from \PWDCNN;
    \addplot[color=black,mark=triangle, mark size= \marksize pt,mark options={solid,fill=black,mark repeat=\markrepeat}]table[x=0,y=1] from \PM;
    \addplot[color=black,mark=pentagon, mark size= \marksize pt,mark options={solid,fill=black,mark repeat=\markrepeat}]table[x=0,y=1] from \AWFS;
    \addplot[color=black,mark=diamond, mark size= \marksize pt,mark options={solid,fill=white,mark repeat=\markrepeat}]table[x=0,y=1] from \APWD; 
    
    \end{axis}
\end{tikzpicture}

%% file: figures/SSIM_48_linear.tex
\pgfplotstableread{figures/linear_array/ssim_pwd_freq_missing_48.txt}\PWD%
\pgfplotstableread{figures/linear_array/ssim_pwd_cnn_freq_missing_48.txt}\PWDCNN%
\pgfplotstableread{figures/linear_array/ssim_pm_freq_missing_48.txt}\PM%
\pgfplotstableread{figures/linear_array/ssim_wfs_freq_missing_48.txt}\WFS%
\pgfplotstableread{figures/linear_array/ssim_awfs_freq_missing_48.txt}\AWFS%
\pgfplotstableread{figures/linear_array/ssim_pwd_apwd_freq_missing_48.txt}\APWD%

\begin{tikzpicture}
	\begin{axis}[
	    enlargelimits=false,
    	legend columns=3,
        legend style={at={(1,1)},anchor=south east},
		ymin=0, ymax=1, 
	  	width=\columnwidth,
        height=.45\columnwidth,
	  	xlabel={Frequency},
	    ylabel={SSIM},
	    x unit =\hertz,
	    grid=major,
	    legend entries={\scriptsize{$\mathrm{MR}$},\scriptsize{$\mathrm{CNN}$},
	    \scriptsize{$\mathrm{PM}$}, 
     \scriptsize{$\mathrm{AWFS}$}, \scriptsize{$\mathrm{AMR}$}},
	    ]
 	
    \newcommand\marksize{2}
    \newcommand\markrepeat{4}
    \addplot[color=black,mark=*,mark size=\marksize pt,mark options={solid,fill=gray,mark repeat=\markrepeat}]table[x=0,y=1] from \PWD;
    \addplot[color=black,mark=square,mark size= \marksize pt,mark options={solid,fill=gray,mark repeat=\markrepeat}]table[x=0,y=1] from \PWDCNN;
    \addplot[color=black,mark=triangle, mark size= \marksize pt,mark options={solid,fill=black,mark repeat=\markrepeat}]table[x=0,y=1] from \PM;
    \addplot[color=black,mark=pentagon, mark size= \marksize pt,mark options={solid,fill=black,mark repeat=\markrepeat}]table[x=0,y=1] from \AWFS;
    \addplot[color=black,mark=diamond, mark size= \marksize pt,mark options={solid,fill=white,mark repeat=\markrepeat}]table[x=0,y=1] from \APWD;

    \end{axis}
\end{tikzpicture}

%% file: figures/NRE_16_circular.tex
\pgfplotstableread{figures/circular_array/nmse_pwd_freq_db_missing_16.txt}\PWD%
\pgfplotstableread{figures/circular_array/nmse_pwd_cnn_freq_db_missing_16.txt}\PWDCNN%
\pgfplotstableread{figures/circular_array/nmse_pm_freq_db_missing_16.txt}\PM%
\pgfplotstableread{figures/circular_array/nmse_wfs_freq_db_missing_16.txt}\WFS%
\pgfplotstableread{figures/circular_array/nmse_awfs_freq_db_missing_16.txt}\AWFS%
\pgfplotstableread{figures/circular_array/nmse_pwd_apwd_freq_db_missing_16.txt}\APWD%

\begin{tikzpicture}
	\begin{axis}[
	    enlargelimits=false,
    	legend columns=3,
        legend style={at={(1,1)},anchor=south east},
	  	width=\columnwidth,
        height=.45\columnwidth,
	  	xlabel={Frequency},
	    ylabel={NRE},
	    y unit=\decibel,
	    x unit =\hertz,
	    grid=major,
	    legend entries={\scriptsize{$\mathrm{MR}$},\scriptsize{$\mathrm{CNN}$},
	    \scriptsize{$\mathrm{PM}$}, 
     \scriptsize{$\mathrm{AWFS}$}, \scriptsize{$\mathrm{AMR}$}},
	    ]
 	
    \newcommand\marksize{2}
    \newcommand\markrepeat{4}
    \addplot[color=black,mark=*,mark size=\marksize pt,mark options={solid,fill=white,mark repeat=\markrepeat}]table[x=0,y=1] from \PWD;
    \addplot[color=black,mark=square,mark size= \marksize pt,mark options={solid,fill=white,mark repeat=\markrepeat}]table[x=0,y=1] from \PWDCNN;
    \addplot[color=black,mark=triangle, mark size= \marksize pt,mark options={solid,fill=white,mark repeat=\markrepeat}]table[x=0,y=1] from \PM;
    \addplot[color=black,mark=pentagon, mark size= \marksize pt,mark options={solid,fill=white,mark repeat=\markrepeat}]table[x=0,y=1] from \AWFS;
    \addplot[color=black,mark=diamond, mark size= \marksize pt,mark options={solid,fill=white,mark repeat=\markrepeat}]table[x=0,y=1] from \APWD; 
    
    \end{axis}
\end{tikzpicture}

%% file: figures/SSIM_16_circular.tex
\pgfplotstableread{figures/circular_array/ssim_pwd_freq_missing_16.txt}\PWD%
\pgfplotstableread{figures/circular_array/ssim_pwd_cnn_freq_missing_16.txt}\PWDCNN%
\pgfplotstableread{figures/circular_array/ssim_pm_freq_missing_16.txt}\PM%
\pgfplotstableread{figures/circular_array/ssim_wfs_freq_missing_16.txt}\WFS%
\pgfplotstableread{figures/circular_array/ssim_awfs_freq_missing_16.txt}\AWFS%
\pgfplotstableread{figures/circular_array/ssim_pwd_apwd_freq_missing_16.txt}\APWD%

\begin{tikzpicture}
	\begin{axis}[
	    enlargelimits=false,
    	legend columns=3,
        legend style={at={(1,1)},anchor=south east},
		ymin=0, ymax=1, 
	  	width=\columnwidth,
        height=.45\columnwidth,
	  	xlabel={Frequency},
	    ylabel={SSIM},
	    x unit =\hertz,
	    grid=major,
	    legend entries={\scriptsize{$\mathrm{MR}$},\scriptsize{$\mathrm{CNN}$},
	    \scriptsize{$\mathrm{PM}$}, 
     \scriptsize{$\mathrm{AWFS}$}, \scriptsize{$\mathrm{AMR}$}},
	    ]
 	
    \newcommand\marksize{2}
    \newcommand\markrepeat{4}
    \addplot[color=black,mark=*,mark size=\marksize pt,mark options={solid,fill=white,mark repeat=\markrepeat}]table[x=0,y=1] from \PWD;
    \addplot[color=black,mark=square,mark size= \marksize pt,mark options={solid,fill=white,mark repeat=\markrepeat}]table[x=0,y=1] from \PWDCNN;
    \addplot[color=black,mark=triangle, mark size= \marksize pt,mark options={solid,fill=white,mark repeat=\markrepeat}]table[x=0,y=1] from \PM;
    \addplot[color=black,mark=pentagon, mark size= \marksize pt,mark options={solid,fill=white,mark repeat=\markrepeat}]table[x=0,y=1] from \AWFS;
    \addplot[color=black,mark=diamond, mark size= \marksize pt,mark options={solid,fill=white,mark repeat=\markrepeat}]table[x=0,y=1] from \APWD;

    \end{axis}
\end{tikzpicture}

%% file: figures/NRE_32_circular.tex
\pgfplotstableread{figures/circular_array/nmse_pwd_freq_db_missing_32.txt}\PWD%
\pgfplotstableread{figures/circular_array/nmse_pwd_cnn_freq_db_missing_32.txt}\PWDCNN%
\pgfplotstableread{figures/circular_array/nmse_pm_freq_db_missing_32.txt}\PM%
\pgfplotstableread{figures/circular_array/nmse_wfs_freq_db_missing_32.txt}\WFS%
\pgfplotstableread{figures/circular_array/nmse_awfs_freq_db_missing_32.txt}\AWFS%
\pgfplotstableread{figures/circular_array/nmse_pwd_apwd_freq_db_missing_32.txt}\APWD%

\begin{tikzpicture}
	\begin{axis}[
	    enlargelimits=false,
    	legend columns=3,
        legend style={at={(1,1)},anchor=south east},
	  	width=\columnwidth,
        height=.45\columnwidth,
	  	xlabel={Frequency},
	    ylabel={NRE},
	    y unit=\decibel,
	    x unit =\hertz,
	    grid=major,
	    legend entries={\scriptsize{$\mathrm{MR}$},\scriptsize{$\mathrm{CNN}$},
	    \scriptsize{$\mathrm{PM}$}, 
     \scriptsize{$\mathrm{AWFS}$}, \scriptsize{$\mathrm{AMR}$}},
	    ]
 	
    \newcommand\marksize{2}
    \newcommand\markrepeat{4}
    \addplot[color=black,mark=*,mark size=\marksize pt,mark options={solid,fill=white,mark repeat=\markrepeat}]table[x=0,y=1] from \PWD;
    \addplot[color=black,mark=square,mark size= \marksize pt,mark options={solid,fill=white,mark repeat=\markrepeat}]table[x=0,y=1] from \PWDCNN;
    \addplot[color=black,mark=triangle, mark size= \marksize pt,mark options={solid,fill=white,mark repeat=\markrepeat}]table[x=0,y=1] from \PM;
    \addplot[color=black,mark=pentagon, mark size= \marksize pt,mark options={solid,fill=white,mark repeat=\markrepeat}]table[x=0,y=1] from \AWFS;
    \addplot[color=black,mark=diamond, mark size= \marksize pt,mark options={solid,fill=white,mark repeat=\markrepeat}]table[x=0,y=1] from \APWD;   
    
    \end{axis}
\end{tikzpicture}

%% file: figures/SSIM_32_circular.tex
\pgfplotstableread{figures/circular_array/ssim_pwd_freq_missing_32.txt}\PWD%
\pgfplotstableread{figures/circular_array/ssim_pwd_cnn_freq_missing_32.txt}\PWDCNN%
\pgfplotstableread{figures/circular_array/ssim_pm_freq_missing_32.txt}\PM%
\pgfplotstableread{figures/circular_array/ssim_wfs_freq_missing_32.txt}\WFS%
\pgfplotstableread{figures/circular_array/ssim_awfs_freq_missing_32.txt}\AWFS%
\pgfplotstableread{figures/circular_array/ssim_pwd_apwd_freq_missing_32.txt}\APWD%

\begin{tikzpicture}
	\begin{axis}[
	    enlargelimits=false,
    	legend columns=3,
        legend style={at={(1,1)},anchor=south east},
		ymin=0, ymax=1, 
	  	width=\columnwidth,
        height=.45\columnwidth,
	  	xlabel={Frequency},
	    ylabel={SSIM},
	    x unit =\hertz,
	    grid=major,
	    legend entries={\scriptsize{$\mathrm{MR}$},\scriptsize{$\mathrm{CNN}$},
	    \scriptsize{$\mathrm{PM}$}, 
     \scriptsize{$\mathrm{AWFS}$}, \scriptsize{$\mathrm{AMR}$}},
	    ]
 	
    \newcommand\marksize{2}
    \newcommand\markrepeat{4}
    \addplot[color=black,mark=*,mark size=\marksize pt,mark options={solid,fill=white,mark repeat=\markrepeat}]table[x=0,y=1] from \PWD;
    \addplot[color=black,mark=square,mark size= \marksize pt,mark options={solid,fill=white,mark repeat=\markrepeat}]table[x=0,y=1] from \PWDCNN;
    \addplot[color=black,mark=triangle, mark size= \marksize pt,mark options={solid,fill=white,mark repeat=\markrepeat}]table[x=0,y=1] from \PM;
    \addplot[color=black,mark=pentagon, mark size= \marksize pt,mark options={solid,fill=white,mark repeat=\markrepeat}]table[x=0,y=1] from \AWFS;
    \addplot[color=black,mark=diamond, mark size= \marksize pt,mark options={solid,fill=white,mark repeat=\markrepeat}]table[x=0,y=1] from \APWD;

    \end{axis}
\end{tikzpicture}

%% file: figures/NRE_48_circular.tex
\pgfplotstableread{figures/circular_array/nmse_pwd_freq_db_missing_48.txt}\PWD%
\pgfplotstableread{figures/circular_array/nmse_pwd_cnn_freq_db_missing_48.txt}\PWDCNN%
\pgfplotstableread{figures/circular_array/nmse_pm_freq_db_missing_48.txt}\PM%
\pgfplotstableread{figures/circular_array/nmse_wfs_freq_db_missing_48.txt}\WFS%
\pgfplotstableread{figures/circular_array/nmse_awfs_freq_db_missing_48.txt}\AWFS%
\pgfplotstableread{figures/circular_array/nmse_pwd_apwd_freq_db_missing_48.txt}\APWD%

\begin{tikzpicture}
	\begin{axis}[
	    enlargelimits=false,
    	legend columns=3,
        legend style={at={(1,1)},anchor=south east},
	  	width=\columnwidth,
        height=.45\columnwidth,
	  	xlabel={Frequency},
	    ylabel={NRE},
	    y unit=\decibel,
	    x unit =\hertz,
	    grid=major,
	    legend entries={\scriptsize{$\mathrm{MR}$},\scriptsize{$\mathrm{CNN}$},
	    \scriptsize{$\mathrm{PM}$}, 
     \scriptsize{$\mathrm{AWFS}$}, \scriptsize{$\mathrm{AMR}$}},
	    ]
 	
    \newcommand\marksize{2}
    \newcommand\markrepeat{4}
    \addplot[color=black,mark=*,mark size=\marksize pt,mark options={solid,fill=white,mark repeat=\markrepeat}]table[x=0,y=1] from \PWD;
    \addplot[color=black,mark=square,mark size= \marksize pt,mark options={solid,fill=white,mark repeat=\markrepeat}]table[x=0,y=1] from \PWDCNN;
    \addplot[color=black,mark=triangle, mark size= \marksize pt,mark options={solid,fill=white,mark repeat=\markrepeat}]table[x=0,y=1] from \PM;
    \addplot[color=black,mark=pentagon, mark size= \marksize pt,mark options={solid,fill=white,mark repeat=\markrepeat}]table[x=0,y=1] from \AWFS;
    \addplot[color=black,mark=diamond, mark size= \marksize pt,mark options={solid,fill=white,mark repeat=\markrepeat}]table[x=0,y=1] from \APWD; 
    
    \end{axis}
\end{tikzpicture}

%% file: figures/SSIM_48_circular.tex
\pgfplotstableread{figures/circular_array/ssim_pwd_freq_missing_48.txt}\PWD%
\pgfplotstableread{figures/circular_array/ssim_pwd_cnn_freq_missing_48.txt}\PWDCNN%
\pgfplotstableread{figures/circular_array/ssim_pm_freq_missing_48.txt}\PM%
\pgfplotstableread{figures/circular_array/ssim_wfs_freq_missing_48.txt}\WFS%
\pgfplotstableread{figures/circular_array/ssim_awfs_freq_missing_48.txt}\AWFS%
\pgfplotstableread{figures/circular_array/ssim_pwd_apwd_freq_missing_48.txt}\APWD%

\begin{tikzpicture}
	\begin{axis}[
	    enlargelimits=false,
    	legend columns=3,
        legend style={at={(1,1)},anchor=south east},
		ymin=0, ymax=1, 
	  	width=\columnwidth,
        height=.45\columnwidth,
	  	xlabel={Frequency},
	    ylabel={SSIM},
	    x unit =\hertz,
	    grid=major,
	    legend entries={\scriptsize{$\mathrm{MR}$},\scriptsize{$\mathrm{CNN}$},
	    \scriptsize{$\mathrm{PM}$}, 
     \scriptsize{$\mathrm{AWFS}$}, \scriptsize{$\mathrm{AMR}$}},
	    ]
 	
    \newcommand\marksize{2}
    \newcommand\markrepeat{4}
    \addplot[color=black,mark=*,mark size=\marksize pt,mark options={solid,fill=white,mark repeat=\markrepeat}]table[x=0,y=1] from \PWD;
    \addplot[color=black,mark=square,mark size= \marksize pt,mark options={solid,fill=white,mark repeat=\markrepeat}]table[x=0,y=1] from \PWDCNN;
    \addplot[color=black,mark=triangle, mark size= \marksize pt,mark options={solid,fill=white,mark repeat=\markrepeat}]table[x=0,y=1] from \PM;
    \addplot[color=black,mark=pentagon, mark size= \marksize pt,mark options={solid,fill=white,mark repeat=\markrepeat}]table[x=0,y=1] from \AWFS;
    \addplot[color=black,mark=diamond, mark size= \marksize pt,mark options={solid,fill=white,mark repeat=\markrepeat}]table[x=0,y=1] from \APWD;

    \end{axis}
\end{tikzpicture}

%% file: figures/NRE_16_circular_radius.tex
\pgfplotstableread{figures/circular_array/nmse_pwd_radius_db_missing_16.txt}\PWD%
\pgfplotstableread{figures/circular_array/nmse_pwd_cnn_radius_db_missing_16.txt}\PWDCNN%
\pgfplotstableread{figures/circular_array/nmse_pm_radius_db_missing_16.txt}\PM%
\pgfplotstableread{figures/circular_array/nmse_wfs_radius_db_missing_16.txt}\WFS%
\pgfplotstableread{figures/circular_array/nmse_awfs_radius_db_missing_16.txt}\AWFS%
\pgfplotstableread{figures/circular_array/nmse_pwd_apwd_radius_db_missing_16.txt}\APWD%

\begin{tikzpicture}
	\begin{axis}[
	    enlargelimits=false,
    	legend columns=3,
        legend style={at={(1,1)},anchor=south east},
	  	width=.85\columnwidth,
        height=.45\columnwidth,
	  	xlabel={$\rho$},
	    ylabel={NRE},
	    y unit=\decibel,
	    x unit =\metre,
	    grid=major,
	    legend entries={\scriptsize{$\mathrm{MR}$},\scriptsize{$\mathrm{CNN}$},
	    \scriptsize{$\mathrm{PM}$}, 
     \scriptsize{$\mathrm{AWFS}$}, \scriptsize{$\mathrm{AMR}$}},
	    ]
 	
    \newcommand\marksize{2}
    \newcommand\markrepeat{4}
    \addplot[color=black,mark=*,mark size=\marksize pt,mark options={solid,fill=black,mark repeat=\markrepeat}]table[x=0,y=1] from \PWD;
    \addplot[color=black,mark=square,mark size= \marksize pt,mark options={solid,fill=gray,mark repeat=\markrepeat}]table[x=0,y=1] from \PWDCNN;
    \addplot[color=black,mark=triangle, mark size= \marksize pt,mark options={solid,fill=black,mark repeat=\markrepeat}]table[x=0,y=1] from \PM;
    \addplot[color=black,mark=pentagon, mark size= \marksize pt,mark options={solid,fill=black,mark repeat=\markrepeat}]table[x=0,y=1] from \AWFS;
    \addplot[color=black,mark=diamond, mark size= \marksize pt,mark options={solid,fill=white,mark repeat=\markrepeat}]table[x=0,y=1] from \APWD;

    \end{axis}
\end{tikzpicture}

%% file: figures/SSIM_16_circular_radius.tex
\pgfplotstableread{figures/circular_array/ssim_pwd_radius_missing_16.txt}\PWD%
\pgfplotstableread{figures/circular_array/ssim_pwd_cnn_radius_missing_16.txt}\PWDCNN%
\pgfplotstableread{figures/circular_array/ssim_pm_radius_missing_16.txt}\PM%
\pgfplotstableread{figures/circular_array/ssim_wfs_radius_missing_16.txt}\WFS%
\pgfplotstableread{figures/circular_array/ssim_awfs_radius_missing_16.txt}\AWFS%
\pgfplotstableread{figures/circular_array/ssim_pwd_apwd_radius_missing_16.txt}\APWD%

\begin{tikzpicture}
	\begin{axis}[
	    enlargelimits=false,
    	legend columns=3,
        legend style={at={(1,1)},anchor=south east},
		ymin=0, ymax=1, 
	  	width=.85\columnwidth,
        height=.45\columnwidth,
	  	xlabel={$\rho$},
	    ylabel={SSIM},
	    x unit =\metre,
	    grid=major,
	    legend entries={\scriptsize{$\mathrm{MR}$},\scriptsize{$\mathrm{CNN}$},
	    \scriptsize{$\mathrm{PM}$}, 
     \scriptsize{$\mathrm{AWFS}$}, \scriptsize{$\mathrm{AMR}$}},
	    ]
 	
    \newcommand\marksize{2}
    \newcommand\markrepeat{4}
    \addplot[color=black,mark=*,mark size=\marksize pt,mark options={solid,fill=black,mark repeat=\markrepeat}]table[x=0,y=1] from \PWD;
    \addplot[color=black,mark=square,mark size= \marksize pt,mark options={solid,fill=gray,mark repeat=\markrepeat}]table[x=0,y=1] from \PWDCNN;
    \addplot[color=black,mark=triangle, mark size= \marksize pt,mark options={solid,fill=black,mark repeat=\markrepeat}]table[x=0,y=1] from \PM;
    \addplot[color=black,mark=pentagon, mark size= \marksize pt,mark options={solid,fill=black,mark repeat=\markrepeat}]table[x=0,y=1] from \AWFS;
    \addplot[color=black,mark=diamond, mark size= \marksize pt,mark options={solid,fill=white,mark repeat=\markrepeat}]table[x=0,y=1] from \APWD; 

    \end{axis}
\end{tikzpicture}

%% file: figures/NRE_32_circular_radius.tex
\pgfplotstableread{figures/circular_array/nmse_pwd_radius_db_missing_32.txt}\PWD%
\pgfplotstableread{figures/circular_array/nmse_pwd_cnn_radius_db_missing_32.txt}\PWDCNN%
\pgfplotstableread{figures/circular_array/nmse_pm_radius_db_missing_32.txt}\PM%
\pgfplotstableread{figures/circular_array/nmse_wfs_radius_db_missing_32.txt}\WFS%
\pgfplotstableread{figures/circular_array/nmse_awfs_radius_db_missing_32.txt}\AWFS%
\pgfplotstableread{figures/circular_array/nmse_pwd_apwd_radius_db_missing_32.txt}\APWD%

\begin{tikzpicture}
	\begin{axis}[
	    enlargelimits=false,
    	legend columns=3,
        legend style={at={(1,1)},anchor=south east},
	  	width=.85\columnwidth,
        height=.45\columnwidth,
	  	xlabel={$\rho$},
	    ylabel={NRE},
	    y unit=\decibel,
	    x unit =\metre,
	    grid=major,
	    legend entries={\scriptsize{$\mathrm{MR}$},\scriptsize{$\mathrm{CNN}$},
	    \scriptsize{$\mathrm{PM}$}, 
     \scriptsize{$\mathrm{AWFS}$}, \scriptsize{$\mathrm{AMR}$}},
	    ]
 	
    \newcommand\marksize{2}
    \newcommand\markrepeat{4}
    \addplot[color=black,mark=*,mark size=\marksize pt,mark options={solid,fill=black,mark repeat=\markrepeat}]table[x=0,y=1] from \PWD;
    \addplot[color=black,mark=square,mark size= \marksize pt,mark options={solid,fill=gray,mark repeat=\markrepeat}]table[x=0,y=1] from \PWDCNN;
    \addplot[color=black,mark=triangle, mark size= \marksize pt,mark options={solid,fill=black,mark repeat=\markrepeat}]table[x=0,y=1] from \PM;
    \addplot[color=black,mark=pentagon, mark size= \marksize pt,mark options={solid,fill=black,mark repeat=\markrepeat}]table[x=0,y=1] from \AWFS;
    \addplot[color=black,mark=diamond, mark size= \marksize pt,mark options={solid,fill=white,mark repeat=\markrepeat}]table[x=0,y=1] from \APWD;

    \end{axis}
\end{tikzpicture}

%% file: figures/SSIM_32_circular_radius.tex
\pgfplotstableread{figures/circular_array/ssim_pwd_radius_missing_32.txt}\PWD%
\pgfplotstableread{figures/circular_array/ssim_pwd_cnn_radius_missing_32.txt}\PWDCNN%
\pgfplotstableread{figures/circular_array/ssim_pm_radius_missing_32.txt}\PM%
\pgfplotstableread{figures/circular_array/ssim_wfs_radius_missing_32.txt}\WFS%
\pgfplotstableread{figures/circular_array/ssim_awfs_radius_missing_32.txt}\AWFS%
\pgfplotstableread{figures/circular_array/ssim_pwd_apwd_radius_missing_32.txt}\APWD%

\begin{tikzpicture}
	\begin{axis}[
	    enlargelimits=false,
    	legend columns=3,
        legend style={at={(1,1)},anchor=south east},
		ymin=0, ymax=1, 
	  	width=.85\columnwidth,
        height=.45\columnwidth,
	  	xlabel={$\rho$},
	    ylabel={SSIM},
	    x unit =\metre,
	    grid=major,
	    legend entries={\scriptsize{$\mathrm{MR}$},\scriptsize{$\mathrm{CNN}$},
	    \scriptsize{$\mathrm{PM}$}, 
     \scriptsize{$\mathrm{AWFS}$}, \scriptsize{$\mathrm{AMR}$}},
	    ]
 	
    \newcommand\marksize{2}
    \newcommand\markrepeat{4}
    \addplot[color=black,mark=*,mark size=\marksize pt,mark options={solid,fill=black,mark repeat=\markrepeat}]table[x=0,y=1] from \PWD;
    \addplot[color=black,mark=square,mark size= \marksize pt,mark options={solid,fill=gray,mark repeat=\markrepeat}]table[x=0,y=1] from \PWDCNN;
    \addplot[color=black,mark=triangle, mark size= \marksize pt,mark options={solid,fill=black,mark repeat=\markrepeat}]table[x=0,y=1] from \PM;
    \addplot[color=black,mark=pentagon, mark size= \marksize pt,mark options={solid,fill=black,mark repeat=\markrepeat}]table[x=0,y=1] from \AWFS;
    \addplot[color=black,mark=diamond, mark size= \marksize pt,mark options={solid,fill=white,mark repeat=\markrepeat}]table[x=0,y=1] from \APWD; 

    \end{axis}
\end{tikzpicture}

%% file: figures/NRE_48_circular_radius.tex
\pgfplotstableread{figures/circular_array/nmse_pwd_radius_db_missing_48.txt}\PWD%
\pgfplotstableread{figures/circular_array/nmse_pwd_cnn_radius_db_missing_48.txt}\PWDCNN%
\pgfplotstableread{figures/circular_array/nmse_pm_radius_db_missing_48.txt}\PM%
\pgfplotstableread{figures/circular_array/nmse_wfs_radius_db_missing_48.txt}\WFS%
\pgfplotstableread{figures/circular_array/nmse_awfs_radius_db_missing_48.txt}\AWFS%
\pgfplotstableread{figures/circular_array/nmse_pwd_apwd_radius_db_missing_48.txt}\APWD%

\begin{tikzpicture}
	\begin{axis}[
	    enlargelimits=false,
    	legend columns=3,
        legend style={at={(1,1)},anchor=south east},
	  	width=.85\columnwidth,
        height=.45\columnwidth,
	  	xlabel={$\rho$},
	    ylabel={NRE},
	    y unit=\decibel,
	    x unit =\metre,
	    grid=major,
	    legend entries={\scriptsize{$\mathrm{MR}$},\scriptsize{$\mathrm{CNN}$},
	    \scriptsize{$\mathrm{PM}$}, 
     \scriptsize{$\mathrm{AWFS}$}, \scriptsize{$\mathrm{AMR}$}},
	    ]
 	
    \newcommand\marksize{2}
    \newcommand\markrepeat{4}
    \addplot[color=black,mark=*,mark size=\marksize pt,mark options={solid,fill=black,mark repeat=\markrepeat}]table[x=0,y=1] from \PWD;
    \addplot[color=black,mark=square,mark size= \marksize pt,mark options={solid,fill=gray,mark repeat=\markrepeat}]table[x=0,y=1] from \PWDCNN;
    \addplot[color=black,mark=triangle, mark size= \marksize pt,mark options={solid,fill=black,mark repeat=\markrepeat}]table[x=0,y=1] from \PM;
    \addplot[color=black,mark=pentagon, mark size= \marksize pt,mark options={solid,fill=black,mark repeat=\markrepeat}]table[x=0,y=1] from \AWFS;
    \addplot[color=black,mark=diamond, mark size= \marksize pt,mark options={solid,fill=white,mark repeat=\markrepeat}]table[x=0,y=1] from \APWD;

    \end{axis}
\end{tikzpicture}

%% file: figures/SSIM_48_circular_radius.tex
\pgfplotstableread{figures/circular_array/ssim_pwd_radius_missing_48.txt}\PWD%
\pgfplotstableread{figures/circular_array/ssim_pwd_cnn_radius_missing_48.txt}\PWDCNN%
\pgfplotstableread{figures/circular_array/ssim_pm_radius_missing_48.txt}\PM%
\pgfplotstableread{figures/circular_array/ssim_wfs_radius_missing_48.txt}\WFS%
\pgfplotstableread{figures/circular_array/ssim_awfs_radius_missing_48.txt}\AWFS%
\pgfplotstableread{figures/circular_array/ssim_pwd_apwd_radius_missing_48.txt}\APWD%

\begin{tikzpicture}
	\begin{axis}[
	    enlargelimits=false,
    	legend columns=3,
        legend style={at={(1,1)},anchor=south east},
		ymin=0, ymax=1, 
	  	width=.85\columnwidth,
        height=.45\columnwidth,
	  	xlabel={$\rho$},
	    ylabel={SSIM},
	    x unit =\metre,
	    grid=major,
	    legend entries={\scriptsize{$\mathrm{MR}$},\scriptsize{$\mathrm{CNN}$},
	    \scriptsize{$\mathrm{PM}$}, 
     \scriptsize{$\mathrm{AWFS}$}, \scriptsize{$\mathrm{AMR}$}},
	    ]
 	
    \newcommand\marksize{2}
    \newcommand\markrepeat{4}
    \addplot[color=black,mark=*,mark size=\marksize pt,mark options={solid,fill=black,mark repeat=\markrepeat}]table[x=0,y=1] from \PWD;
    \addplot[color=black,mark=square,mark size= \marksize pt,mark options={solid,fill=gray,mark repeat=\markrepeat}]table[x=0,y=1] from \PWDCNN;
    \addplot[color=black,mark=triangle, mark size= \marksize pt,mark options={solid,fill=black,mark repeat=\markrepeat}]table[x=0,y=1] from \PM;
    \addplot[color=black,mark=pentagon, mark size= \marksize pt,mark options={solid,fill=black,mark repeat=\markrepeat}]table[x=0,y=1] from \AWFS;
    \addplot[color=black,mark=diamond, mark size= \marksize pt,mark options={solid,fill=white,mark repeat=\markrepeat}]table[x=0,y=1] from \APWD; 

    \end{axis}
\end{tikzpicture}

%% file: figures/NRE_12_real.tex
\pgfplotstableread{figures/real_array/nmse_pwd_freq_db_missing_12.txt}\PWD%
\pgfplotstableread{figures/real_array/nmse_pwd_cnn_freq_db_missing_12.txt}\PWDCNN%
\pgfplotstableread{figures/real_array/nmse_pm_freq_db_missing_12.txt}\PM%
\pgfplotstableread{figures/real_array/nmse_wfs_freq_db_missing_12.txt}\WFS%
\pgfplotstableread{figures/real_array/nmse_awfs_freq_db_missing_12.txt}\AWFS%
\pgfplotstableread{figures/real_array/nmse_pwd_apwd_freq_db_missing_12.txt}\APWD%

\begin{tikzpicture}
	\begin{axis}[
	    enlargelimits=false,
    	legend columns=3,
        legend style={at={(1,1)},anchor=south east},
	  	width=.95\columnwidth,
        height=.45\columnwidth,
	  	xlabel={Frequency},
	    ylabel={NRE},
	    y unit=\decibel,
	    x unit =\hertz,
	    grid=major,
	    legend entries={\scriptsize{$\mathrm{MR}$},\scriptsize{$\mathrm{CNN}$},
	    \scriptsize{$\mathrm{PM}$}, 
     \scriptsize{$\mathrm{AWFS}$}, \scriptsize{$\mathrm{AMR}$}},
	    ]
 	
    \newcommand\marksize{2}
    \newcommand\markrepeat{4}
    \addplot[color=black,mark=*,mark size=\marksize pt,mark options={solid,fill=white,mark repeat=\markrepeat}]table[x=0,y=1] from \PWD;
    \addplot[color=black,mark=square,mark size= \marksize pt,mark options={solid,fill=white,mark repeat=\markrepeat}]table[x=0,y=1] from \PWDCNN;
    \addplot[color=black,mark=triangle, mark size= \marksize pt,mark options={solid,fill=white,mark repeat=\markrepeat}]table[x=0,y=1] from \PM;
    \addplot[color=black,mark=pentagon, mark size= \marksize pt,mark options={solid,fill=white,mark repeat=\markrepeat}]table[x=0,y=1] from \AWFS;
    \addplot[color=black,mark=diamond, mark size= \marksize pt,mark options={solid,fill=white,mark repeat=\markrepeat}]table[x=0,y=1] from \APWD;   
    
    \end{axis}
\end{tikzpicture}

%% file: figures/NRE_28_real.tex
\pgfplotstableread{figures/real_array/nmse_pwd_freq_db_missing_28.txt}\PWD%
\pgfplotstableread{figures/real_array/nmse_pwd_cnn_freq_db_missing_28.txt}\PWDCNN%
\pgfplotstableread{figures/real_array/nmse_pm_freq_db_missing_28.txt}\PM%
\pgfplotstableread{figures/real_array/nmse_wfs_freq_db_missing_28.txt}\WFS%
\pgfplotstableread{figures/real_array/nmse_awfs_freq_db_missing_28.txt}\AWFS%
\pgfplotstableread{figures/real_array/nmse_pwd_apwd_freq_db_missing_28.txt}\APWD%

\begin{tikzpicture}
	\begin{axis}[
	    enlargelimits=false,
    	legend columns=3,
        legend style={at={(1,1)},anchor=south east},
	  	width=.95\columnwidth,
        height=.45\columnwidth,
	  	xlabel={Frequency},
	    ylabel={NRE},
	    y unit=\decibel,
	    x unit =\hertz,
	    grid=major,
	    legend entries={\scriptsize{$\mathrm{MR}$},\scriptsize{$\mathrm{CNN}$},
	    \scriptsize{$\mathrm{PM}$}, 
     \scriptsize{$\mathrm{AWFS}$}, \scriptsize{$\mathrm{AMR}$}},
	    ]
 	
    \newcommand\marksize{2}
    \newcommand\markrepeat{4}
    \addplot[color=black,mark=*,mark size=\marksize pt,mark options={solid,fill=white,mark repeat=\markrepeat}]table[x=0,y=1] from \PWD;
    \addplot[color=black,mark=square,mark size= \marksize pt,mark options={solid,fill=white,mark repeat=\markrepeat}]table[x=0,y=1] from \PWDCNN;
    \addplot[color=black,mark=triangle, mark size= \marksize pt,mark options={solid,fill=white,mark repeat=\markrepeat}]table[x=0,y=1] from \PM;
    \addplot[color=black,mark=pentagon, mark size= \marksize pt,mark options={solid,fill=white,mark repeat=\markrepeat}]table[x=0,y=1] from \AWFS;
    \addplot[color=black,mark=diamond, mark size= \marksize pt,mark options={solid,fill=white,mark repeat=\markrepeat}]table[x=0,y=1] from \APWD;   
    
    \end{axis}
\end{tikzpicture}

%% file: figures/NRE_44_real.tex
\pgfplotstableread{figures/real_array/nmse_pwd_freq_db_missing_44.txt}\PWD%
\pgfplotstableread{figures/real_array/nmse_pwd_cnn_freq_db_missing_44.txt}\PWDCNN%
\pgfplotstableread{figures/real_array/nmse_pm_freq_db_missing_44.txt}\PM%
\pgfplotstableread{figures/real_array/nmse_wfs_freq_db_missing_44.txt}\WFS%
\pgfplotstableread{figures/real_array/nmse_awfs_freq_db_missing_44.txt}\AWFS%
\pgfplotstableread{figures/real_array/nmse_pwd_apwd_freq_db_missing_44.txt}\APWD%

\begin{tikzpicture}
	\begin{axis}[
	    enlargelimits=false,
    	legend columns=3,
        legend style={at={(1,1)},anchor=south east},
	  	width=.95\columnwidth,
        height=.45\columnwidth,
	  	xlabel={Frequency},
	    ylabel={NRE},
	    y unit=\decibel,
	    x unit =\hertz,
	    grid=major,
	    legend entries={\scriptsize{$\mathrm{MR}$},\scriptsize{$\mathrm{CNN}$},
	    \scriptsize{$\mathrm{PM}$}, 
     \scriptsize{$\mathrm{AWFS}$}, \scriptsize{$\mathrm{AMR}$}},
	    ]
 	
    \newcommand\marksize{2}
    \newcommand\markrepeat{4}
    \addplot[color=black,mark=*,mark size=\marksize pt,mark options={solid,fill=white,mark repeat=\markrepeat}]table[x=0,y=1] from \PWD;
    \addplot[color=black,mark=square,mark size= \marksize pt,mark options={solid,fill=white,mark repeat=\markrepeat}]table[x=0,y=1] from \PWDCNN;
    \addplot[color=black,mark=triangle, mark size= \marksize pt,mark options={solid,fill=white,mark repeat=\markrepeat}]table[x=0,y=1] from \PM;
    \addplot[color=black,mark=pentagon, mark size= \marksize pt,mark options={solid,fill=white,mark repeat=\markrepeat}]table[x=0,y=1] from \AWFS;
    \addplot[color=black,mark=diamond, mark size= \marksize pt,mark options={solid,fill=white,mark repeat=\markrepeat}]table[x=0,y=1] from \APWD;   
    
    \end{axis}
\end{tikzpicture}